\documentclass[twocol]{ametsocV6.1}

\usepackage{lineno}
\usepackage[greek, english]{babel}
\usepackage[  
    colorlinks=true,
    linkcolor=red
]{hyperref}
\usepackage{academicons}
\usepackage{amsmath}
\usepackage{graphicx}
\newcommand{\ZH}{\ensuremath{Z_{\mathrm{H}}}}
\newcommand{\ZDR}{\ensuremath{Z_{\mathrm{DR}}}}
\newcommand{\KDP}{\ensuremath{K_{\mathrm{DP}}}}
\newcommand{\rhv}{\ensuremath{\rho_{\mathrm{HV}}}}
\newcommand{\VR}{\ensuremath{V_{\mathrm{R}}}}
\newcommand{\VSR}{\ensuremath{V_{\mathrm{SR}}}}
\newcommand{\Vrot}{\ensuremath{V_{\mathrm{rot}}}}
\newcommand{\zrad}{\ensuremath{\zeta_{\mathrm{rad}}}}
\newcommand{\BRM}{B2K\ensuremath{_{\mathrm{RM}}}}
\newcommand{\ms}{m\,s\ensuremath{^{-1}}}
\newcommand{\mss}{m\ensuremath{^{2}}\,s\ensuremath{^{-2}}}
\newcommand{\Jkg}{J\,kg\ensuremath{^{-1}}}
\newcommand{\si}{s\ensuremath{^{-1}}}
\newcommand{\dg}{\ensuremath{^{\circ}}}
\definecolor{orcidlogocol}{HTML}{A6CE39}

\title{Localized Tornado Outbreak at the Upstream of a Tropical Easterly Wave in Camarines Norte, Philippines (13 September 2025)}

\authors{Generich H. Capuli\href{https://orcid.org/0000-0003-1253-7043}{\includegraphics[scale=0.5]{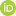}}\aff{a,b}\correspondingauthor{Generich H. Capuli, genecapuli@gmail.com / generich.capuli@pagasa.dost.gov.ph}
}
\affiliation{\aff{a}{Numerical Modeling Section, Research $\&$ Development and Training Division, Department of Science and Technology - Philippine Atmospheric, Geophysical, and Astronomical Services Administration, Brgy. Central, Quezon City 1100, Metro Manila, Philippines}\\
\aff{b}{Project Severe Weather Archive of the Philippines, Quezon City, Philippines}\\
}

\abstract{On 13 September 2025 at around 22 UTC, a localized tornado outbreak affected Camarines Norte in the eastern Philippines. The event developed within an atypical easterly severe weather regime, with warm, moist southeasterly low-level flow and modest low-level shear ahead of an approaching easterly wave. The upstream segment of the inverted trough focused convective initiation within a zone of vorticity convergence that may have enhanced low-level rotation. MUCAPE reached 2167 \Jkg, LCLs were below 200 m, and the moist layer was deep, accompanied by modestly curved but strongly streamwise hodographs that favored supercells and tornadogenesis. At least five vortices were identified. Three were tornadoes produced by supercells: the Magang tornado was confirmed through a ground damage survey and Doppler radar analysis, while the Cahabaan and Napilihan tornadoes were identified primarily through radar signatures. On the International Fujita (IF) Scale, the Magang tornado was rated IF2.5 along a $\sim$1.4 km path, and the Cahabaan ($\sim$2.1 km path) and Napilihan tornadoes were each rated IF1, while the remaining two vortices were rated IF0. The three tornadoes occurred near-simultaneously from semi-discrete supercells in a common mesoscale environment, and the Magang and Cahabaan storms were close enough that inflow--outflow interactions may have aided tornadogenesis. Dual-polarization radar revealed \ZDR\ and \KDP\ columns. The Magang tornado showed a velocity couplet co-located with a cross-correlation ratio (\rhv) depression consistent with a tornadic debris signature, while the Cahabaan supercell displayed a well-defined bounded weak echo region. Despite the limitations of the available observations and the incompleteness of the national tornado record, integrating multiple meteorological datasets with a ground damage survey allows this study to document what is, to our knowledge, the first well-observed tornado outbreak with simultaneous tornadic supercells reported in the Philippines within an easterly severe weather regime.
}

\begin{document}

\maketitle

\statement
\begin{itemize}
\item Documents the first well-observed Philippine tornado outbreak from near-simultaneous supercells.
\item Formed in an atypical easterly severe weather regime with warm, moist flow and modest low-level wind shear.
\item Local terrain interactions may have enhanced low-level convergence and vorticity, aiding tornadogenesis.
\item Sufficient instability, very low LCLs and a modestly curved but strongly streamwise hodograph supported tornadic supercells.
\item Dual-polarization radar revealed BWER, both \ZDR\ and \KDP\ columns, and tornado debris signatures.
\end{itemize}

\section{INTRODUCTION}

Tornado-producing storms occur annually in the Philippine Archipelago; particularly within the Greater Metro Manila Region (GMMR), Western Visayas, and Cotabato provinces. These hazards are by-products of the combination of ample vertical wind shear and atmospheric instability, including lifting mechanism (i.e., forcing) either from plain-to-mountain flow\footnote{Especially for GMMR and Cotabato provinces as they straddle between two mountainous ranges.} or lowering of the low- and mid-level heights \citep[e.g., mountains, tropical cyclones, easterly waves;][and references therein]{Kirshbaum2018,Capuli2024,Capuli2026}. Although previous climatological analyses using documented severe weather reports have emphasized the spatial and temporal occurrences of these severe weather events \citep[hereafter SWE;][]{Capuli2024}, the propensity of these severe convective storms remains poorly understood in the region. Still, the documentation of tornadoes remains serendipitous, occurring typically when tornadoes strike cities amidst the mostly sparsely populated areas \citep{Anderson2007,SilvaDias2011,Oliveira2022}. There is evidence, though, that tornadic events in rural lands and forests occur regularly but are rarely reported, especially when the damage is slight or restricted to vegetation with few damage indicators or low population density \citep{Dyer1988,Wagner2019,Wurman2021}\footnote{Also known as the Population Density Effect.}.

Many studies have addressed the relationship between severe thunderstorms capable of producing tornadoes or tornado outbreaks and associated environmental variables \citep[e.g.,][]{Thomspon2012,Lee2013,Coniglio2020,BAGAGLINI2022,Pilguj2022b}. This multi-decadal research interest emerged because tornadic storms are among the most devastating natural phenomena, and while local, they cause immense social and material impacts \citep[e.g.,][]{Antonescu2017,AndersonFrey2021,Pucik2024}. However, limited tornado documentation in the Philippines has left a clear gap in our understanding of the environments and storm characteristics associated with tornadic events \citep{Capuli2024a,Capuli2024}. This gap stands in contrast to regions where tornado environments have been more comprehensively described, including the United States \citep{Evans2001,Bunkers2006,Boustead2013} and Europe \citep{Groenemeijer2014,Antonescu2017,Antonescu2019}. More recently, Mexico has also received increased attention, offering a useful point of comparison given its tropical setting and temporal variation in SWEs \citep{LeonCruz2022,LEONCRUZ2025}.

Key ingredients for severe storm development include: (1) sufficient atmospheric instability, (2) strong vertical wind shear, (3) abundant moisture in both low-levels and boundary layer, and (4) a forcing mechanism to trigger deep moist convection, e.g., an orographic feature, frontal system, dryline, or low-level convergence zone \citep{Rasmussen2003,Thompson2003,Mercer2012,Sherburn2014}. Furthermore, both \citet{markowski2010} and \citet{DAVIESJONES2015} asserted that tornadic thunderstorms that evolve from supercells are more likely to form in environments characterized by high values of convective available potential energy (CAPE) and sufficient deep-layer shear (DLS; refers to the magnitude of the vector difference at the surface and 6 km AGL) which governs the amount of tropospheric instability and mid-level rotation supporting long-lived supercells, respectively \citep{Rasmussen1998,Brooks2003,Coffer2019}. However, environments conducive to tornadogenesis are not exclusively limited to the isolated supercells. Some Mesoscale Convective Systems (hereafter MCSs) might produce tornadoes as revealed by observations \citep{Nielsen2020} and numerical simulations \citep{Nielsen2018}.

To better understand SWEs, it is often necessary to draw on established case studies that have applied similar approaches. Numerous investigations worldwide demonstrate how synoptic-scale and mesoscale environments shape tornado-producing storms. As an example, \citet{Thompson2000} analyzed the 3 May 1999 event in the U.~S. Great Plains that spawned the Bridge-Creek, Oklahoma tornado, \citet{Lopes2025} documented the 11--12 June 2018 tornado outbreak event in the Argentina-Brazil border, and \citet{Choo2018} examined the 6 May 2012 significant tornado in Japan. All aforementioned authors concluded that tornadic storms can initiate under sufficient or weak synoptic forcing (i.e., frontal systems, upper-level vortices, or jet streaks), favorable combinations of atmospheric instability and vertical wind shear. Moreover, complex topography and mesoscale boundaries, such as orographically-induced lifting, channeling effects, and localized outflow or storm interactions (e.g., nudging, merging etc.) can enhance vertical motion and aid in tornadogenesis \citep{Kirshbaum2018,Nixon2024}. In tropical settings, these mechanisms are reinforced by strong diurnal land-sea thermal contrasts and by the limited horizontal scale of island terrain, so that convergence lines produced by sea-breeze fronts, nocturnal land-breeze outflow, and orographic channeling are ubiquitous and frequently collocate with convective initiation \citep{Qian2008,Houze2012}. Radar signatures of tornadic supercells have also been examined in recent work, including studies by \citet{Clark2024} and \citet{Wakimoto2025}. Through dual-polarimetric parameters such as differential reflectivity (\ZDR) and cross-correlation coefficient (CC, also known as \rhv), these authors found \ZDR\ column within the tornado debris signature (TDS), where lowered \rhv\ values ($<$ 0.80) are superimposed with high reflectivities. Complementary research by \citet{Yuan2024} also revealed decreased polarimetric signatures within the low-level mesocyclone of 14 May 2021 Jiangsu tornadic supercell.

Recent efforts to systematically document severe storm phenomena and develop severe storm report databases have begun in the Philippines through Project Severe Weather Archive of the Philippines \citep[SWAP;][]{Capuli2024}. Still, there is a scarcity of tornado damage surveys that have been conducted. Recent examples of tornado-focused research in the Philippines include those presented by \citet{Capuli2024a} as a thesis, and \citet{Capuli2026}. Among these studies, \citet{Capuli2026} described the synoptic and mesoscale conditions (including radar features such as bounded weak echo region/BWER and hook echo feature) present during the development of a semi-discrete supercell storm that spawned an EF2 tornado in the rural area of Candating in Arayat, Pampanga back on 27 May 2024, located at the favorable area for tornadic storms in GMMR. The authors showed that the tornado developed in a modest synoptic forcing characterized by lowering of 500-hPa mid-level heights as induced by the migratory TC. This synoptic pattern where mid-level height falls are in-place tend to be associated with major tornado outbreaks found in the U.~S. Great Plains' Tornado Alley due to the intense and rapid low-level atmospheric response \citep{Mercer2012,jiang2025}. The lower troposphere exhibited a moist boundary layer, leading to low lifting condensation levels (LCLs), and sufficient southwesterly windflow (or Southwest Monsoon) that generated ample near-surface and deep-layer wind shear (LLS and DLS) in both speed and direction manifested as a curvature in the hodograph profile. It has been long established that ``sickle-shaped'' hodographs\footnote{Hook shaped; where a layer of strong 0--1 km LLS sits beneath an abrupt clockwise turning of the shear vector with height.} have become accepted as favorable for significant ($\geqslant$ EF2) tornadoes along cyclonic, right-moving supercells \citep{Weisman1986,Thompson2000,Nixon2022}, especially those that feature strong shear near the surface \citep{Coffer2020}.

In the early morning hours of 14 September 2025 (in UTC: 13 September 2025), a severe thunderstorm developed over the coastal municipality of Daet, Camarines Norte along ongoing convection associated with an easterly wave. This storm subsequently produced a significant tornado that touched down in and impacted the suburb of Magang, located within the aforementioned municipality. The tornado damaged or destroyed buildings along its path: 10 houses were totally destroyed and 111 partially damaged, affecting 121 families (605 people). Two people were killed, a 37-year-old woman and her 13-year-old niece, and two others were injured. These figures were confirmed by the Municipal Disasters Risk Reduction and Management Office of Daet (MDRRMO-Daet), Department of Social Welfare and Development's Disaster Response Operations Management, Information and Communication (DSWD-DROMIC, Report No. 2 as of 16 September 2025), and local media. In addition to the Magang tornado, other tornadic events occurred on the same day (three tornadoes and one potential landspout), which are also examined in this study.

Collectively, this localized tornado outbreak; and particularly the Magang event, represents a rare example of tornadic activity in the Philippines as it occurred outside the country's commonly identified tornado hotspots, offering a valuable opportunity to investigate the meteorological processes governing tornado initiation and development in a tropical environment. Although the atmospheric ingredients required for tornadic supercell formation are the same in any part of the world, typical conditions that lead to severe convective storms in this area of the Philippines may exhibit characteristics that differ from their well-documented U.~S. Great Plains or European severe storm environments \citep{Taszarek2020a,Taszarek2020b}. Thus, an improved documentation of how these ingredients are brought together, in the moist tropical environment of the Philippines, can add to the overall knowledge of the distinct atmospheric regimes that are capable of producing tornadic supercells.

The initial reports and media were scattered across various social platforms, Project SWAP facilitated the systematic collection, verification, and archiving of these materials (particularly in the 4th Data Release). However, given the fact that it occurred just before the sun rises, photographs and videos of the tornado were scarce, but the damage reports were extensive and compiled to form a comprehensive dataset, presented in Section~\ref{sec3}, providing essential context for assessing the tornado's impact. To our knowledge, no previous study has documented a tornado outbreak associated with an easterly wave in the Philippines, nor has any Philippine tornado case been examined using dual-polarization radar observations at close range. Prior Philippine tornado studies have concerned single events within the recognized hotspots and under synoptic regimes dominated by conventional synoptic mechanism \citep{Capuli2024a,Capuli2026}, leaving the environmental conditions and storm-scale processes governing tornadogenesis in the tropical easterly regime essentially unconstrained. This is a consequential gap, as the kinematic pathways by which tornadic rotation is achieved in easterly wave environments cannot be assumed to follow those established for midlatitude regimes. This study aims to examine the meteorological conditions that led to tornadogenesis, encompassing the synoptic, mesoscale, and dynamic environmental aspects associated with the parent storm. More importantly, compared to \citet{Capuli2026}, a nearby dual-polarization radar was also available at that time (about 5--10 km away from the outbreak) allowing microphysical and refined damage indicator analysis using dual-polarization parameters. Supported by extensive observational data and evaluated through sounding parameters, this investigation serves as a foundational study from which future research in tornadoes that occurred in the tropics can be built.

The paper is organized as follows; Section~\ref{sec2} outlines the materials and methods, including the reanalysis dataset, atmospheric profile parameters, satellite, radar, and lightning data used, in the study. Section~\ref{sec3} presents a comprehensive analysis of the visual-damage, environmental context (from synoptic-scale to mesoscale), and storm dynamics associated with the event. Finally, Section~\ref{sec4} provides the summary of findings and an extended discussion of the results.

\section{METHODOLOGY}\label{sec2}

\subsection{Environmental Reanalysis Dataset}\label{sec21}

In order to understand the synoptic and mesoscale setting associated with the 13 September 2025 tornado event, this study utilized reanalysis data to assess the potential for severe convective storm (SCS) development. Reanalysis datasets are widely employed to diagnose atmospheric conditions conducive to severe weather such as tornadoes, damaging winds, and large hail. They are particularly useful for calculating instability indices and other parameters essential to SCS forecasting \citep{Taszarek2020b}.

Environmental reanalysis data for this case were obtained from the fifth-generation European Centre for Medium-Range Weather Forecast (ECMWF) reanalysis (ERA5) accessed through the Climate Data Store \citep[CDS;][]{Hersbach2020}. ERA5 has demonstrated good performance in depicting vertical profiles of convective environments, especially across the United States and Europe \citep{Coffer2020,Taszarek2021a,Pilguj2022}. However, some known biases persist, particularly in the boundary layer. These include discrepancies in low-level parcel characteristics and vertical shear parameters, especially near surface boundaries \citep{King2019}. Such biases are influenced by geographic location and surface elevation. Despite these limitations, ERA5 remains among the most reliable and accessible datasets for investigating severe convective environments, globally \citep{Coffer2020,Taszarek2021a,Taszarek2021b}. For large-scale synoptic analysis, the study domain extended from 4\dg--22\dg\,N latitude and 110\dg--135\dg\,E longitude. The mesoscale sector, which also served for satellite visualization, covered 11\dg--17\dg\,N and 119\dg--126\dg\,E. Both domains used a spatial resolution of 0.25\dg\ $\times$ 0.25\dg. An hourly temporal resolution was used to allow detailed analysis of the atmospheric evolution surrounding the event, which occurred around 22 UTC. The time steps selected for in-depth analysis were 21, 22, and 23 UTC, enabling a focused examination of pre-, during, and post-storm conditions.

Although ERA5 provides several convective parameters, including CAPE, it has been shown to substantially underestimate the severity of convective environments. \citet{Capuli2025} demonstrated this for a Philippine hailstorm case in which native ERA5 CAPE of $\sim$2000--2500 \Jkg\ rose to nearly 5000 \Jkg\ when recomputed from the model-level profile with a surface-based (SB) parcel; the difference between a marginally severe environment and one capable of the large hail observed. The bias likely stems from the ERA5 formulation, which lifts candidate parcels from a series of model levels within the lowest $\sim$350 hPa rather than from the near surface, so the selected parcel is frequently more dilute than the true boundary-layer parcel. To address this, both the ERA5 single-level and 137 hybrid sigma-pressure model-level fields were combined and re-gridded onto a common grid, and CAPE was recalculated using most-unstable (MU) parcel which is the parcel yielding the largest CAPE among candidates lifted from levels within the lowest 300 hPa, with the lowest candidate taken from the 2 m level. Likewise, Storm-relative Helicity (SRH) was recomputed for the 0--1 and 0--3 km layers using the \textit{xcape} codebase \citep{lepore2021}. All \textit{xcape}-derived thermodynamic quantities (MUCAPE, SBCIN), including those underlying the ERA5 spatial maps, assume liquid-only pseudoadiabatic ascent, with condensate removed on formation and ice processes neglected as chosen for computational efficiency and consistency across the domain. Kinematic parameters, such as SRH\footnote{Represented by the vertically integrated storm-relative flux of streamwise vorticity into an updraft.}, the \citet{Bunkers2000} internal-dynamics storm motion (B2K), and the resulting storm-relative wind (\VSR), are computed via the right-moving vector (\BRM), consistent with standard practice for the Northern Hemisphere. This assumes that the storm moves to the right of the non-pressure-weighted mean wind in the 0--6 km layer.

The interpretation of all reanalysis-derived parameters (including derived through the Skew-T Hodograph in Section~\ref{sec22}) is nonetheless constrained by the resolution of ERA5 relative to the scale of the setting. Spatial averaging across a land--sea boundary and across smoothed terrain tends to dilute local thermodynamic extremes and to remove the small-scale flow deformation that enhances low-level rotation, so that ERA5-derived (and the recalculated from previous section) CAPE and SRH are more likely to over or underestimate the near-storm environment. The parameters reported here are accordingly interpreted as characterizing the mesoscale environment within which the storms developed; a conservative description of the conditions available, rather than as measurements of the inflow actually ingested.

\subsection{Skew-T Hodograph}\label{sec22}

Because of the lack of observed proximity sounding ($<$ 30--50 km) in the area of interest, a model-derived profile was extracted from ERA5 data using a combination of single-level and 137 hybrid sigma-pressure level fields. The reanalysis-based soundings were analyzed to assess the thermodynamic and kinematic conditions preceding, at the time, and after the tornadic event.

While the ERA5 recalculation described in Section~\ref{sec21} uses a liquid-only pseudoadiabatic ascent, the parcel profiles shown on the Skew-T diagrams were computed assuming a non-entraining, irreversible mixed-phase adiabatic process following \citet{Peters2022}. Under pseudoadiabatic ascent, all condensate is assumed to fall out of the parcel immediately \citep{Emanel1994}; the present formulation instead retains condensate and represents the mixed-phase layer between the 0\,\dg C level and the level of homogeneous freezing ($\sim$ $-$40\,\dg C), where liquid and ice coexist. Because it enforces conservation of energy rather than of moist entropy, it avoids the biases that arise when entropy-based formulations are applied across the mixed-phase region. Support for retaining condensate in a tropical context comes from \citet{Xu1989}, who found that mean tropical soundings are nearly neutral with respect to reversible moist adiabatic ascent and are therefore only weakly conditionally unstable, indicating that the tropical environmental temperature profile is more closely matched by an adiabatic than by a pseudoadiabatic parcel. We note that their analysis employed a reversible liquid-only parcel including water loading rather than the mixed-phase formulation adopted here. Thus, cited as physical motivation for retaining condensate, not as direct validation of the present parcel model.

Furthermore, an irreversible adiabatic ascent with entrainment was also employed to compute the Entraining CAPE \citep[ECAPE;][readers are referred to the paper for the full derivation]{Peters2023b}. ECAPE accounts for dilution of updraft plumes due to environmental entrainment, particularly under conditions of limited storm-relative inflow and/or mid- to upper-tropospheric dryness. Thus, providing a more realistic estimate of updraft intensity than the conventional, undiluted CAPE, which can overestimate buoyancy in environments where entrainment is significant. All undiluted buoyancy profiles are calculated with virtual temperature correction. Depending on meteorological context, either MU or SB parcels were selected for key thermodynamic parameters. CAPE and the level of free convection (LFC) were computed using MU parcel profiles, while convective inhibition (CIN) and LCL were determined using SB profiles. Although mixed-layer parcels are often preferred for their robustness, SB and MU parcels were preferred in this study due to their better representation of near-surface instability (or stability) and lowest cloud base heights. These factors are critical in distinguishing surface-based from elevated convection, particularly in tornadic environments.

The hodograph was used to understand the kinematic properties associated with the tornado event. In a conventional, ground-relative hodograph, no transformation is made to the modeled $u$ and $v$ wind components \citep{Markowski2003,Nowotarski2018}. However, this method cannot capture how the wind profile changes from the perspective of the storm or assess for diagnosing convective storm dynamics. To place the kinematic analysis in the frame of the storm rather than the ground, the wind profile is expressed in storm-relative coordinates. The transformation is a rigid translation of the hodograph by the storm motion vector $\mathbf{c} = (c_u, c_v)$:
\begin{equation}
\mathbf{V}_{SR}(z) = \mathbf{V}(z) - \mathbf{c}
\end{equation}
where $\mathbf{V}(z) = [u(z), v(z)]$ is the ground-relative wind. Because $\mathbf{c}$ is independent of height, the vertical shear vector and the horizontal vorticity are unchanged by the transformation; only the origin of the hodograph moves. Here, the $\mathbf{c}$ used is the \BRM\ similar to the prior section.

The kinematic parameters were calculated from the interpolated model output of the $u$ and $v$ wind components. The bulk wind difference (BWD) was calculated across several vertical layers to assess shear magnitude. These included the layers from 0--500 m (BWD$_{500}$), 0--1 km (BWD$_{01}$), 0--3 km (BWD$_{03}$), 0--6 km (BWD$_{06}$; DLS), 1--3 km (BWD$_{13}$), and 1--6 km (BWD$_{16}$). SRH is of the greatest dynamical importance in this regard \citep{DaviesJones1984}. The SRH was computed for several layers: 0--500 m (SRH$_{500}$), 0--1 km (SRH$_{01}$), 0--3 km (SRH$_{03}$), and 1--3 km (SRH$_{13}$). For the near-ground layer (0--1 km), the mean \VSR\ was calculated relative to the \BRM. This analysis also included the mean streamwise vorticity integrated over 0--500 m and 0--1 km layer denoted as $\omega_{s500}$, and $\omega_{s01}$, respectively. These quantities were calculated as:
\begin{equation}
\omega_s = \frac{\left(\nabla \times (\mathbf{v}-\mathbf{c})\right) \cdot (\mathbf{v}-\mathbf{c})}{\|\mathbf{v}-\mathbf{c}\|}
\end{equation}
where $\mathbf{v}$ is the three-dimensional environmental wind and $\mathbf{c}$ is the storm motion vector under the assumption of the B2K method. Furthermore, the vertical vorticity along the low-level mesocyclone was calculated. Based on the first-order approximation of tilting efficiency proposed \citet{Coffer2023}, the vertical vorticity generated from the tilting of streamwise vorticity in a purely streamwise environment can be expressed as;
\begin{equation}
\zeta_{LLM} = \omega_s \frac{w}{V_{SR}}
\end{equation}
where $\omega_s$ is the streamwise vorticity\footnote{In this case, angle-average and its maximum $\omega_s$ in the lowest 1 km.}, $w$ is the vertical velocity of the updraft\footnote{CAPE$_{03}$ is used as a proxy for the vertical velocity of the low-level rotating updraft via the parcel-theory relation $w = \sqrt{2\,\mathrm{CAPE}_{03}}$. This expression is a strict upper bound.}, and \VSR\ is the storm-relative wind in the lowest 1 km. This formulation highlights how the conversion of environmental streamwise vorticity into vertical vorticity depends on the interaction and balance between vertical and horizontal motions \citep{DaviesJones1984}. It provides a physically intuitive approximation for assessing the potential for mesocyclone development and subsequent tornadogenesis.

The ``streamwiseness'' ($\widetilde{\omega}_s$) of horizontal vorticity was also calculated by dividing the integrated streamwise component of horizontal vorticity by the integrated total magnitude of the horizontal vorticity in a given layer;
\begin{equation}
\widetilde{\omega}_s = \frac{\omega_s}{\omega}
\end{equation}
The use of $\widetilde{\omega}_s$ is justified for two reasons. First, near-zero shear conditions are rare in tornado environments \citep{Coffer2020}. Second, the parameter offers intuitive interpretation; for example, a value of 0.9 indicates 90\% of the horizontal vorticity is streamwise in nature. Nonetheless, the degree to which vorticity is streamwise remains a fundamental aspect of supercell dynamics and tornadogenesis \citep{Coffer2017}.

Lastly, the Deviant Tornado Motion (DTM) was calculated based on the method developed by \citet{Nixon2021}. This parameter evaluates the potential deviation of tornado motion from both the low-level advective flow and the parent storm motion, and offers insight into tornado track behavior relative to ambient wind conditions. All reanalysis-derived soundings associated in this event were analyzed using SounderPy (v3.1.0) by \citet{Gillett2025}. A list of the evaluated kinematic and thermodynamic parameters, along with their definitions and computation methods, is presented in Table~\ref{Table_2}.

\subsection{Satellite Data}

\subsubsection{HIMAWARI-9 Satellite Data}

High spatio-temporal resolution data from the Advanced Himawari Imager (AHI) aboard Himawari-9 \citep{Bessho2016} were used to identify and characterize the convective system associated with the tornadic event. The AHI comprises three visible bands (central $\lambda$ 0.47--0.64 $\mu$m), three near-infrared bands (0.86--2.3 $\mu$m) and ten infrared bands (3.9--13.3 $\mu$m). This study used the visible bands and the 10.4 $\mu$m infrared atmospheric window band (Band 13, B13) to monitor deep convection. B13 is particularly useful for detecting cold cloud-top regions [brightness temperature (BT) $<$ 233 K ($-$40\,\dg C)] while minimizing interference from water vapor \citep{Hirose2019}, and is widely adopted for the rapid identification of convective features associated with extreme rainfall at short lag time \citep{Wang2022}. The visible bands have a spatial resolution of 500 m--1 km, whereas B13 is coarser at 2 km. Full-disk B13 scenes at 10-minute intervals were obtained on the GEOS141 fixed grid (2 km at nadir) for 2140--2230 UTC 13 September 2025.

Because the study area lies well off the sub-satellite point (140.7\dg E), the ground footprint of a pixel is appreciably larger than its nadir value. At the location of the storm (14.1\dg N, 122.9\dg E) the satellite zenith angle is 26.3\dg\ and the slant range 36,350 km, giving a footprint of approximately 2.03 km $\times$ 2.27 km, or 4.61 km$^2$, compared with 4.00 km$^2$ at nadir. The area of every pixel was therefore computed individually by inverting the geostationary fixed-grid projection and projecting the four pixel corners onto the WGS-84 ellipsoid, rather than assuming a uniform nadir value; using the nadir value would underestimate cloud areas by $\sim$15\% and equivalent radii by $\sim$7\% at this location.

The scale of the convective system was quantified by its equivalent radius ($r$). Applying an IR threshold of $-$40\,\dg C, the cold-cloud shield was delineated by connected-component labeling of all contiguous pixels colder than that threshold, seeded at the storm location, so that the system is defined by its own continuity rather than by a predefined search window. The analysis domain was kept substantially larger than the region of interest so that the shield was not truncated. The equivalent radius is the radius of a circle of the same area as the labeled shield is calculated as;
\begin{equation}
r = \sqrt{N\left(\frac{A}{\pi}\right)}
\end{equation}
where $N$ is the number of pixels and $A$ is the area of the pixel. As a rule of thumb, MCSs are defined as convective systems with radii $>$ 100 km, a mean cloud temperature $<$ $-$30\,\dg C for the entire system, and at least one overshooting top \citep[BT $<$ $-$65\,\dg C;][]{Carvalho2001}; smaller storms with radii between 50 and 100 km and cloud temperatures $<$ $-$30\,\dg C are categorized as weak and disorganized convective systems (WDCSs), the most frequent convective system in the Tropics \citep{Rossow2013,Tan2013}.

Penetrative overshooting tops (OTs) were examined as signatures of severe convection. OTs are local minima in cloud-top brightness temperature that mark updrafts strong enough to penetrate the equilibrium level and intrude into the lower stratosphere, and their occurrence is frequently linked to intense storm dynamics and severe weather, including large hail and tornadoes \citep{Bedka2010}. In this case the anvil was itself colder than $-$80\,\dg C over most of its area, so an absolute threshold of the magnitude normally used for OT screening does not isolate individual overshoots. Candidate OTs were therefore identified within the storm's deep-convective region, defined as the contiguous area with BT $<$ $-$70\,\dg C, whose brightness-temperature distribution was closely reproducible between scenes. An OT was defined as a contiguous group of pixels that: (i) is colder than $-$86\,\dg C i.e., within the coldest $\sim$1\% of the anvil, (ii) covers at least 4 km$^2$, and (iii) is at least 5 K colder than the median BT of a surrounding 15--30 km annulus, following the anvil-relative approach of \citet{Bedka2010}. Groups whose centroids lay within 15 km of a colder group were merged and each OT was located at its coldest pixel.

\subsubsection{PlanetScope Observations}

Remote sensing provides innovative technologies for assessing tornado debris \citep{HOQUE2017}. These technologies have shown to be effective in damage assessment following disasters, as demonstrated by numerous research \citep{Xie2016}. Aerial images \citep{schaefer2020}, a form of airborne remote sensing technology, are one important source of data for storm damage assessment.

PlanetScope (PS) is a CubeSat 3U form factor (10 cm $\times$ 10 cm $\times$ 30 cm) satellite constellation operated by Planet Labs, Inc. and was utilized in this study. While in sun-synchronous orbit, these cubesats provide daily 4-band (RGB and NIR) imagery of the Earth at 3-m spatial resolution \citep{PlanetARPS2024}. Currently in orbit are the Dove Classics, the improved Dove-Rs with more spectral and radiometric precision and sharpness, and the hyperspectral SuperDoves. Most images used in the study were daily Dove Classic imagery, or the newer generation Dove-R images (\url{https://www.planet.com/}). These have a Ground Sampling Distance (GSD) of 3--4 m at nadir and a positional accuracy of $<$ 10 m RMSE. We used the level 3B Surface Reflectance (SR) product, which is orthorectified and atmospherically corrected by Planet Labs with their in-house processing. These provide more consistency across time and location localized atmospheric conditions while minimizing uncertainty in the spectral response.

\begin{figure}[!t]
\includegraphics[width=\columnwidth]{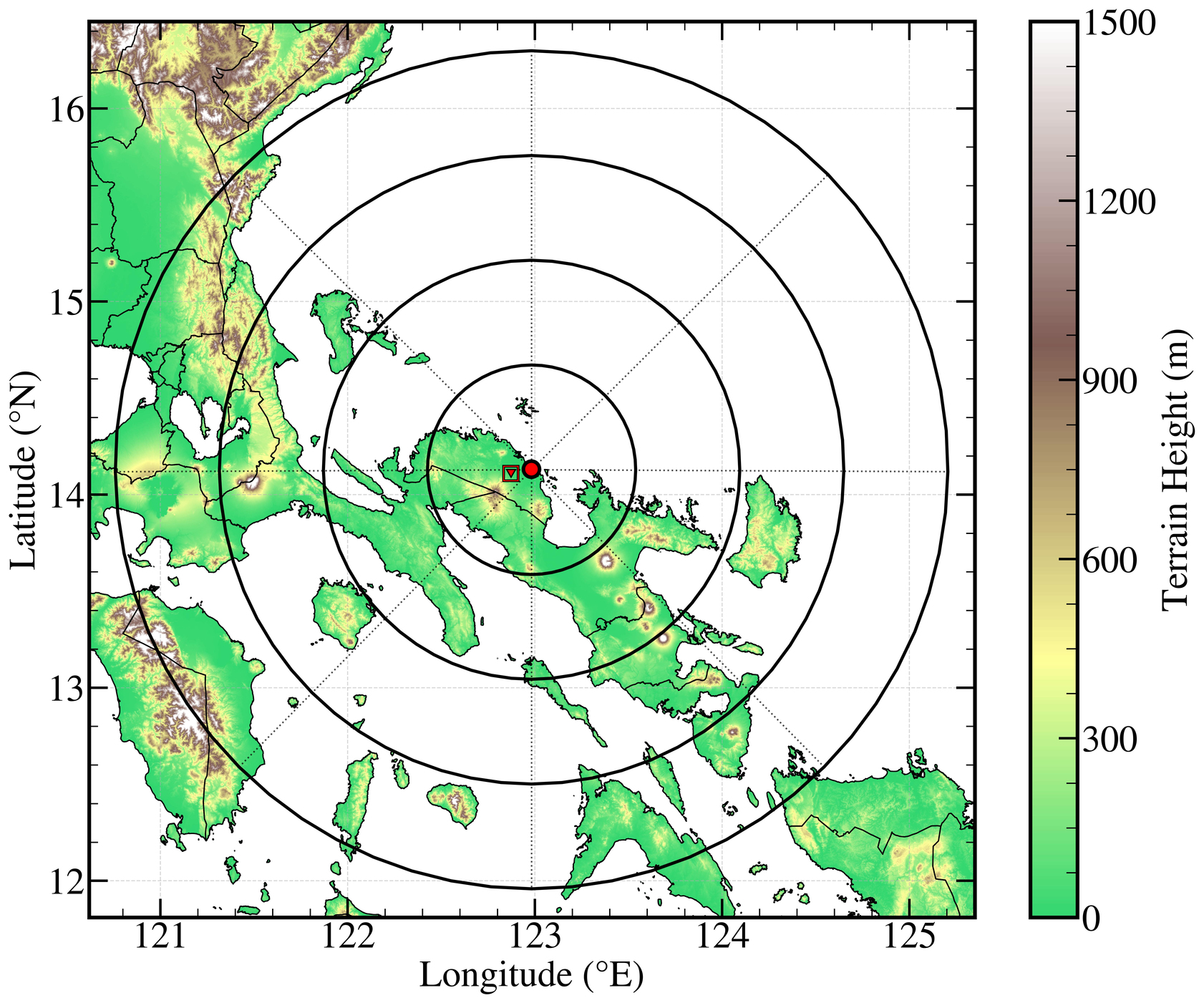}
\caption{Elevation map encompassing S-DAT (red circle) and its 240 km radar coverage with increments of 60 km each. The red triangle enclosed by a box is the location of the tornado outbreak event.}
\label{fig_1}
\end{figure}

The high-resolution satellite images that were chosen for this analysis are nearly clear images (cloudless, hazeless, and smokeless) within the areas of interest for this analysis of before (04 August 2025)\footnote{Typically, it is better to attain a pre-event imagery in the same month i.e., in September, but observations for it were heavily obscured by clouds.} and after the event (28 September 2025). Upon inspecting the raw data along the investigation area, it contains some thin-to-thick clouds. Therefore, we apply spatial masking using a sensor-agnostic deep learning model that segments clouds and cloud shadow. This is applied through OmniCloudMask \citep[OCM;][]{WRIGHT2024}, a python library that consists of two ensembled U-Net \citep{Ronnberger2015} models. These constituent models leverage the use of Fastai Dynamic U-Net \citep{Howard2020}, which allows any CNN to be used as a backbone model. The models chosen for this purpose were the \textit{``RegnetY 004''} and the \textit{``ConvNextV2 nano''}, supplied with pretrained weights sourced from the timm library \citep{rw2019timm}. With the masked data, the PS satellite imagery can be used to detect the damage from a tornado event using Normalized Difference Vegetation Index (NDVI).

Following the creation of the before and after from the PS with clouds removed, the NDVI value is calculated using the pre-tornado imagery (NDVI$_{\text{BEFORE}}$), as well as using the post tornado imagery (NDVI$_{\text{AFTER}}$). These values are then used to measure the percent change in NDVI between the before and after images. This normalized percent change is an empirical method of calculating the change in plant health and accounts for the starting health of the vegetation. Previous studies used the difference between the two NDVI images rather than the percent change equation due to the similarity in vegetation conditions as they analyzed singular events. \citet{Burow2020} took the NDVI values from post event imagery and subtracted them from the pre-event imagery to create an NDVI difference image. The normalized percent change in NDVI value is calculated as;
\begin{equation}
NDVI_{change} = \frac{NDVI_{AFTER} - NDVI_{BEFORE}}{NDVI_{BEFORE}} \times 100\%
\end{equation}

\subsection{Observational Data from DOST-PAGASA}

\subsubsection{Radar Data}

DOST-PAGASA operates a nationwide network of 17 weather radars, 7 of which are dual-polarization (dual-pol) C-band radars, 3 dual-pol S-band, and 7 single-polarization (single-pol) S-band radars. These systems support the agency's capacity for early detection and monitoring of weather systems across the archipelago. The 13 September 2025 tornadic supercell tracked through the domains of the Daet S-band radar (hereafter S-DAT), one of the 3 dual-pol S-band radar strategically located on near the coast of Daet, Camarines Norte (14.128562\dg N, 122.982693\dg E); along the Bicol Region (Fig.~\ref{fig_1}). The S-DAT radar, which operationally started by 2018, provides extensive coverage over this region, an area that typically sees tropical cyclone activity. S-DAT employs a Modified Volume Coverage Pattern 11 (VCP11) which optimizes radar sampling across both low and high elevation angles. This configuration is particularly well-suited to tropical environments. For this study, the Doppler mode scanning radius was set to 240 km. Technical radar specifications are described in Table~\ref{Table_1}.

\begin{table}[t]
\caption{Technical specifications of S-DAT.}
\label{Table_1}
\centering
\resizebox{\columnwidth}{!}{%
\begin{tabular}{ll}
\hline\hline
Frequency Band & S-band, 2.851 GHz ($\lambda$ = 10.52 cm) \\
Polarization & Dual-polarization (Dual-pol) \\
Position & 14.128562\dg N, 122.982693\dg E \\
Altitude & 34 mASL \\
Pulsing Scheme & 4:3 dual-PRF on the lowest six tilts (0.5--5.3\dg); \\
 & single PRF on the upper eight tilts (6.2--19.5\dg) \\
Low / high PRF & 600 / 800 Hz (1.667 / 1.250 ms) \\
Nyquist Velocity & $\pm$15.77 \ms\ (600 Hz), $\pm$21.03 \ms\ (800 Hz); \\
 & extended $\pm$63.1 \ms\ on dual-PRF tilts \\
Maximum Range / Doppler Range & 250 km / 240 km \\
Beam Width / Range Resolution (Gate Spacing) & 1.69\dg\ ($h$ and $v$) / 250 m \\
Azimuthal Sampling & 1.0\dg\ (360 rays per sweep) \\
Elevation Angles (14) & 0.5\dg, 1.5\dg, 2.4\dg, 3.4\dg, 4.3\dg, 5.3\dg, 6.2\dg, \\
 & 7.5\dg, 8.7\dg, 10\dg, 12\dg, 14\dg, 16.7\dg, 19.5\dg \\
Volume Cycle Intervals & 10 minutes \\
Start of Operation & 2018 \\
\hline
\end{tabular}%
}
\end{table}

\begin{table*}[!t]
\caption{Environmental Sounding Parameters evaluated in this case.}
\label{Table_2}
\centering
\resizebox{\textwidth}{!}{%
\begin{tabular}{lllccc}
\hline\hline
Parameter & Definition & Notes & 21 UTC ERA5 & 22 UTC ERA5 & 23 UTC ERA5 \\
\hline
CAPE & Convective Available Potential Energy & MU parcel; undiluted & 1336 \Jkg & 2167 \Jkg & 2387 \Jkg \\
CAPE$_{03}$ & 0--3 km CAPE & MU parcel; undiluted & 87 \Jkg & 115 \Jkg & 131 \Jkg \\
CAPE$_{06}$ & 0--6 km CAPE & MU parcel; undiluted & 356 \Jkg & 516 \Jkg & 563 \Jkg \\
ECAPE & Entraining CAPE & MU parcel; entraining & 1071 \Jkg & 1534 \Jkg & 1766 \Jkg \\
CIN & Convective Inhibition & SB parcel; undiluted & 0 \Jkg & 0 \Jkg & 0 \Jkg \\
LCL & Lifted Condensation Level & SB parcel; undiluted & 157 m & 180 m & 197 m \\
LFC & Level of Free Convection & MU parcel; undiluted & 157 m & 180 m & 197 m \\
LR$_{03}$ & 0--3 km Lapse Rate & Low-level Lapse Rate & 5.52\,\dg C km$^{-1}$ & 5.63\,\dg C km$^{-1}$ & 5.74\,\dg C km$^{-1}$ \\
LR$_{36}$ & 3--6 km Lapse Rate & Mid-level Lapse Rate & 5.39\,\dg C km$^{-1}$ & 5.84\,\dg C km$^{-1}$ & 5.88\,\dg C km$^{-1}$ \\
RH$_{13}$ & Mean 1--3 km Relative Humidity & & 0.90 / 90\% & 0.86 / 86\% & 0.84 / 84\% \\
RH$_{16}$ & Mean 1--6 km RH & & 0.83 / 83\% & 0.77 / 77\% & 0.76 / 76\% \\
BWD$_{01}$ & 0--1 km Bulk Wind Difference & Low-level Shear (LLS) & 13.82 kt / 7.12 \ms & 12.85 kt / 6.61 \ms & 11.83 kt / 6.09 \ms \\
BWD$_{03}$ & 0--3 km BWD & Mid-level Shear (MLS) & 15.46 kt / 7.95 \ms & 18.52 kt / 9.53 \ms & 16.10 kt / 8.28 \ms \\
BWD$_{06}$ & 0--6 km BWD & Deep-layer Shear (DLS) & 13.16 kt / 6.77 \ms & 15.77 kt / 8.11 \ms & 16.54 kt / 8.50 \ms \\
BWD$_{13}$ & 1--3 km BWD & & 7.58 kt / 3.90 \ms & 7.22 kt / 3.71 \ms & 6.45 kt / 3.32 \ms \\
BWD$_{16}$ & 1--6 km BWD & & 3.84 kt / 1.98 \ms & 7.20 kt / 3.70 \ms & 6.02 kt / 3.10 \ms \\
\BRM & Bunkers RM Storm Motion & RM: Right Mover & 15.23 kt / 7.84 \ms & 12.10 kt / 6.22 \ms & 17.47 kt / 8.99 \ms \\
\VSR & Mean 0--1 km Storm-relative Wind & Using \BRM & 18.64 kt / 9.58 \ms & 16.66 kt / 8.57 \ms & 19.34 kt / 9.95 \ms \\
SRH$_{500}$ & 0--500 m Storm-relative Helicity & Using \BRM & 43 \mss & 41 \mss & 48 \mss \\
SRH$_{01}$ & 0--1 km SRH & Using \BRM & 75 \mss & 68 \mss & 71 \mss \\
SRH$_{03}$ & 0--3 km SRH & Using \BRM & 96 \mss & 100 \mss & 93 \mss \\
SRH$_{13}$ & 1--3 km SRH & Using \BRM & 21 \mss & 32 \mss & 21 \mss \\
$\omega_{s500}$ & 0--500 m Streamwise Vorticity & Using \BRM & 0.009 \si & 0.009 \si & 0.009 \si \\
$\widetilde{\omega}_{s500}$ & 0--500 m Streamwiseness & Using \BRM & 0.9184 / 91.84\% & 0.9035 / 90.35\% & 0.9672 / 96.72\% \\
$\omega_{s01}$ & 0--1 km Streamwise Vorticity & Using \BRM & 0.008 \si & 0.008 \si & 0.007 \si \\
$\widetilde{\omega}_{s01}$ & 0--1 km Streamwiseness & Using \BRM & 0.9261 / 92.61\% & 0.9290 / 92.90\% & 0.9089 / 90.89\% \\
$\zeta_{LLM}$ & 0--1 km Low-level Vertical Vorticity & MU parcel and \BRM & 0.016 \si & 0.020 \si & 0.016 \si \\
DTM & Deviant Tornado Motion & Using \BRM & 7.11 kt / 3.66 \ms & 5.39 kt / 2.77 \ms & 8.24 kt / 4.24 \ms \\
WMAXSHEAR & Undiluted Updraft Velocity $\times$ BWD$_{06}$ & MU parcel; undiluted & 494.91 \mss & 755.06 \mss & 830.58 \mss \\
SCP & Supercell Composite Parameter & & 1.67 & 2.64 & 2.54 \\
STP & Significant Tornado Parameter & & 0.37 & 0.60 & 0.56 \\
\hline
\multicolumn{6}{l}{*Note: LCL = LFC especially for surface-based thunderstorms.}\\
\end{tabular}}
\end{table*}

\subsubsection{Quality Control}

S-DAT's radar data underwent quality control based on a fuzzy logic approach designed to detect and eliminate interference patterns. This method draws on prior algorithm assessments \citep[e.g.,][and references therein]{Kilambi2018,Overeem2020} and was adapted for both single and dual-pol radar products following the framework described in \citet{Lin2021}. Two distinct velocity artifacts were present in the Daet volumes and were corrected sequentially. The radar applies 4:3 dual-PRF processing to the lowest six tilts (0.5--5.3\dg), yielding an extended Nyquist velocity of $\pm$63.1 \ms\ from base Nyquist velocities of $\pm$15.77 and $\pm$21.03 \ms, while the remaining eight tilts (6.2--19.5\dg) operate at a single PRF with a Nyquist velocity of $\pm$15.77 \ms. The dual-PRF tilts are therefore subject to processor errors, in which the unfolding of a gate between the two PRF estimates fails and the retrieved velocity is displaced by an integer multiple of twice one of the base Nyquist velocities; the single-PRF tilts are instead subject to conventional aliasing.

Dual-PRF processor errors were corrected first, following \citet{Alford2022}. Candidate errors were identified as departures from a spatially smoothed velocity field exceeding three standard deviations in both the radial and azimuthal dimensions that corresponded to integer multiples of twice the low or high Nyquist velocity, and were corrected by removing the multiple. The corrected fields were then dealiased using UNRAVEL \citep[UNfold RAdar VELocity;][]{Louf2020}, an open-source modular algorithm applied here to volumes read with the Python ARM Radar Toolkit \citep[Py-ART;][]{Helmus2016}. UNRAVEL requires no external reference wind profile; it establishes continuity from the least-aliased portion of the field and propagates it outward through a sequence of increasingly permissive continuity tests, operating first within each sweep in azimuth and range (unfolding gates that violate continuity by more than $\alpha$ = 0.8 of the Nyquist interval) and subsequently in three dimensions across adjacent elevation angles. Because the Nyquist velocity varies between tilts in this volume, per-sweep Nyquist velocities were supplied explicitly rather than inferred from the volume as a whole.

Given the radar's S-band wavelength and its regular maintenance by DOST-PAGASA radar operators, no specific attenuation correction (including systematic bias and wet-radome effects) was applied, as attenuation at S-band is negligible for the geometry of this case. Following the ITU-R P.838-3 specific attenuation model \citep{ITU2005}, one-way rain attenuation at 2.8--3.0 GHz is $\sim$0.015 dB km$^{-1}$ at 50 mm hr$^{-1}$ and $\sim$0.03--0.04 dB km$^{-1}$ at rain rates approaching 100 mm hr$^{-1}$, roughly a factor of twenty smaller than at C-band and two orders of magnitude smaller than at X-band \citep{Doviak1993,Diedrich2015}. The tornadic storms analyzed here occurred approximately 5--10 km from the radar site. Even under the physically extreme assumption that such rain rates were sustained along the entire path, the accumulated two-way attenuation would be $\sim$0.6--0.8 dB, and for the more realistic case of a heavy-rain core a few kilometers deep it remains below $\sim$0.3 dB. Both values are comparable to or smaller than the calibration uncertainty of the system, and the differential attenuation affecting \ZDR\ is smaller again by roughly an order of magnitude, so attenuation is without consequence for the qualitative polarimetric signatures examined in this study. We note that this argument rests on the short path lengths of this case and would not hold at long ranges or at shorter wavelengths.

The proximity of the event also places it well within effective radar coverage, with the storms sampled at low elevation angles throughout their convective phase. Statistics and additional information regarding the \VR\ correction are attached in the Supplementary File of this paper, particularly for the volume scans of interest.

\subsubsection{Mesocyclone Detection Algorithm}

To identify potential rotation signatures within the dealiased \VR\ field of the S-DAT radar, we used the Mesocyclone Detection Algorithm (MDA) and its detection thresholds as developed by \citet{Stumpf1998} and following modern applications of this MDA stipulated in \citet{Hengstebeck2018} and \citet{Feldmann2021}. These parameters involved the maximum tangential (rotational) velocity ($>$ 10 \ms), vorticity ($>$ 10$^{-2}$ \si), horizontal scale, and duration of the feature. In particular, the rotational velocity (per nomenclature; \Vrot) is calculated using the maximum outbound velocity ($V_{max,out}$) and the maximum inbound velocity ($V_{max,in}$) determined from the radial segments in the 2D feature:
\begin{equation}
V_{rot} = \frac{V_{max,out} - V_{max,in}}{2}
\end{equation}
The vertical vorticity, as measured through the radar (\zrad), can be determined using the equation below:
\begin{equation}
\zeta_{rad} = 2\,\frac{V_{max,out} - V_{max,in}}{\Delta X}
\end{equation}
where $\Delta X$ is the cartesian distance (m) between the velocity maxima. Couplet peaks were identified after correcting residual dual-PRF errors and masking isolated gates departing by more than 10 \ms\ from their 3 $\times$ 3 median, so that no single gate sets \Vrot.

While we apply these mesocyclone detection techniques and thresholds, we emphasize that they are largely adapted from studies conducted in midlatitude regions, where radar configurations, storm structures, and environmental conditions differ substantially from those in the tropics. Tropical maritime supercells developing under weaker DLS can be smaller in horizontal scale and shallower than their midlatitude counterparts, so that \Vrot\ may fall below the nominal thresholds even where a coherent mesocyclone is present. Conversely, the strong LLS characteristic of the tropical marine boundary layer can produce velocity couplets that satisfy the thresholds in the absence of organized rotation. \citet{Capuli2026} found for a tornadic supercell that rotational velocities of only $\sim$8--9 \ms\ were already indicative of a mesocyclonic signature (potential/developing), well below the $>$ 10 \ms\ criterion applied here. The thresholds used in this study should therefore be regarded as provisional, and may either under- or over-detect mesocyclones in this environment. Establishing thresholds calibrated for tropical maritime supercells will require a larger sample of radar-observed cases than presently exists for this region. We therefore encourage the development of regionally calibrated detection algorithms suited to the broader Philippine radar network, as conducted by \citet{Feldmann2021} for the Alpine region.

\subsubsection{Radar Techniques}

Horizontal winds were retrieved from the dealiased \VR\ with the Velocity Azimuth Display technique \citep[VAD;][]{Browning1968}. For a sweep at elevation $\theta$, we sampled a ring at every fourth range gate between 4 and 90 km and fitted the \VR\ around each ring by least squares to:
\begin{equation}
V_r(\phi) = a_0 + (u\sin\phi + v\cos\phi)\cos\theta
\end{equation}
where $\phi$ is azimuth, $u$ and $v$ are the eastward and northward-wind components, and $a_0$ absorbs the contributions of divergence and of the vertical motion of the scatterers. A ring was fitted only if it had at least 20 valid gates covering at least 18\% of the circle, no azimuthal gap wider than 200\dg, and a normal-equation condition number of 40 or less. Gates more than 2.5 standard deviations from the fit were removed iteratively, in up to three passes. A fit was kept only if at least 12 gates remained, the residual root-mean-square (RMS) was $\leq$ 6 \ms, and the retrieved speed was $\leq$ 60 \ms. Each ring was assigned the beam-center height from the 4/3 effective-Earth-radius model \citep{Doviak1993}:
\begin{equation}
h = \sqrt{r^2 + (k_e a)^2 + 2 r k_e a \sin\theta} - k_e a, \qquad k_e a = \tfrac{4}{3} \times 6371~\mathrm{km}
\end{equation}
where $r$ is the slant range. Ring fits from all sweeps were pooled into 250-m height bins. Each fit was weighted by the inverse square of its residual RMS (with a minimum RMS of 0.5 \ms), and a bin was kept only if at least two rings contributed. From the VAD wind profiles, the radar hodographs were derived. These hodographs, particularly at 2210 and 2220 UTC, were constructed in storm-relative manner and from the corresponding VAD profile. Each profile was linearly interpolated to 100-m levels from the surface to the highest retrieved level, with heights above radar level (ARL) taken as heights above ground (below the lowest retrieved level, the wind was held constant).

Meanwhile, the motions of the tornadic storms were estimated from the 2220 UTC volume by following each storm's reflectivity core upward through five successive tilts (sweeps 1--5). On each tilt, the reflectivity signature within 14 km of S-DAT was first remapped onto a Cartesian grid with 0.25-km spacing. Where several gates fell within one grid box, the highest reflectivity was retained together with the time at which that gate was sampled. The core was then identified as a contiguous area of grid boxes exceeding a reflectivity threshold, with boxes that touch at an edge or a corner treated as connected, and the search was confined to a fixed radius around a reference point. On the lowest selected tilt, this reference point was a manually selected seed where in each higher tilt, it was the core position found on the tilt below, so that the same core was followed with height. When more than one qualifying area fell within the search radius, the one nearest the reference point was retained. The core position was defined as the centroid of this area weighted by the linear reflectivity factor ($Z_H = 10^{\mathrm{dBZ}/10}$), so that the most intense echo dominates the estimate, as in centroid-based storm trackers such as Thunderstorm Identification, Tracking, Analysis, and Nowcasting \citep[TITAN;][]{Dixon1993}. The core time was the mean sampling time of its grid boxes. At thresholds below 54 dBZ, isolated grid boxes scattered around the storm can exceed the threshold and bias the centroid. At these thresholds, the thresholded area was therefore first filtered with a morphological opening, which removes echo features too small to contain a 3 $\times$ 3 block of grid boxes (0.75 $\times$ 0.75 km) while leaving larger echo areas intact. This filter was not applied at higher thresholds, where it would also have removed the small but real reflectivity cores being tracked.

As the readers will see in the subsequent section, the two storms formed a single connected echo at $\sim$50 dBZ (a discussion of storm interaction), thresholds and search radii were set separately for each storm. TS1 used 62 dBZ, falling back to 50 dBZ, within 2.0 km, with a minimum core area of 0.5 km$^2$. TS2 used 57 dBZ, falling back to 55 dBZ, within 1.5 km, with a minimum core area of 0.2 km$^2$; 55 dBZ is the lowest threshold at which TS2 separates from TS1. The five core positions (A--E) define four legs, each with a displacement, an elapsed time and a bearing; the net A--E displacement was computed as well. The legs were combined as the vector mean of their velocities.

Both storms lay close to the radar (between 5--10 km), which makes this tilt-by-tilt analysis particularly informative. At these ranges the beam center rises by only $\sim$0.16 km per degree of elevation, so the five tilts sample a shallow and nearly contiguous layer of the lower storm. In addition, the 1\dg\ azimuthal sampling corresponds to a spacing of only $\sim$0.12--0.16 km, so each core is well resolved on every tilt. The chain of centroids follows a single coherent low-level core rather than disparate parts of the storm. Because successive tilts are only $\sim$30 s apart, the tilt-to-tilt centroid shift still contains the lean of the core with height, thus per-leg speeds may under- or over-estimate the storm speed. In contrast, the bearing remains robust as the tilt-to-tilt displacements are consistent in direction and a core that leans downshear leans approximately in the direction of storm motion. Thus, we took the direction of motion from this analysis using the vector mean of the four tilt-to-tilt velocities and were used as the reference motions for the storm-relative VAD hodographs and kinematic parameters, and were assumed to remain representative at both 2210 and 2220 UTC.

\subsubsection{Lightning Data}

Lightning data was obtained from DOST-PAGASA Lightning Detection Network (PLDN), which consists of 28 strategically deployed lightning sensors distributed throughout the country. These sensors are primarily located within PAGASA observing stations and support real-time thunderstorm monitoring. Since its operational inception in 2018, the PLDN has provided continuous coverage for both intra-cloud (IC) and cloud-to-ground (CG) lightning discharges. It also enables early detection of rapid increases in lightning activity, known as `lightning jumps', which often precede the onset of severe weather \citep{Schultz2011}. For this case study, lightning data spanning from 21 UTC of 13 September 2025 to 00 UTC of 14 September 2025 were used.

To ensure data quality, all lightning strokes were aggregated into 1-minute intervals. For each interval, we calculated the total number of strokes, along with average and median peak current values, separately for positive and negative polarities. This time-resolved approach enabled a detailed characterization of lightning intensity, polarity distribution, and temporal evolution in relation to the supercell's development and maturity.

\section{RESULTS}\label{sec3}

\subsection{Tornado Damage Paths and Visual Observations}\label{sec31}

\begin{figure*}[!t]
\centering
\includegraphics[width=\textwidth]{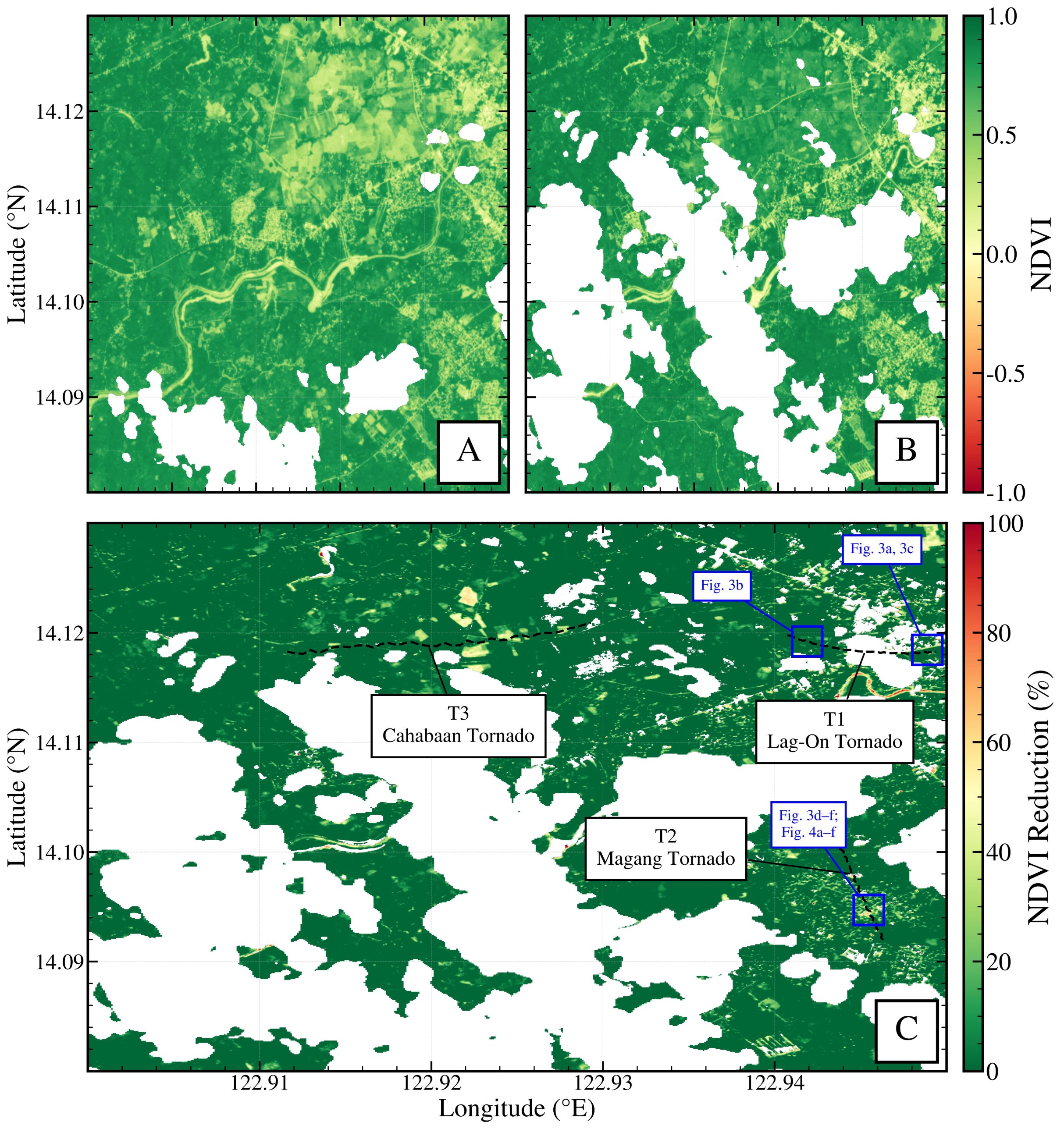}
\caption{Isopleths of delineated tornado tracks based on PlanetScope observations. (a) NDVI$_{\text{BEFORE}}$ at 04 August 2025, (b) NDVI$_{\text{AFTER}}$ at 28 September 2025, and (c) NDVI Change between 04 August and 28 September 2025. For Panel C, the tornado tracks (black dashed lines) are labeled accordingly. Blue boxes correspond to the identified damage indicators and visual observations of the tornadoes.}
\label{fig_2}
\end{figure*}

The damage paths of the tornadoes that struck several areas of Camarines Norte were examined through change-detection analysis using pre- and post-event PlanetScope observations. Figure~\ref{fig_2} shows NDVI before and after the event together with their difference. The study area is a predominantly rural landscape comprising bare soil, residential structures, paved surfaces and vegetation.

The pre- and post-event scenes were separated by 55 days (4 August and 28 September 2025), a closer pre-event image being unavailable owing to cloud contamination. Over such an interval, NDVI differences also reflect vegetation phenology, harvest and other agricultural activity, and unrelated environmental change. The NDVI fields are therefore used to assess the spatial continuity and orientation of damage swaths whose existence was established by other evidence, and are interpreted alongside eyewitness accounts, the ground survey and the radar analysis rather than as independent confirmation. The weight the NDVI signal carries differs between the three tracks, as set out below.

\begin{figure*}[!t]
\includegraphics[width=\textwidth]{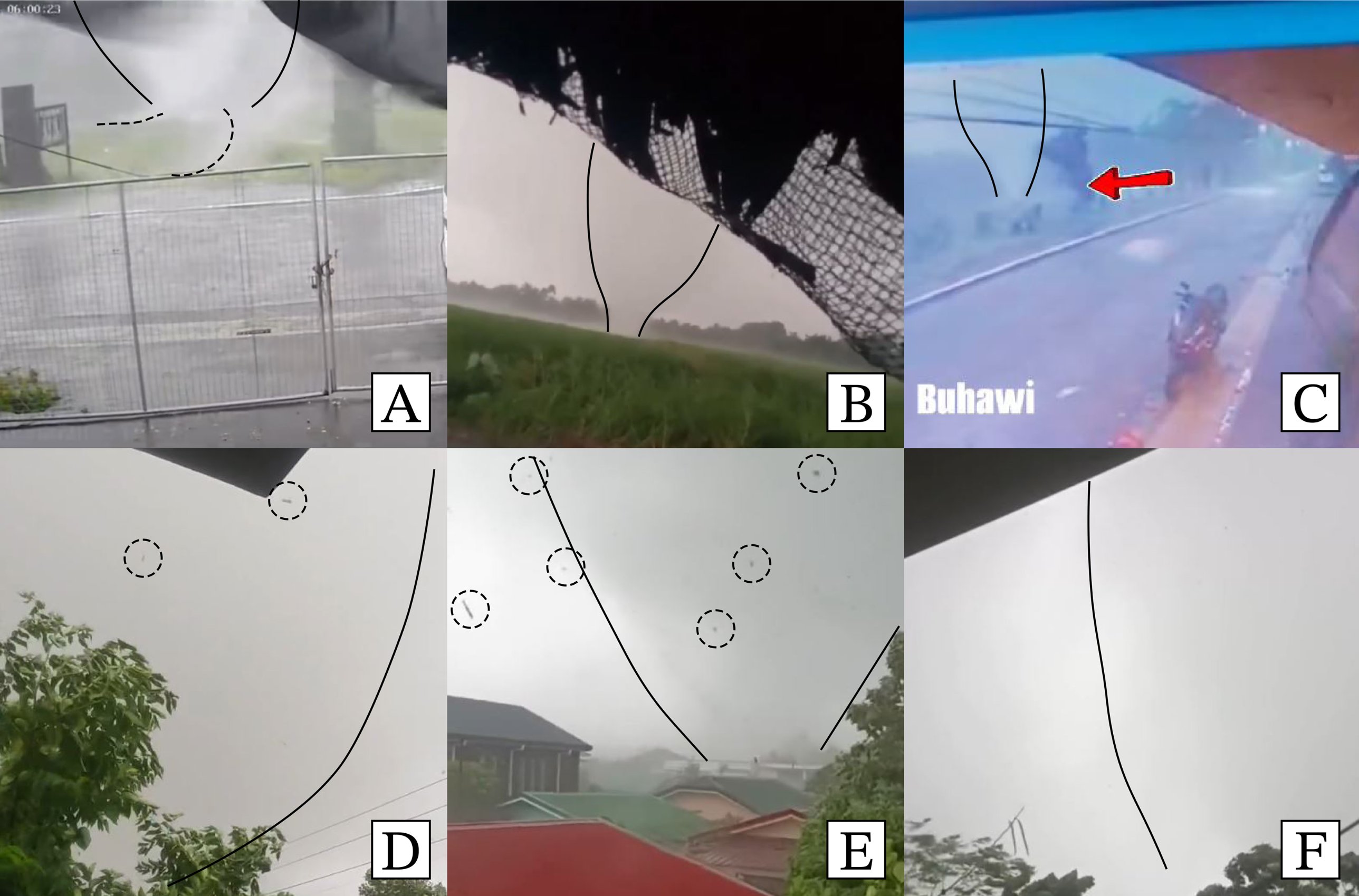}
\caption{Video surveillance of T1 and T2 (as labeled in Figure~\ref{fig_2}) that impacted Brgy. Lag-on and Magang in Daet, Camarines Norte, respectively. Courtesy of (a) Dong Medina Dela Cruz, (b) Vivian Clavite, (c) Marcelyn Centeno Carillo, (d) Wilson Obog Pajares, (e) Albert Elnar Tagolino, and (f) Chai Gallardo. Dashed circles in 3d and 3e are debris lofted by T2-Magang Tornado.}
\label{fig_3}
\end{figure*}

\begin{figure*}[!t]
\includegraphics[width=\textwidth]{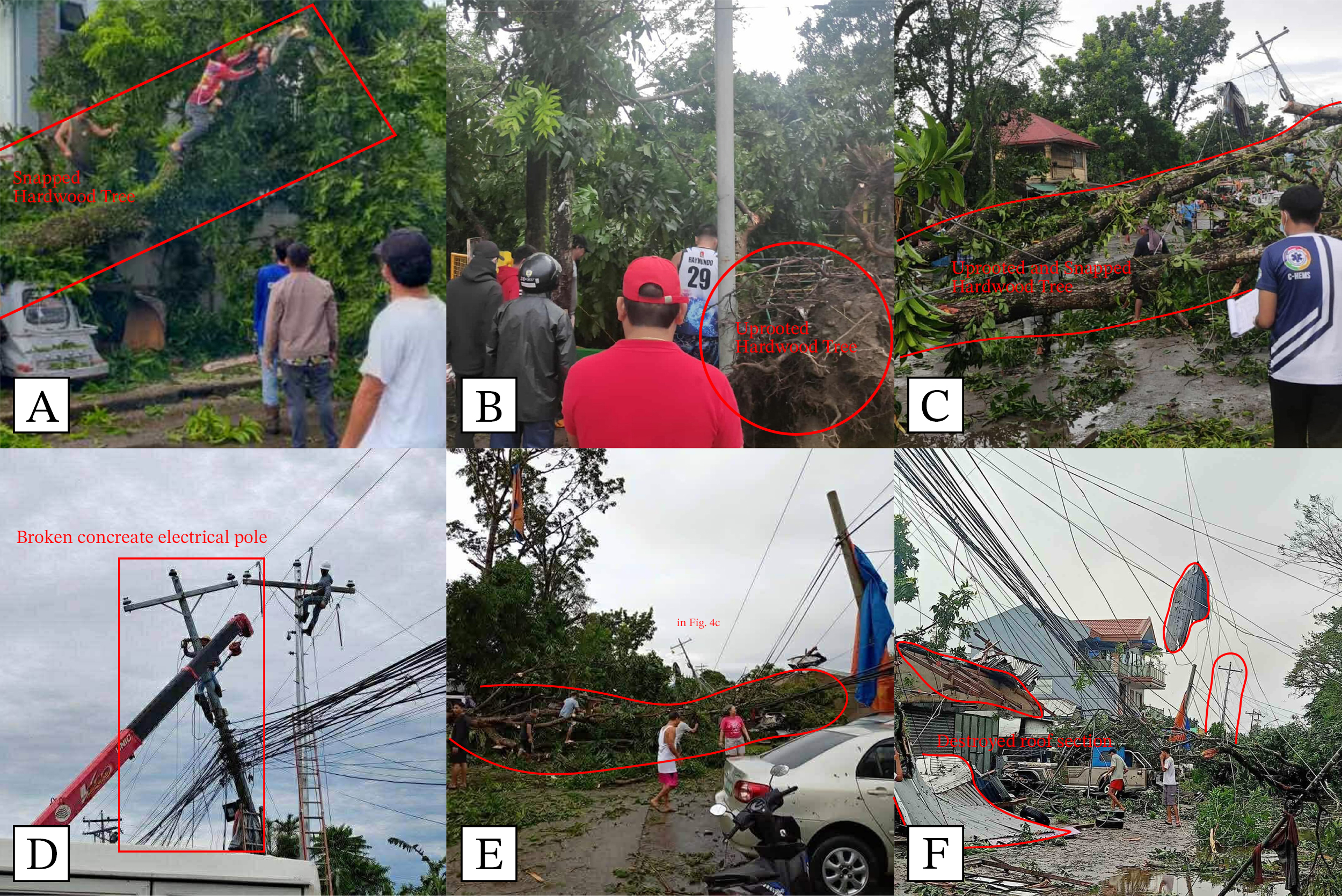}
\caption{Ground photos of the damages brought by T2 (as labeled in Figure~\ref{fig_2}; otherwise known as the Magang event).}
\label{fig_4}
\end{figure*}

Each isopleth (black dashed lines) is labeled according to its associated tornado signature. T1, the Lag-on tornado, is supported by multiple eyewitness accounts confirming a tornado on the ground (Fig.~\ref{fig_3}), and is classified as confirmed. T2, the Magang tornado, is supported by a ground damage survey along a clearly defined swath in the Fairview 2 area of Magang, Daet (Fig.~\ref{fig_4}), and is likewise classified as confirmed. A third damage track, designated T3 and attributed to the Cahabaan tornado, was identified through examination of the PlanetScope imagery together with the radar analysis presented later. Unlike the preceding cases, this event was not reported in official documentation and was not widely documented across online platforms, including social and news media. Therefore, it is classified as identified rather than confirmed.

Normalized percent change in NDVI was computed with pixels of pre-event NDVI below 0.30 masked to avoid the numerical instability that arises where the denominator is small, and statistics evaluated without clipping within 200 m corridors centered on each track. In Fig.~\ref{fig_2}c, T1 appears as a well-defined east--west elongated swath originating near Lag-on and extending into adjacent rice and crop fields, T2 as a southeast--northwest-oriented swath originating over agricultural fields and extending toward the more densely populated area of Magang, and T3 as a generally east--west-oriented zone comprising multiple patches of change. The relative reduction within the corridors was 23.6\% for T1, 28.4\% for T2 and 32.3\% for T3, against 16.1\% for the scene as a whole, indicating that a minority of pixels along each track lost vegetation cover relative to the background. These values describe the magnitude of change within a corridor and not its spatial organization, and the two differ for T3. Although T3 carries the largest reduction, its change pattern is the least spatially continuous of the three, comprising discrete patches rather than a complete swath. Discontinuous damage is consistent both with intermittent tornado contact and with unrelated field-scale agricultural change, and the NDVI statistics alone cannot distinguish between them. Therefore, the T3 track is established by the radar evidence presented in subsequent sections rather than by the imagery. Track lengths were approximately 1 km for T1, $\sim$1.4 km for T2 and $\sim$2.1 km for T3.

For T2 (the Magang tornado), the ground survey documented impacts to one- and two-story residential structures, several electrical transmission lines, a church, and a junior/senior high school building. Based on the corresponding Damage Indicators (DI) and Degrees of Damage (DoD) observed across multiple locations, these are consistent with an intensity rating of IF2--IF2.5 on the International Fujita (IF) Scale. The IF Scale, developed by the European Severe Storms Laboratory, rates intensity using instantaneous three-dimensional winds at the height of the damage rather than a 3-s averaged horizontal wind at 10 m AGL \citep{Groenemeijer2023}.\footnote{IF Scale Guidelines (v1.0e): \url{https://www.essl.org/cms/wp-content/uploads/IF-scale_v1.0e.pdf}} It is adopted here because its damage indicator inventory is not calibrated exclusively to the CONUS construction scheme, admits half-step ratings, and it provides explicit criteria for leaving an event unrated. This severity rating corresponds approximately to Rating No. 2 to 3 in the Enhanced Fujita (EF; EF2--EF3) scale, but the comparison is indicative only as the two scales differ in reference height, wind-averaging convention, and indicator inventory, and no formal conversion exists.

Meanwhile, T1 and T3 tracked entirely across open farmland, so neither could be rated through the full DI/DoD framework. T1 passed over open terrain without noticeable damage to structures. Following Section 4.2 of the IF Scale, which permits an IF0 rating where a tornado passes over vulnerable objects that are not damaged, T1 is assigned IF0. Along the T3 path, crops were likely flattened intermittently, consistent with the patchy NDVI change, while the radar debris signature described in subsequent sections indicates that vegetative debris was lofted. Because the IF Scale inventory contains no damage indicator for standing crops, this rating cannot be derived from a DoD and is instead assigned by expert judgment as a lower bound: T3 is rated IF1. The absence of a crop damage indicator is a recognized limitation of applying damage-based intensity scales in agricultural landscapes, where crops rather than structures dominate the exposed surface, and ratings for tornadoes confined to farmland will systematically tend toward the low end of the scale.

To support the IF2.5 classification of the Magang tornado, a comprehensive damage survey was conducted using the adopted framework, focusing on the most severe structural and environmental impacts observed in the field. Indicators of three different types reached IF2 at several separate locations along the track. The snapping of large, strong trees (TR-S, DoD 5; $\geq$ IF2, $\geq$ 60 \ms, $\geq$ 220 km h$^{-1}$) was a recurring indicator of high-intensity winds, as seen in the vegetation surrounding public transport and private vehicles (Figs.~\ref{fig_4}a, \ref{fig_4}c, \ref{fig_4}f-unhighlighted, S2, S3, S5, S9 and S10). Building indicators (BR-D, DoD 2; $\geq$ IF2) revealed instances where strong roof structures were completely blown over or destroyed (Figs.~\ref{fig_4}f, S2, S11). And the overturning of a three-wheeled public transport vehicle (VH-C, DoD 2; IF2, 60 \ms) provided a mobile indicator independent of vegetation condition and construction quality (Fig. S1). The full inventory, giving the damage indicator, degree of damage and rating assigned to each photograph, is provided in Table S1 of the supplementary document.

Two entries in Table S1 carry a rating adjusted above the tabulated value, and the basis of each is stated here. Several tree entries in the IF Scale, including TR-S DoD 3 and DoD 5, are expressed as open lower bounds (e.g., $\geq$ IFx). Hence, the tabulated value is the minimum wind consistent with the observed failure rather than a fixed assignment as a higher rating within the same class is admissible where the condition of the specimen or the combination of failure modes observed on it, indicates stronger loading. No closed entry, such as VH-C DoD 2 = IF2, was adjusted. The first adjustment concerns a large, old tree that was uprooted and caused the two fatalities (TR-S, DoD 3; tabulated $\geq$ IF1.5, $\geq$ 50 \ms; Figs.~\ref{fig_4}b and S6). The root plate lifted intact together with its attached soil mass, indicating failure of a well-developed root system rather than of a shallow-rooted specimen, and the rating was accordingly adjusted to IF2 (60 \ms). We note the opposing consideration that the soil had been saturated by the concurrent heavy rainfall, which lowers the wind required to uproot a tree while leaving the snapping threshold of a sound stem unaffected. This entry is therefore conservative relative to the snapping indicators recorded nearby. The fatalities are reported as an impact of the event and are not used as evidence of intensity. The second adjustment yields the rating of the event. A strong, well-rooted tree was both snapped and stripped of a substantial part of its large branches (Fig. S16). Trunk snapping alone establishes $\geq$ IF2 for a category S tree, while major debranching of a strong tree with the trunk intact (TR-S, DoD 6) is tabulated at IF2.5 (70 \ms). A specimen exhibiting both failure modes therefore lies between DoD 5 and DoD 6, and the entry is recorded as DoD 5 with the rating adjusted to IF2.5 (Item \#31, Table S1).

We emphasize that this value rests on this single indicator, whereas at least IF2 intensity along the track is supported by the three independent indicator types described above. A single indicator can legitimately carry a rating under the IF Scale, as the maximum intensity of an event is the highest-rated combination of DI and DoD found along the path. We recognize that the framework advises care where that indicator is a tree, since strength class, health and rooting can be judged only from the specimen's appearance: the category S classification here rests on photographic evidence of a mature, sound and well-rooted specimen, internal decay cannot be excluded by external inspection, and the position of the tree relative to the vortex center is unknown. Together with the approximately $\pm$20\% range that the framework attaches to its central values (IF2 spanning $\sim$48--72 \ms), these considerations place the adjustment to IF2.5 within the stated precision of the framework.

\subsection{Synoptic Conditions}\label{sec32}

\begin{figure*}[!t]
\includegraphics[width=\textwidth]{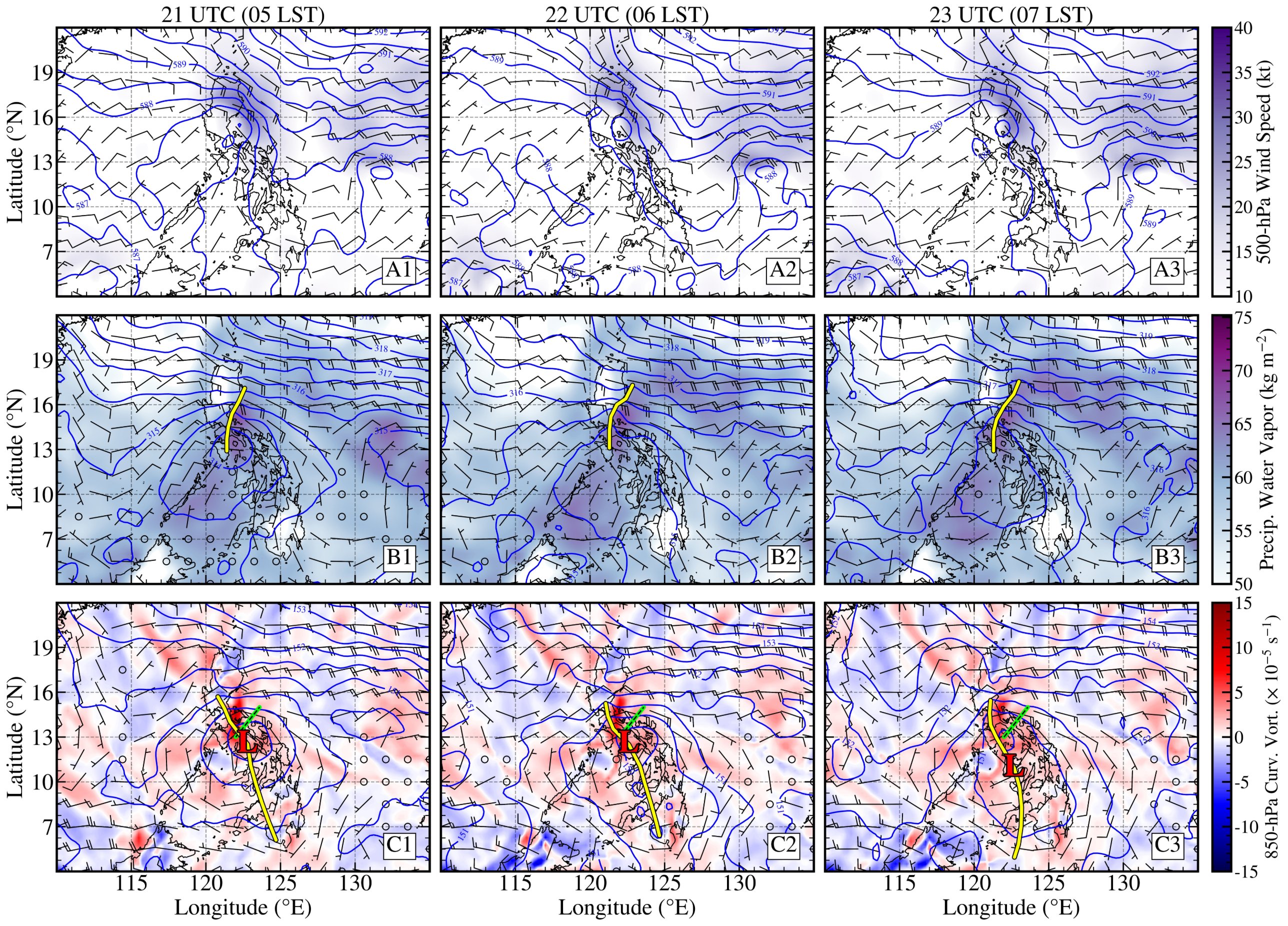}
\caption{Synoptic environment of the Philippine archipelago before (1; 21 UTC/05 LST), during (2; 22 UTC/06 LST), and after (3; 23 UTC/07 LST) the event. (a) 500-hPa Geopotential heights (blue; dam) and winds (kt). Contours are wind speeds $>$ 10 kts. (b) 700-hPa Geopotential heights (blue; dam), precipitable water vapor (contour), and winds (kt). (c) 850-hPa Geopotential heights (blue; dam), curvature vorticity ($\times$ 10$^{-5}$ \si), and winds (kt). Yellow lines correspond to the axis of the inverted trough. The dashed black line (green outline) corresponds to the cross section derived in the subsequent figure.}
\label{fig_5}
\end{figure*}

To investigate the mechanism that facilitated convective initiation, we analyzed the synoptic environment on multiple constant-pressure levels (Fig.~\ref{fig_5}). At 500 hPa (Fig.~\ref{fig_5}a1--a3), a weak mid-level height perturbation ($\lesssim$1 dam deep) was located just off the eastern coast of Luzon, near the Polillo Islands, with 15--25-kt southeasterly flow along its southeastern flank over the event area. In contrast to the EF2 tornado that struck Candating, Arayat, on 27 May 2024 \citep{Capuli2026}, no 500-hPa height falls accompanied this event as the 12-hr height change at the event location was weakly positive (+0.2 -- +1.3 dam) during 2100--2300 UTC indicative of little synoptic-scale mid-level forcing for ascent (lowering of 500-hPa heights). Instead, the upstream/upward flow may have provided persistent ascent and moisture transport; especially in the low-levels, which in turn can promote atmospheric destabilization. Regardless of whether the mid-level perturbation appeared relatively weak, it favored discrete, if not semi-discrete, development of convective cells \citep{Schumann2010}, whereas stronger synoptic forcing typically supports mixed or linear convective modes e.g., quasi-linear convective systems \citep[QLCS;][]{Bunkers2006,Dial2010}\footnote{The Oklahoma--Kansas tornado outbreak of 3 May 1999 was an example of weak-ill defined forcing \citep{Thompson2000}. QLCS is a subset of MCS and is synonymous to a Squall Line system.}.

At 700-hPa (Fig.~\ref{fig_5}b1--b3), the height perturbation becomes more pronounced (yellow line), exhibiting a relatively broad center over the Visayas region. This perturbation likely reflects the low-level easterly trough axis and is accompanied by precipitable water (PWAT) values $>$ 60 kg m$^{-2}$ during the analysis period along the area of interest, indicating a deeply moist lower troposphere. Such a structure is consistent with the classical description of synoptic-scale wave disturbances in the equatorial western Pacific by \citet{Reed1971}, who documented a deep moist layer and maximum convective activity in the vicinity of the trough, and with subsequent characterizations of tropical-depression-type disturbances in this basin \citep[e.g.,][]{Takayabu1993,Lau1990}. Analogous moisture structures have been described for easterly waves in the eastern Pacific ITCZ \citep{Serra2008}, where barotropic and baroclinic conversions act as eddy energy sources, although the applicability of those energetics to the western Pacific is not assumed here.

Complementing the 700-hPa analysis, the 850-hPa fields (Fig.~\ref{fig_5}c1--c3), representative of the planetary boundary layer, reveal a tropical easterly wave characterized by an elongated low-pressure system (circulation center marked by a red ``L''), with its wave axis extending from Central Luzon to Mindanao. The easterly inverted trough is identified using 850-hPa curvature vorticity ($\zeta_{CV}$), a component of relative vorticity, where enhanced positive $\zeta_{CV}$ \citep[$>$ 1.0 $\times$ 10$^{-5}$ \si;][]{Brannan2019,Elless2019} along the base of the open wave indicates low-level cyclonic circulation over the area of interest. \citet{Serra2008} further noted that the convective maximum in such low-pressure systems is typically located behind (i.e., upstream of) the trough axis. In this case, the collocation of the vorticity maximum, modest moisture advection, and sufficiently high PWAT values suggests that the tilted structure of the tropical wave promoted low-level convergence over eastern Luzon. Therefore, large-scale advective processes associated with the tropical wave and its trough axis likely played a crucial role in establishing an environment favorable for severe thunderstorm initiation, including supercells capable of tornadogenesis \citep{Doswell2001}.

\begin{figure}[!t]
\includegraphics[width=\columnwidth]{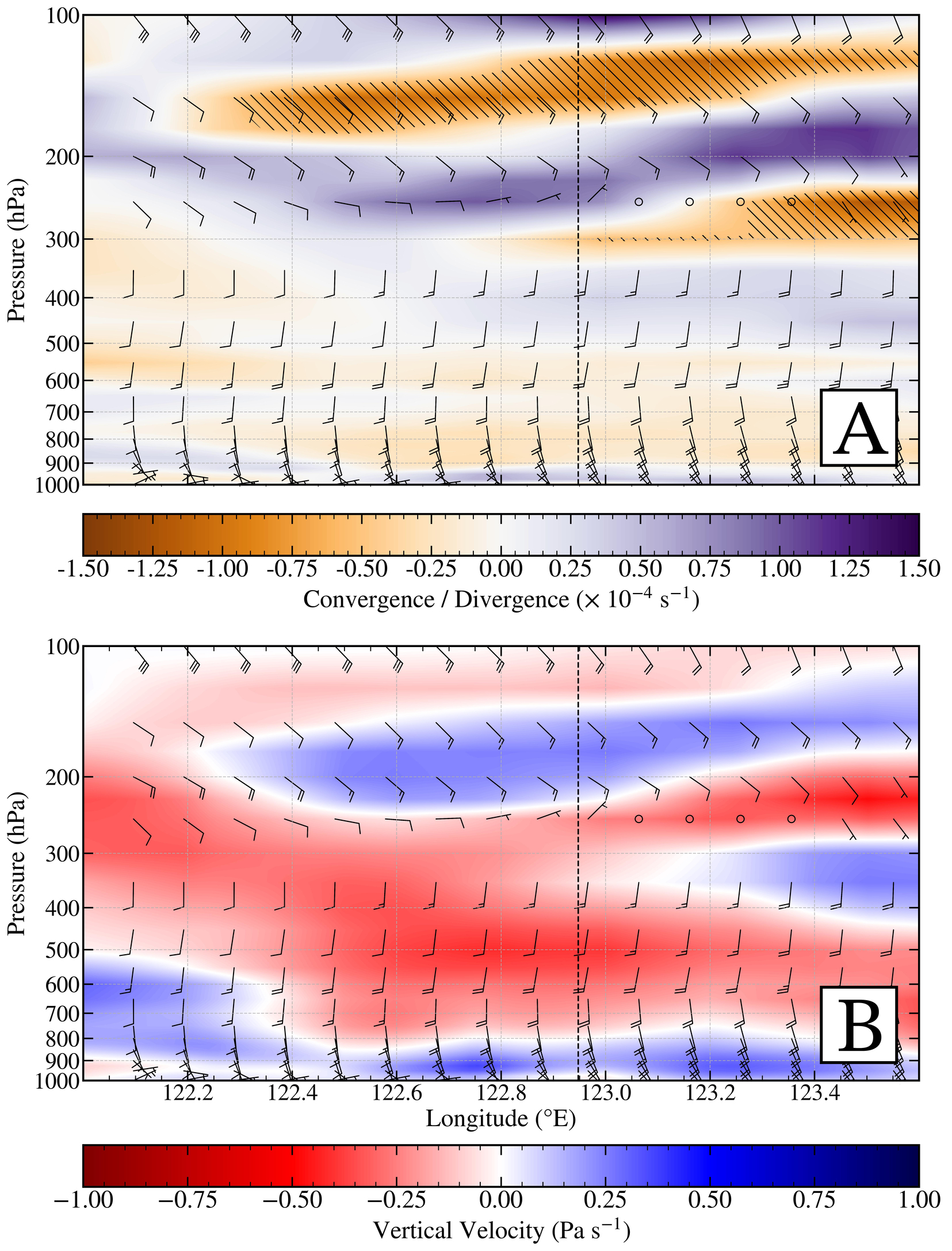}
\caption{22 UTC/06 LST ERA5 Vertical Cross Section of; (a) Convergence / Divergence (\si) with hatched area of Convergence $<$ $-$0.50 $\times$ 10$^{-5}$ \si\ and (b) Vertical Velocity (Pa s$^{-1}$) with hatched area of Vertical Velocity $<$ $-$0.50 Pa s$^{-1}$. Both cross sections are accompanied by tangential and normal components (kt).}
\label{fig_6}
\end{figure}

Several dynamic cross sections across the regions of interest are shown in Figure~\ref{fig_6} at 22 UTC (06 LST). In particular, negative values of the convergence ($\nabla\cdot\mathbf{F} < 0$) are found in the low-levels up to mid-levels (500-hPa) in Figure~\ref{fig_6}a are indicative of forcing for ascent is still present in the environment especially along the upstream section of the inverted trough. This was complemented by negative vertical velocity anomalies reaching $\sim$$-$0.3--0.5 Pa s$^{-1}$ in the lower troposphere, particularly between 800 to 300-hPa pressure levels (Fig.~\ref{fig_6}b). These upward vertical motions serve as an important ingredient for initiation and sustaining deep convective storms and its updrafts \citep{Wakimoto2004}. These dynamical and synoptic processes may have been further modulated by local topography via thermal winds i.e., slope and mountain-induced circulations, given the proximity of the outbreak sequence in the Camarines Norte Mountain Range ($>$ 300 mASL) with the highest being Mt. Labo ($\sim$1,540 mASL) and elevated terrain of Bicol Natural Park and its Mt. Colasi (also known as Mt. Balagbag; $\sim$960 mASL) to the west and south of Daet, Camarines Norte, respectively. Such orographic influence may have caused the enhancement of convergence and vorticity as southeasterly-to-easterly winds impinged on the windward slopes of higher terrain, likely enhancing low-level confluence and horizontal shear along the foothills and coastal plain near the area of interest, thus promoting lifting of warm, moist air parcels and initiating deep convection capable of tornadogenesis \citep{Kirshbaum2018}. We discuss further this terrain enhancement feature in the subsequent section.

\subsection{Mesoscale Environment}\label{sec33}

\begin{figure*}[!t]
\includegraphics[width=\textwidth]{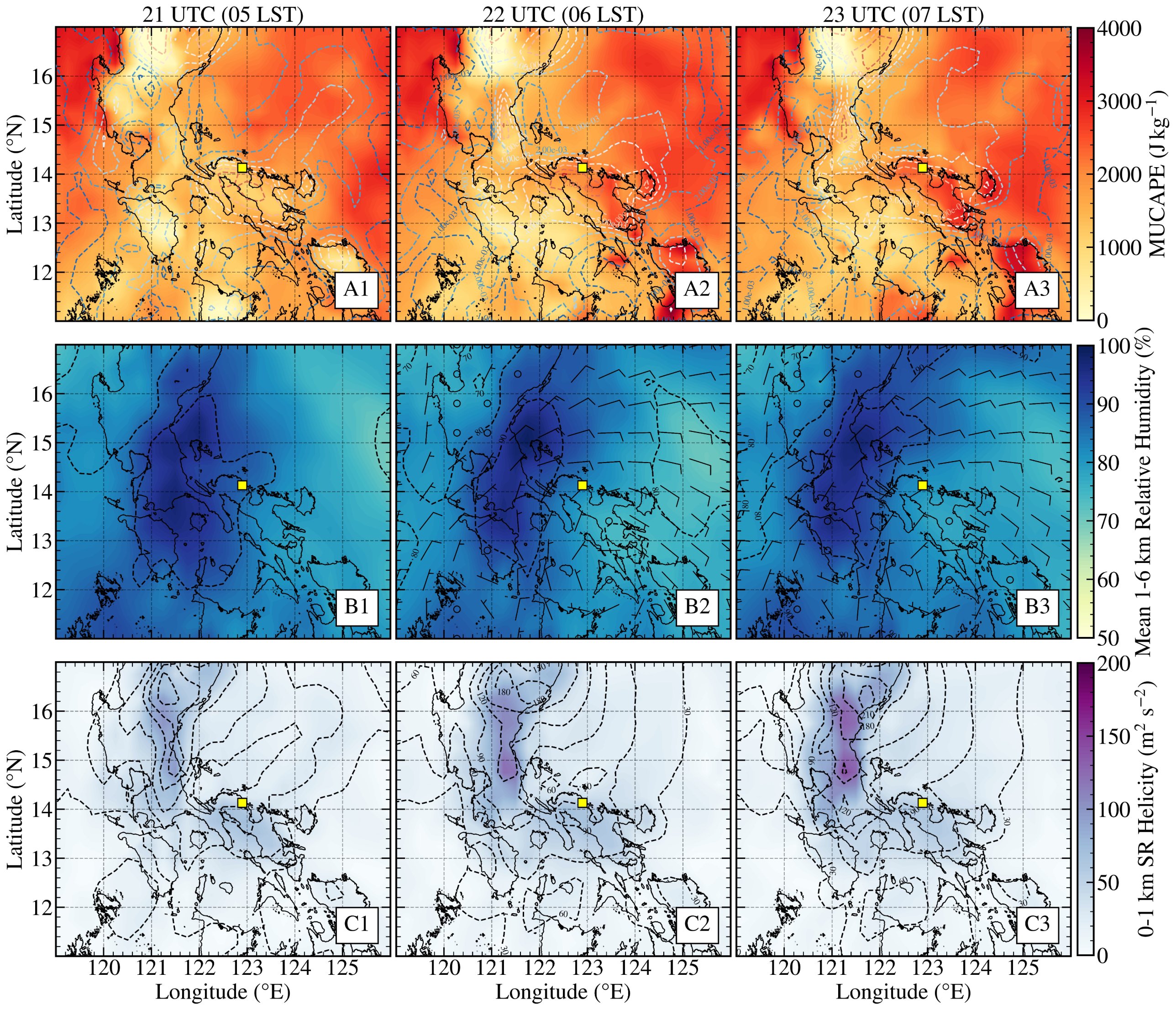}
\caption{Convective and Kinematic environment of the Luzon landmass before (1; 21 UTC/05 LST), during (2; 22 UTC/06 LST), and after (3; 23 UTC/07 LST) the event. (a) MUCAPE (\Jkg) and 0--1 km Streamwise Vorticity (gradient dashed; $\times$ 10$^{-3}$ \si). (b) Average 1--6 km Relative Humidity (\%), including Average 1--3 km Relative Humidity (black dashed; \%), and 10-m Winds (kt). (c) 0--1 km and 0--3 km SRH (black dashed; \mss). The case area is demarcated in yellow box.}
\label{fig_7}
\end{figure*}

Analyzing the atmospheric conditions during the early morning hours is essential for understanding the background environment that influenced storm development. The mesoscale convective conditions are illustrated in Figure~\ref{fig_7} using ERA5 reanalysis from 21 to 23 UTC (05 to 07 LST).

As shown in Figures~\ref{fig_7}a1--a3, CAPE measurements between 1500--2000 \Jkg\ were present in the area of interest indicative of ample atmospheric instability. This also coincides with ambient near-surface vorticity supportive of tornado-producing supercells. The overlap of these conditions is commonly observed in the Great Plains' Tornado Alley during spring-season severe weather outbreaks. In such regimes, the low-level shear vector often exhibits pronounced clockwise curvature with height, forming a `sickle/hook shaped' profile that is favorable for the development of tornadic supercells \citep{Weisman1986,Thompson2000,Coffer2020,Nixon2022}. In Figures~\ref{fig_7}b1--b3, the low-level and mid-level RH fields show a moist environmental profile, particularly before the event, with RH$_{13}$ $>$ 80\% across the Daet, Camarines Norte region (highlighted in the yellow box). To the west of the storm initiation zone, RH above the cloud base exhibited RH$_{13}$ $>$ 90\%, which may have contributed to the maintenance of additional convective development. \citet{Peters2022a,Peters2022b} emphasized that free-tropospheric moisture is fundamental to convection initiation, governing the likelihood of convective updraft formation. This moist environment was superimposed with modest 0--1 km and 0--3 km SRH exceeding 70--100 \mss\ within both layers, as illustrated in Figures~\ref{fig_7}c1--c3. These conditions indicate that the combination of thermodynamic instability and kinematics was enhanced by the approach of the inverted trough and the advection of moist air via the prevailing southeasterly-to-easterly flow. This in turn facilitated increased CAPE and provided sufficient buoyancy to sustain robust convective updraft, capable of ingesting vorticity for tornadogenesis \citep{Pucik2015}.

\begin{figure*}[!t]
\includegraphics[width=\textwidth]{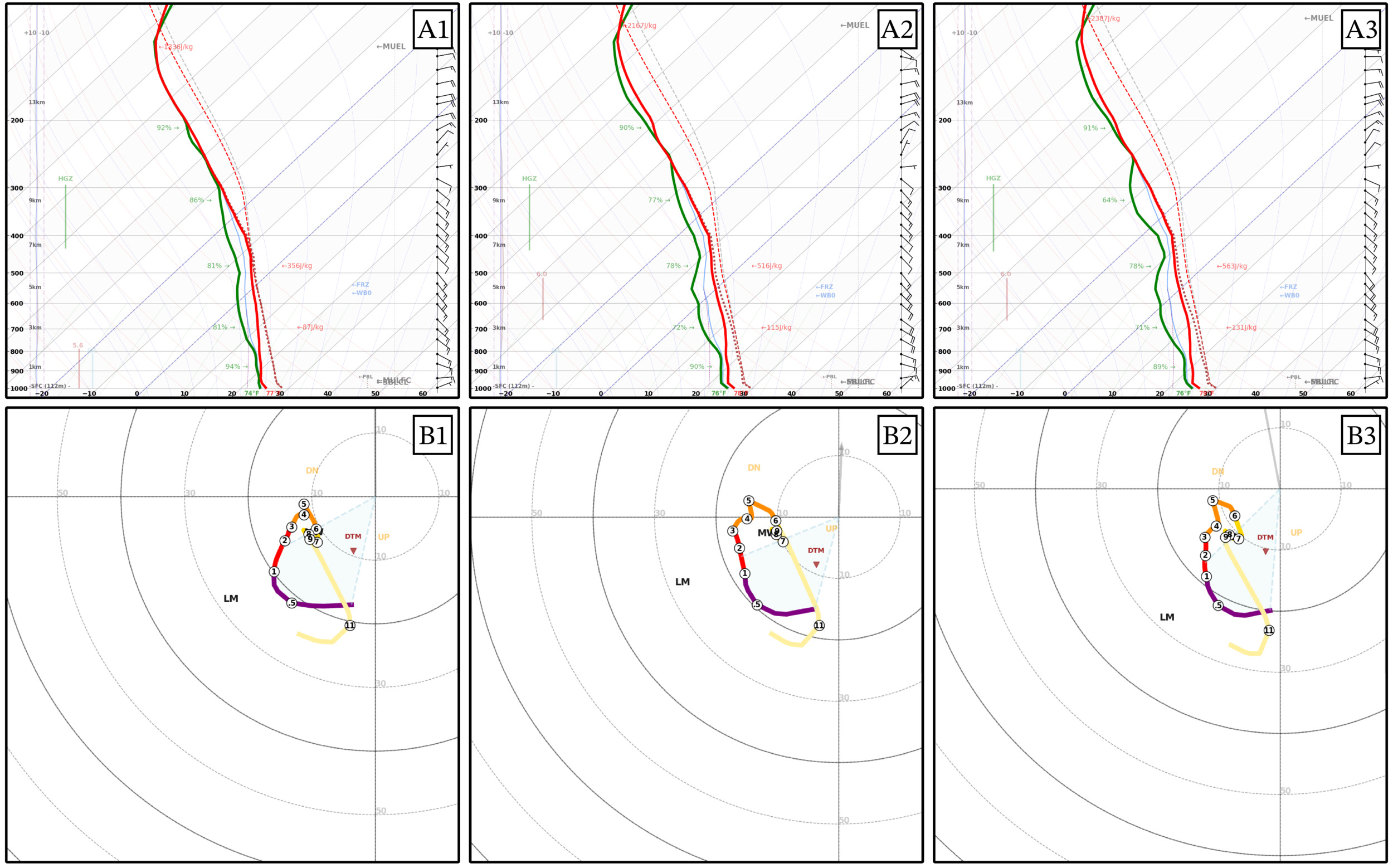}
\caption{ERA5 Sounding Profile before (1; 21 UTC/05 LST), during (2; 22 UTC/06 LST), and after (3; 23 UTC/07 LST) the event. (a) Thermodynamic profile along the area of interest. Annotations to the thermodynamic profiles include the SBLCL, PBL, MULFC, FRZ, and MUEL. (b) Storm-relative Hodographs associated with the thermodynamic profile above. Annotations to the wind profiles include the individual Bunkers' Storm Motion, Deviant Tornado Motion, and Corfidi's components.}
\label{fig_8}
\end{figure*}

To further examine the evolving thermodynamic and kinematic environment, Figure~\ref{fig_8} presents the derived model sounding and hodographs at 21, 22, and 23 UTC (05, 06, and 07 LST 13 September 2025) representing the before, during, and after the tornado event. The model soundings (Fig.~\ref{fig_8}a1--a3) exhibit a moist tropical profile characterized by a small dewpoint depression\footnote{Dewpoint depression means the difference between the temperature and dewpoint temperature. Lower T--T$_\text{D}$ spreads mean moist profile.} from the near-surface to around 800-hPa, the absence of a significant inhibition layer (CIN $\sim$ 0 \Jkg), and low LCLs ranging from 150--200 m AGL. Additionally, the profiles indicate a moist PBL with RH$_{13}$ $>$ 80\%, with mid-level lapse rates improved from 5.39\,\dg C km$^{-1}$ to close to 6\,\dg C km$^{-1}$ by the time of the tornado occurrence. The moisture quality between the lower and mid-tropospheric levels (represented by RH$_{16}$) only decreased subtly due to the start of surface heating, but this subtle decrease in mid-level moisture aided in the thermodynamic environment to be favorable for convective development. Before the event (21 UTC/05 LST), the profile is mostly defined by narrow (`skinny') CAPE profile resulting in undiluted CAPE of 1336 \Jkg\ with most of the instability layer confined above 400-hPa. Due to the nature of the thermodynamic profile, CAPE$_{03}$ were also minimal $<$ 100 \Jkg\ thereby initially limiting parcel buoyancy in the low-levels. However, at 22 UTC (06 LST), coinciding with the tornado occurrence, CAPE exceeds 2167 \Jkg\ with low-level CAPE breaching $>$ 100 \Jkg\ threshold (115 \Jkg\ to be exact) due to the subtle decrease of mid-level RH while steepening of MLLR. The abundance of low-level instability in the environment is important in the tornadogenesis process as it can lead to higher vertical velocities near the surface. \citet{markowski2010} stated that once a tornado becomes established, tilting of the surface-layer horizontal vorticity by the extreme vertical velocity gradient associated with the updraft itself probably contributes to the near-ground vertical vorticity in a significant way. This substantial increase in instability, coupled with low LCLs, highlights a thermodynamic environment that favored surface-based severe storms capable of producing tornadoes \citep{GRUNWALD2011,Matsui2016}.

One defining feature of the modeled hodographs associated with this event was their unusual orientation (Fig.~\ref{fig_8}b1--b3). The storm-relative hodograph exhibited a low-level clockwise rotation of the wind profile where winds originate from the east-southeast diverging from the classical configuration (south-southwesterly), indicating a unique shear environment. This wind profile produced a DLS vector that supported supercells with northeastward motion, with right-moving (RM) supercells veering further northward. Such hodograph shape is unusual and has not been documented elsewhere wherein easterly windflow manifested as hook-shape/curved hodographs and eventually produced tornadoes. Unlike in the warm-season North American Monsoon regime \citep[NAM;][]{Adams1997,Blanchard2011} where low-level southwesterlies accompanied by northeasterly winds aloft produces hodographs that can favor hook echo signatures in the northwestern quadrant of supercells, this east-southeasterly-driven hodograph favors hook echo located on the southern to southeastern side of the storm (i.e., lower-right flank of the storm). In this case, RM storm motions were directed toward $\sim$355--360\dg\ at speeds of 12--17 kt (6--9 \ms) with a vector-averaged mean motion of $\sim$150\dg\ at 16 kt (8.23 \ms). Although the presence of hodograph curvature is evident, the LLS remains weak (between 6--7 \ms) as compared to tornadic supercells recorded in the Great Plains with median LLS at around 11 \ms\ \citep{Nixon2022}. Regardless, this still affirms work by \citet{Coffer2020} that significant tornadoes tend to have sufficient concentrations of SRH in the low-levels (between half kilometer to 1 km). Modest \VSR\ remained at $\sim$17--19 kt ($\sim$8.5--10 \ms) which helped mitigate the negative impacts of entrainment dilution on supercell's updraft. An ECAPE-to-CAPE ratio of 71--80\% was computed throughout the sounding analysis, indicating that the storm's updraft realizes most of the undiluted CAPE. This suggests that most of the available convective energy was effectively converted into updraft kinetic energy \citep{Romps2010,Peters2019}.

Additional kinematic parameters derived from the ERA5-derived storm-relative hodographs further support a favorable tornadic environment. The ambient near-surface horizontal vorticity is sufficient, with $\omega_{s500}$ $\sim$0.009 \si, and is closely aligned with the storm-relative flow near the time of tornadogenesis at 22 UTC (06 LST), with $\widetilde{\omega}_{s500}$ $>$ 90\%. In the lowest 1 km the streamwise vorticity remains comparable in magnitude, 0.008 \si\ at the time of the event, and similarly streamwise throughout the analysis period ($\widetilde{\omega}_{s01}$ $>$ 90\%). Although modest, these magnitudes approach the $\sim$0.01 \si\ values characteristic of environments supporting tornadic supercells and exceed the $\sim$0.005 \si\ lower threshold \citep{Coffer2017,Coffer2019}. The two quantities are complementary as the magnitude sets the vorticity available to the updraft, whereas the streamwise fraction sets how much of it is usable, since vorticity aligned with the storm-relative inflow is ingested, tilted, and stretched into the vertical rather than being carried around the storm's updraft.

Although the model soundings did not resolve enhanced low-level shear (LLS $<$ 10 \ms), given its coarse resolution, topographically induced modifications were likely present. We therefore offer the following as a hypothesis rather than a result. Specifically, this can be attributed to terrain-driven processes where low-level southeasterly flow impinged on the surrounding high terrain and was subsequently deflected/channeled along the foothills and coastal plain, favoring a quasi-stationary low-level confluence boundary that concentrates horizontal shear and cyclonic vertical vorticity i.e., a terrain-anchored convergence--vorticity zone, analogous in mechanism to the Denver Convergence Vorticity Zone (DCVZ)\footnote{DCVZ is a well-documented mesoscale phenomena of enhanced low-level convergence and cyclonic vorticity that forms east of Denver as moist southeasterly upslope flow interacts with the Palmer Divide and terrain-modified flow from the Rocky Mountain foothills, focusing convective initiation.}, can precondition the environment for convective initiation and locally enhanced rotation via convergence-forced ascent and vorticity stretching \citep{Szoke1991,Wilczak1992,Kirshbaum2018,Trier2019}. Such terrain influences are consistent with the numerical simulation results of \citet{Bosart2006} for the Great Barrington, Massachusetts F3 tornado, who showed that tornadogenesis becomes more likely when local terrain-driven perturbations to the low-level flow enhance near-storm convergence and vertical vorticity. When this enhancement occurs in an environment characterized by ample streamwise vorticity in the lowest half to a kilometer, collocated with ample CAPE$_{03}$ and upper-bound vertical vorticity in the low-level mesocyclone ($\zeta_{LLM}$) exceeding 0.01--0.02 \si\ throughout the analysis period, convective-scale stretching and tilting processes can proceed efficiently, increasing the likelihood that a supercell produces a tornado \citep{Coffer2023}.

The composite parameter WMAXSHEAR, which combines instability i.e., CAPE and DLS, exceeded 755 \mss\ near the time of the tornadic event, representing a significant increase from the initial sounding of $<$ 500 \mss. This magnitude falls well within the climatological range associated with environments favorable for severe convection and tornadogenesis \citep{Taszarek2020b}. In addition, the Supercell Composite Parameter (SCP) reached 2.6, while the Significant Tornado Parameter (STP) was 0.6. Although modest, these values fall within the lower tail of the climatological distributions for significant tornadoes reported by \citet{Thompson2003,Thomspon2012}, in which a fraction of significant tornadoes occurred with STP $<$ 1. Hence, it does not preclude significant tornadoes and the modest value here is not in tension with the IF2.5 rating, which rests on damage evidence rather than on the environment. Overall, these composite parameters suggest that the background environment was favorable for tornadic supercells, despite the parameters not being able to account for terrain-modified flow and/or surface boundaries. Moreover, additional mesoscale influences, particularly terrain-induced modifications (as discussed above), may have enhanced the ambient kinematics and buoyant instability available for storm development. However, the persistence, spatial extent, and tornadic nature of these storms may also be attributed to internal storm-scale dynamics, which will be addressed in the following subsection. A summary of key thermodynamic and kinematic indices derived from the ERA5 model sounding is provided in Table~\ref{Table_2} for comparison and reference.

\subsection{Storm Development and Life Cycle}\label{sec34}

\subsubsection{Supercells' Characteristics}\label{sec341}

\begin{figure*}[!t]
\includegraphics[width=\textwidth]{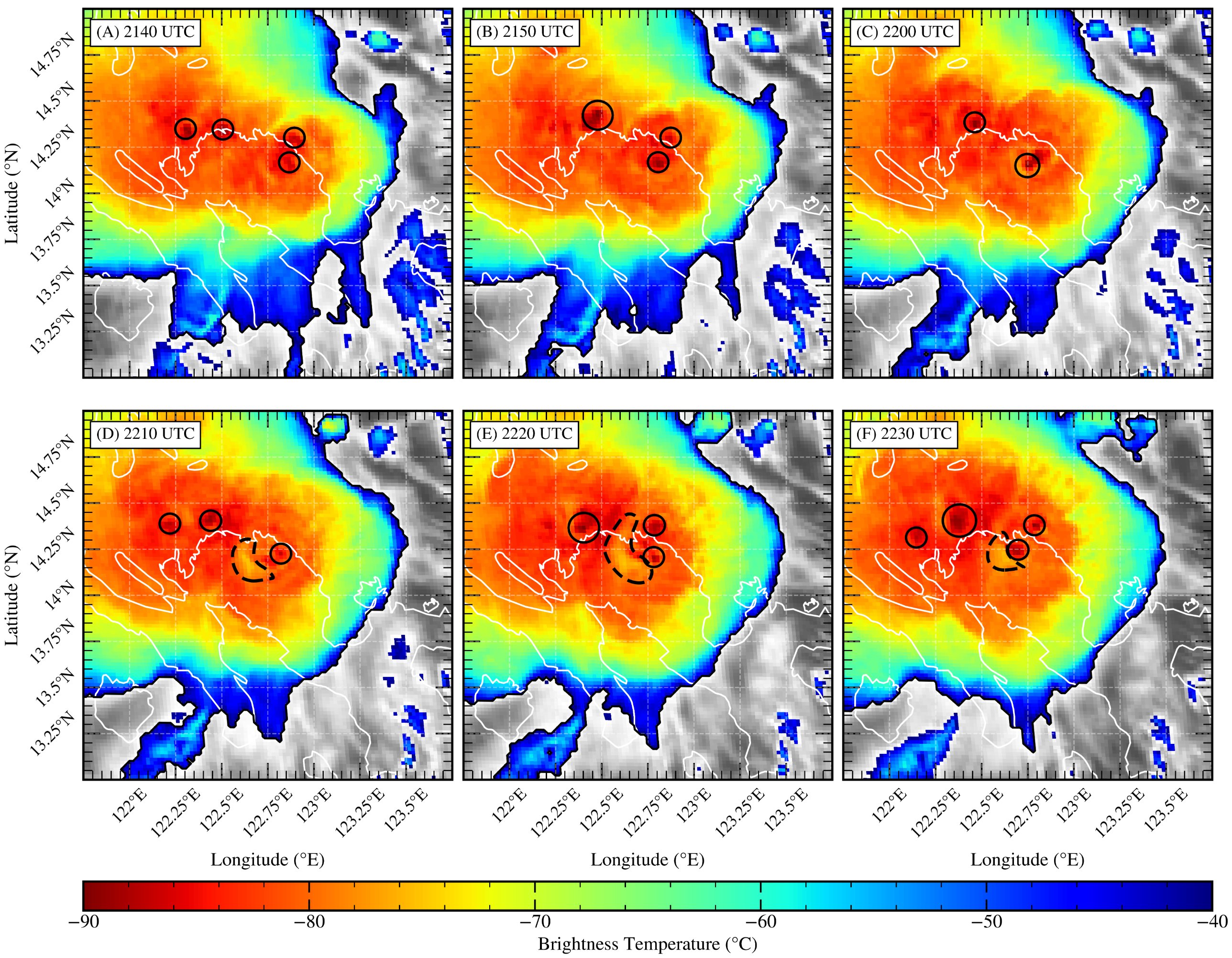}
\caption{HIMAWARI-9 AHI 10.4 $\mu$m B13 depicting MCS evolution of the 13 September 2025 tornadic supercell. Each scan is every 10 minutes, starting at 2140--2230 UTC (0540--0630 LST; a--f). Black circles are areas of identified OTs, while the dashed lines are the AACPs.}
\label{fig_9}
\end{figure*}

The temporal evolution of the convective storm is illustrated in Figure~\ref{fig_9} using 10-minute interval scans from HIMAWARI-9 between 2140--2230 UTC (0540--0630 LST). While satellite data prior to 2140 UTC and beyond 2230 UTC are not visualized, the relevant information outside this period is discussed to provide a descriptive context of the storm's life cycle.

At 2100 UTC (05 LST), the Southern Luzon Regional Services Division of DOST-PAGASA (otherwise known as PAGASA-SLPRSD), issued its Regional Weather Forecast\footnote{RWF: \url{https://www.facebook.com/share/p/18jTVcruRd/}} within its domain area and asserted that cloudy skies and scattered rainshowers caused by thunderstorms is possible to affect the Camarines Norte accompanied by light to moderate surface winds generally from the east. The aforementioned office also issued a Heavy Rainfall Warning No. 3\footnote{HRW No. 3: \url{https://www.facebook.com/share/p/1JUc5TR4yo/}} focused along Camarines Norte due to the low pressure system affecting the area of interest, which can also cause flooding and landslides in higher terrain.

\begin{figure*}[!t]
\includegraphics[width=\textwidth]{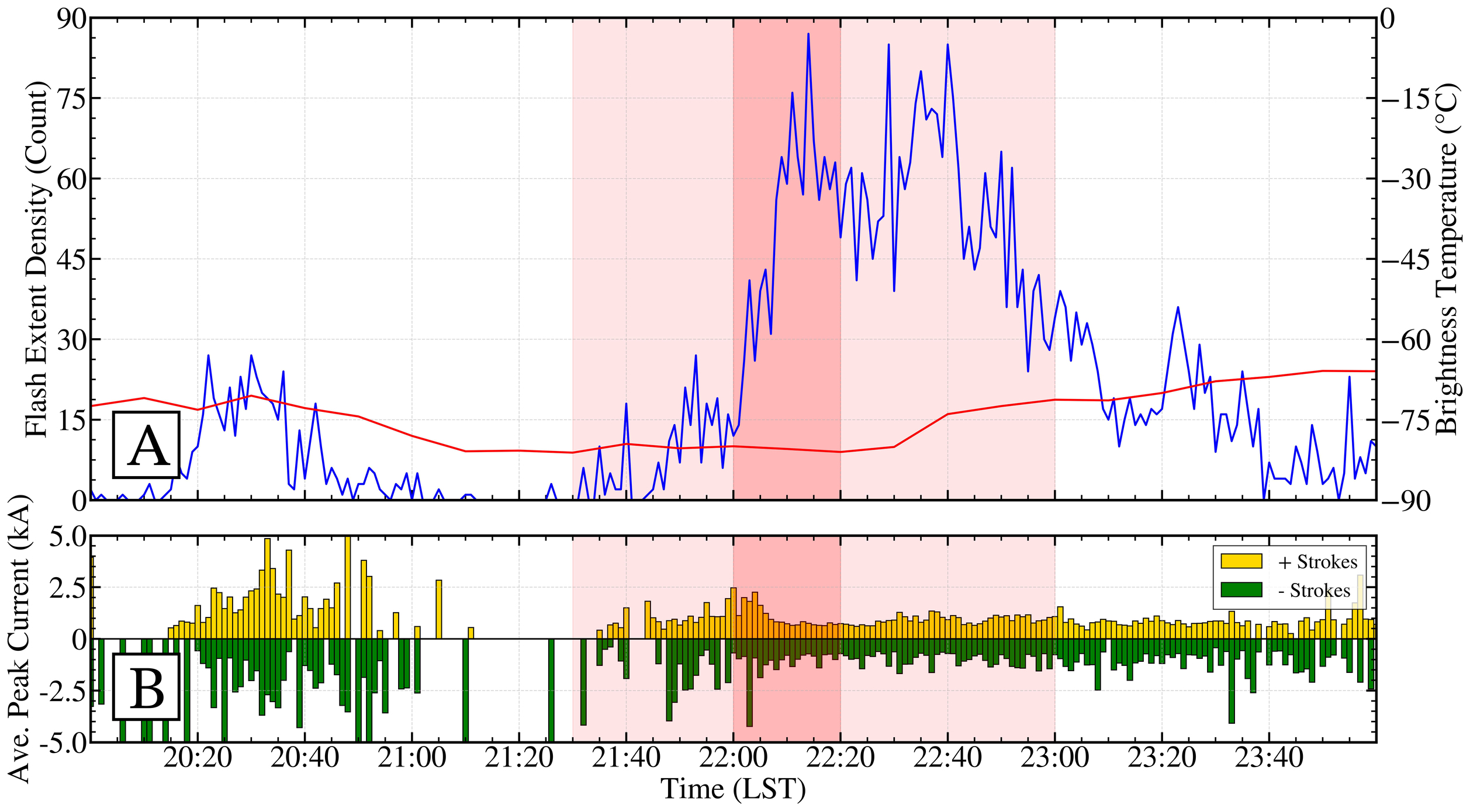}
\caption{Time series of (a) Minimum 10.4 $\mu$m brightness temperature of the convective storm (red; \dg C), maximum flash extent density per minute (blue; counts or flashes min$^{-1}$). (b) Average Peak Current (kA) per minute of recorded Positive (yellow) and Negative Strokes (green) within the supercell. Highlighted in the time series are the estimated severe thunderstorm occurrence (light red) and tornadic event (dark red).}
\label{fig_10}
\end{figure*}

Satellite IR imagery taken by HIMAWARI-9 shows the severe convective storm already persisting and affecting the coastal areas of Southern Luzon, primarily along Camarines Norte (Fig.~\ref{fig_9}). The convection is depicted with anvil plume exceeding $-$75\,\dg C -- $-$80\,\dg C or colder, presence of above anvil cirrus plumes (AACPs), and multiple OTs ($-$88.6 to $-$90.5\,\dg C) all due to intense updrafts penetrating the tropopause. These satellite features are tell-tale signs of severe weather in the region capable of producing significant wind, hail, and/or tornadoes \citep{Schmit2013,Bedka2015,Bedka2018}. The severe storm also produced intense lightning activity ahead of the tornado event (Fig.~\ref{fig_10}). Of the 5932 CG flash strokes detected, 4328 (73\%) were classified as positive and 1604 (27\%) as negative. However, 4057 of the +CG strokes (94\%) had peak currents below 10 kA, and low-current positive detections of this kind are frequently misclassified intracloud discharges \citep{Cummins2009}. If these are treated as intracloud, $-$CG strokes (1604) outnumber +CG strokes (271) by about six to one, so the storm is not described here as +CG-dominant. The $-$CG strokes also carried the stronger peak currents, and the $-$CG rate reached 133 flashes min$^{-1}$ at one point in the storm's life cycle. A summary of the severe thunderstorm's characteristics and lightning property is in Table~\ref{Table_3}. Upon further satellite inspection, the calculated $r$ from the B13 ranges from $\sim$136 km to $\sim$155 km with consistent mean BT of $\sim$$-$63\,\dg C for this storm. Given its large spatial coverage satisfying the storm scale thresholds \citep[e.g., $>$ 100 km;][]{Carvalho2001}, mean cloud-top BT, and the presence of multiple OTs, this can be classified as a tropical MCS with embedded supercells within the cloud head. Embedded supercells within the greater storm complex have been documented many times in past studies, such as in CONUS, they have attributed 13.8\% to 21\% of all tornadoes to MCS, particularly QLCS convective modes \citep{Smith2012,Ashley2019}.

\begin{table*}[!t]
\caption{Overall satellite and lightning characteristics of the storm cell.}
\label{Table_3}
\centering
\footnotesize
\begin{tabular}{lcccccc}
\hline\hline
Property / Timestamp (UTC) & 2140 & 2150 & 2200 & 2210 & 2220 & 2230 \\
\hline
Equivalent $r$ (km) & 136.710 & 147.067 & 148.020 & 149.495 & 154.931 & 140.122 \\
Mean cloud-top BT (\dg C) & $-$62.968 & $-$63.009 & $-$63.228 & $-$63.272 & $-$63.050 & $-$64.772 \\
Minimum BT (\dg C) & $-$90.525 & $-$89.650 & $-$89.463 & $-$88.650 & $-$89.650 & $-$89.460 \\
\hline
Storm Cell Duration & \multicolumn{6}{c}{$>$ 3 hours (embedded)} \\
$-$CG Flashes & \multicolumn{6}{c}{1604 flashes} \\
$-$CG Flashes with peak currents $<$ $-$10 kA & \multicolumn{6}{c}{413 flashes} \\
+CG Flashes & \multicolumn{6}{c}{4328 flashes} \\
+CG Flashes with peak currents $>$ 10 kA & \multicolumn{6}{c}{271 flashes} \\
Average peak current for $-$CG (kA) (median) & \multicolumn{6}{c}{$-$7.910 kA ($-$3.842 kA)} \\
Average peak current for $-$CG ($<$ $-$10 kA) (median) & \multicolumn{6}{c}{$-$20.320 kA ($-$16.927 kA)} \\
Average peak current for +CG (kA) (median) & \multicolumn{6}{c}{4.290 kA (2.787 kA)} \\
Average peak current for +CG ($>$ 10 kA) (median) & \multicolumn{6}{c}{19.128 kA (14.185 kA)} \\
Maximum Total Flash Rate (min$^{-1}$) & \multicolumn{6}{c}{173 flashes min$^{-1}$} \\
Maximum $-$CG Rate (min$^{-1}$) & \multicolumn{6}{c}{133 flashes min$^{-1}$} \\
Maximum +CG Rate (min$^{-1}$) & \multicolumn{6}{c}{59 flashes min$^{-1}$} \\
\hline
\end{tabular}
\end{table*}

\begin{figure*}[!t]
\includegraphics[width=\textwidth]{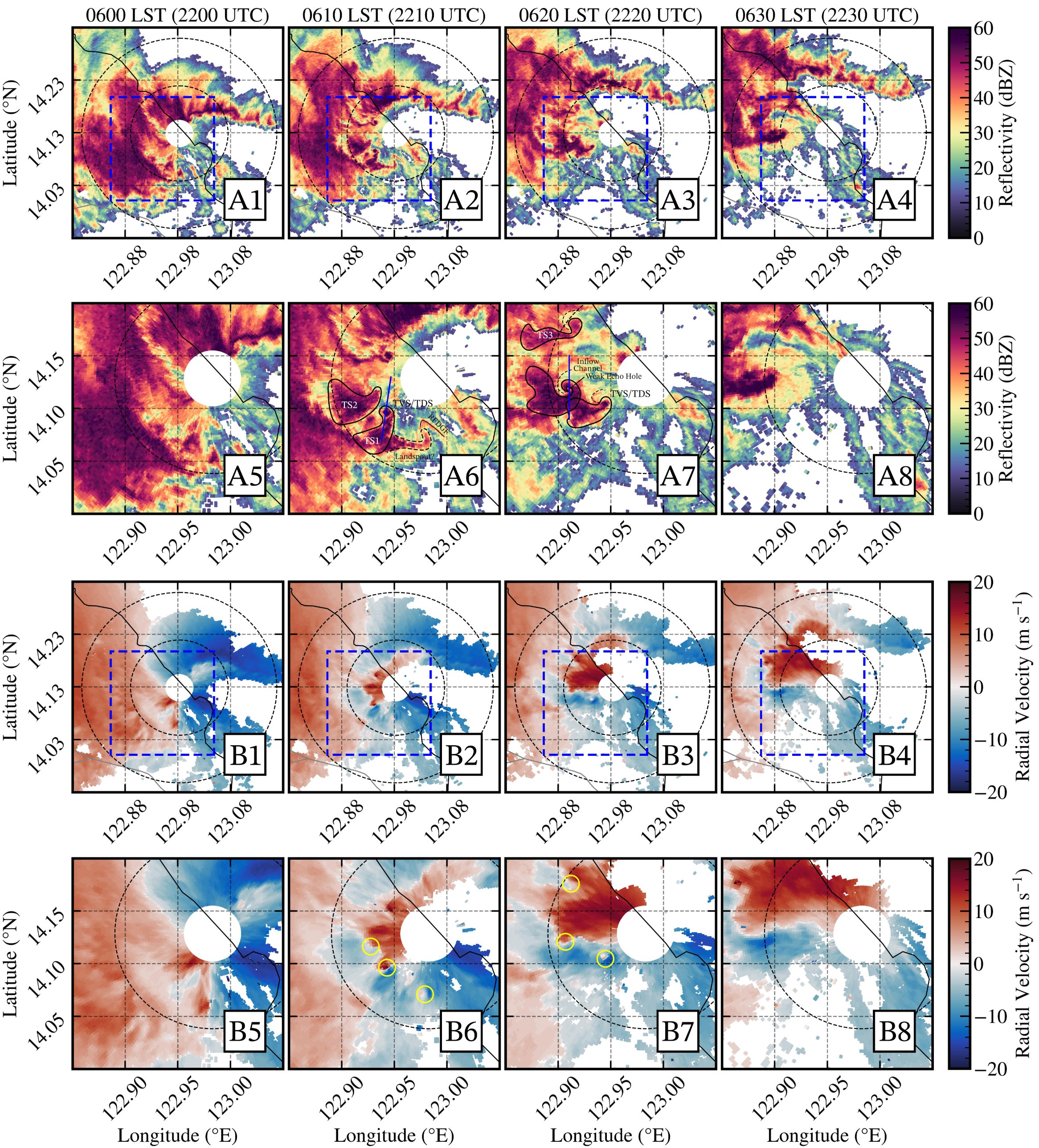}
\caption{S-DAT 4th elevation scans (3.4\dg) of the tornadic supercell for every 10 minutes. Radar variables include (a) Reflectivity (dBZ) and (b) Radial Velocity (\ms). Time stamps start from 2200--2230 UTC (0600--0630 LST). Both a5--a8 and b5--b8 sub-figures are `zoomed in' radar scans to their respective time and radar products (as in dashed blue box). Blue lines in panels A6 and A7 correspond to the radar cross sections conducted in subsequent figures. Annotations in the reflectivity field are included e.g., Tornadic Supercell (TS), Tornado Vortex Signature (TVS), while the yellow circles are the inherent velocity couplets.}
\label{fig_11}
\end{figure*}

\begin{figure*}[!t]
\includegraphics[width=\textwidth]{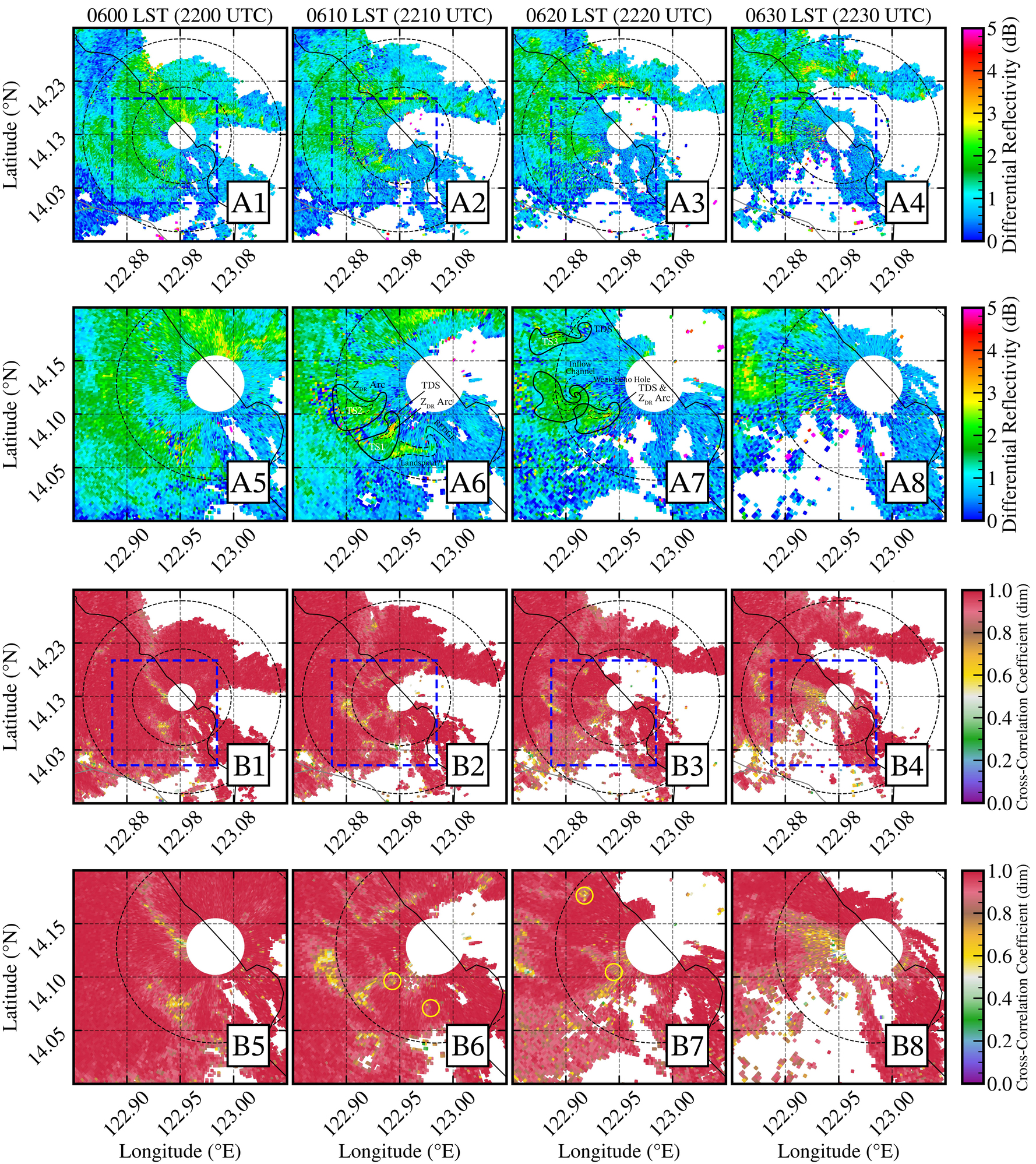}
\caption{Similar to Figure~\ref{fig_11} but (a) Differential Reflectivity (dB) and (b) Cross-Correlation Ratio (dimensionless). Time stamps start from 2200--2230 UTC (0600--0630 LST). Both a5--a8 and b5--b8 sub-figures are `zoomed in' radar scans to their respective time and radar products (as in dashed blue box). Annotations in the differential reflectivity field are included.}
\label{fig_12}
\end{figure*}

To complement the satellite-based analysis, radar scans from the S-DAT were analyzed. Both Figures~\ref{fig_11} and \ref{fig_12} represent the evolution of the tornadic supercells from 2200--2230 UTC (0600--0630 LST) at 10-minute intervals. Figure~\ref{fig_11} is composed of both \ZH\ and \VR\ scans, while Figure~\ref{fig_12} introduces the polarimetric variables such as \ZDR\ and CC using the 4th radar elevation angle (at 3.4\dg\ elevation; roughly 300--500 m above radar level within 10 km coverage). Both figures are accompanied by a `zoomed in' version from subfigures no. 5 to 8 e.g., Fig.~\ref{fig_11}a5--a8, therefore, will be used in this discussion, which is highlighted subfigures no. 1 to 4 as dashed blue lines.

Reflectivity (\ZH) scans (Fig.~\ref{fig_11}a5--a8) depict a complex of SCS in the area. By the 2210 and 2220 UTC (0610 and 0620 LST) volume scans, three distinct supercells had become apparent, each outlined in black. These are labeled as `TS1' (Tornadic Supercell No. 1), identified as the Magang tornado-producing storm; `TS2', associated with the Cahabaan tornado in Talisay; and `TS3', linked to the Napilihan tornado in Vinzons which was likely brief. Unfortunately, the Lag-on tornado identified earlier was not resolved by the radar given its close proximity to the radar site's cone of silence. At a beam height of 306 m above radar level (mARL) and 487 mARL, both TS1 and TS2 exhibited maximum \ZH\ $>$ 60 dBZ (at the storm's core), respectively, coinciding with the satellite-identified OTs and indicative of persistent and vigorous updraft activity. In particular, TS1 also exhibited a coherent TDS \citep[also coincident with the tornado vortex signature or TVS, a radial velocity couplet;][]{Ryzhkov2005,Brown2012} collocated with \ZH\ $>$ 65 dBZ. Additionally, a small-thin convective element (dashed outline) was located along the rear flank of the parent TS1 storm. The \ZH\ structure is unusual for a supercell morphology, but seems to have its own hook echo signature and a rear flank downdraft gust front (RFDGF) at the tip of this radar signature. We labeled this as a potential landspout vortex or may have been a representation of non-supercellular landspout processes.

At the rear flank of TS1, TS2 also convected due to the present volatile environment. Eventually, at 2220 UTC (0620 LST), both tornadic supercells were still present and in close proximity to one another with TS1 still accompanied by a strong TDS only separated by 3.2 km from their respective reflectivity cores. The storm-relative configuration aligns with the results of \citet{Nixon2024}, which suggests that in supercell pairs, both the head and rightmost supercells, have the higher chance to become tornadic by achieving inflow-outflow balance thereby decelerating pre-tornadic vertical advection; a known precursor to tornadogenesis under weak storm relative inflow \citep{Dowell2002}. Thus, the 17--19 kt of \VSR\ may imply that only a small perturbation to the outflow strength would be required to shift the storm into, or out of, inflow--outflow balance. In fact, TS2 eventually became tornadic as the presence of a weak-echo hole \citep[WEH;][and references therein]{Fujita1981,Wakimoto1996,Wurman1996,Wakimoto2011,Wurman2000,Bluestein2007b,Wurman2013,Wakimoto2015,Wakimoto2025} with lowered \ZH\ signal of 46 dBZ and inflow channel which is consistent with the development of the streamwise vorticity current \citep[SVC;][and references therein]{Klemp1983,Orf2017,Schueth2021} was depicted. The formation of WEH within the tornado core is likely caused by the centrifuging of hydrometeors and debris \citep{Dowell2005}.

A key feature observed in the \VR\ scans was the emergence of several compact cyclonic velocity couplets between 2210 and 2220 UTC (0610--0620 LST), marked in Figures~\ref{fig_11}b2--b3 and \ref{fig_11}b6--b7 (yellow circles). Applying the MDA criteria described in the previous section at the 3.4\dg\ tilt, TS1 exhibited an initial \Vrot\ of 15.42 \ms\ with a radar-derived vertical vorticity (\zrad) of 0.105 \si\ at 2210 UTC ($\Delta X$ $\sim$ 588 m, $\sim$340 mARL). This exceeds the nominal detection thresholds and is indicative of a tornado-scale vortex embedded within the broader parent mesocyclone. \Vrot\ for TS1 remained above 10 \ms\ in the succeeding scan, at 10.94 \ms\ with \zrad\ of 0.086 \si\ ($\Delta X$ $\sim$ 510 m). Meanwhile, TS2 registered \Vrot\ of 7.71 \ms\ and \zrad\ of 0.053 \si\ at 2210 UTC ($\Delta X$ $\sim$ 587 m), strengthening to 10.20 \ms\ with \zrad\ of 0.036 \si\ by 2220 UTC ($\Delta X$ $\sim$ 1130 m). Because TS2's circulation broadened over this interval, its vorticity decreased even as its \Vrot\ increased. TS2's inbound and outbound peaks lay at the widest separation the detection resolved, so this $\Delta X$ is a minimum and the corresponding \zrad\ an upper estimate\footnote{It should be noted that for TS2, because the velocity maxima lie on discrete rays, $\Delta X$ changes in steps of one azimuthal sample ($\sim$80--140 m at these ranges), so each \zrad\ is uncertain by $\sim$$\pm$15--30\%.}. Furthermore, TS2's initial \Vrot\ is below the $>$ 10 \ms\ criterion. However, the threshold was derived from midlatitude continental supercells observed by the WSR-88D network, and \citet{Capuli2026} found that \Vrot\ of 8--9 \ms\ is already indicative of a developing mesocyclone in a Philippine tornadic supercell. TS2's initial values approach this range and increase with height, from 5.97 \ms\ at 1.5\dg\ to 9.17 \ms\ at 6.2\dg. Thus, TS2's initial measurements can be interpreted as an incipient mesocyclone on the combined evidence of its couplet, its \ZH\ structure, and its polarimetric signatures (to be discussed later) rather than on the nominal threshold alone. Over the same interval, TS1 weakened while remaining tornadic as its \Vrot\ averaged over the six lowest tilts decreased by about 28\%, from 14.9 to 10.7 \ms, while its TDS persisted at the lowest tilts. A cyclonic couplet was also evident at TS3 in the 2220 UTC volume, coincident with the Napilihan tornado, but was no longer apparent by 2230 UTC. As the circulation was captured in a single volume only, its \Vrot\ and \zrad\ were not derived.

Because a velocity couplet observed at a single elevation angle can also result from non-supercellular processes, such as the candidate landspout vortex described above, identifying TS1 and TS2 as mesocyclonic circulations required additional evidence. That evidence was vertical coherence across successive elevation angles and temporal persistence between volume scans, and both circulations satisfied these criteria. Each couplet was detected at all six tilts from 1.5\dg\ to 6.2\dg\ ($\sim$140--800 mARL) in both volume scans. Moreover, their velocity couplets coincided spatially and temporally with the development of a hook echo. In the case of TS1, they also coincided with a TDS, which provides polarimetric evidence of a tornado independent of the velocity field. The coincidence of the TS1 couplet and TDS with the visual and photographic evidence presented in Fig.~\ref{fig_3} further supports its identification with the Magang tornado.

Furthermore, the \Vrot\ values of each vortex discussed above are detections of radar-resolved rotation rather than measurements of tornado wind speed, but they place a lower bound on tornado intensity. The velocity difference across a couplet is $\Delta V = 2V_{rot}$, giving 30.84 \ms\ for TS1 at 2210 UTC and 20.40 \ms\ for TS2 at 2220 UTC, and both are lower bounds. At the 4.6--6.6 km range of TS1 and 5.4--8.1 km of TS2, the 1.69\dg\ radar beam subtended $\sim$135--240 m, comparable to or larger than the expected tornado-core diameter of roughly 100--200 m. \citet{Wood1997} demonstrated that when the radar beam is comparable to or larger than the vortex core, beam averaging and the position of the beam relative to the vortex can substantially reduce the observed peak \VR\ and alter the apparent vortex structure. In addition, the couplets were sampled at $\sim$140--800 mARL, above the near-surface layer where the strongest tornadic winds occur. $\Delta V$ also excludes the translational motion of the storm, which adds to ground-relative winds on one flank of the vortex. Thus, the measured $\Delta V$ values should not be interpreted as the actual maximum velocity difference within the tornado. No quantitative correction is applied, because the magnitude of the radar sampling bias depends on the vortex-to-beam size ratio and on the relative position of the vortex within the radar volume, neither of which is known here. Specifically, \citet{Wood1997} showed that these effects can produce substantial scan-to-scan variability in the observed TVS magnitude.

Dual-polarimetric parameters were also evaluated in the presence of these embedded supercell storms. Figure~\ref{fig_12} showcases \ZDR\ (\ref{fig_12}a1--a8) and \rhv\ (\ref{fig_12}b1--b8) signatures of the storms of interest, with subfigures no. 5 to 8 being the zoomed in version (similar to Figure~\ref{fig_11}). Starting with the parent storms, for TS1, the \ZDR\ field displayed values at around 2--3 dB along the forward-flank co-located with high reflectivities. Meanwhile, TS2's \ZDR\ field also showed values between 1.75--2.5 dB also collocated with high reflectivities in the FFD all the way to the hook echo. Furthermore, at 2210 UTC, an increased signal of \ZDR\ with high \ZH\ was located at the intersection of the FFD and the inflow notch of the low-level circulation. This feature is known as the \ZDR\ arc \citep{Kumjian2008}. The \ZDR\ arc marked the zone of maximum gradient in echo intensity along the FFD region where measurements reached 3--4 dB for TS1, while $\sim$2.25--2.8 dB for TS2, with varying unique placement of the \ZDR\ arc signatures to each storm of interest. For TS1, the polarimetric feature extends rearward very close to the TDS, while the \ZDR\ arc of TS2 is located mainly in the FFD extending toward the developing hook echo. Recent studies on tornadic supercells have demonstrated that additional polarimetric signatures can provide novel insights into the storm-scale processes required for tornadogenesis and maintenance \citep{Kumjian2009,French2015,Homeyer2020,DenBroeke2020,Clark2024}. Thus, \ZDR\ arc features of the tornadic supercells could be an indication of enhanced SRH, and more importantly an indicator of storm severity since the ERA5 model sounding cannot capture completely the storm-scale evolution of SRH (Max SRH$_{03}$ = 100 \mss\ at the onset of tornado), but still gives a depiction of the environment conducive to severe weather. Such as that the \ZDR\ arc tends to extend back into the inflow notch of supercells (as seen in TS1 and TS2) leading up to the strengthening of the low-level mesocyclone and potential tornadogenesis. Furthermore, TS2 also exhibited an increase in the areal extent of the \ZDR\ field which is indicative of an increase in the low-level \ZH\ by 10--15 minutes \citep[e.g.,][]{Picca2010}.

Other storm traits were evident at the time of the tornado events. TS1 exhibited very low \rhv\ (single-gate minimum $\sim$ 0.34) and negative \ZDR\ ($-$1.7 dB) collocated with the TDS and the radial velocity couplet discussed above, values strongly indicative of lofted debris. For TS2, the signature was weaker and confined to the lowest elevation. At 2220 UTC, at the 1st tilt (0.5\dg, Sweep 0; $\sim$70 mARL), \rhv\ fell to 0.736 with \ZDR\ = $-$2.5 dB, transitioning to higher \rhv\ and positive \ZDR\ at successive elevations, and no \rhv\ depression was apparent at the 4th tilt (3.4\dg, $\sim$487 mARL). Because \rhv\ values of 0.70--0.75 are not by themselves diagnostic (wet hail, or a mixture of hydrometeors and debris, can produce comparable reductions), the interpretation rests on three additional lines of evidence. First, the reduction lies below the $\sim$0.8--0.85 that hail alone typically produces at S-band, and \ZDR\ is strongly negative ($-$2.5 dB), whereas melting or water-coated hail yields near-zero to positive \ZDR\ because the water coating stabilizes the stones and makes them more oblate. Strongly negative \ZDR\ instead indicates non-Rayleigh scatterers with a common alignment, characteristic of lofted debris \citep{Ryzhkov2005,Kumjian2008}. Second, the signature is confined to the lowest tilt. All tilts sampled the storm below the 0\,\dg C level, where a hail shaft is vertically continuous from aloft to the surface and would reduce \rhv\ at successive tilts as well, not only in a shallow near-surface layer. Third, the depression is collocated with the WEH, where \ZH\ falls to 46 dBZ, whereas wet hail produces enhanced rather than reduced \ZH, so a \rhv\ minimum within a local \ZH\ minimum is more consistent with centrifuging of debris and hydrometeors from the vortex core \citep{Dowell2005}.

The weaker depression for TS2 relative to TS1 is consistent with a smaller quantity of less radar-reflective debris. The T3 track lay entirely over farmland, so the available debris was predominantly vegetative and of lower dielectric contrast and smaller size than the structural debris lofted along the T2 track. Precipitation entrainment into the circulation, as documented by \citet{Bluestein2007a} for high-precipitation supercells, and slightly elevated \rhv\ within the minimum-echo eye arising from storm interaction \citep{Schwarz2011Xband}, may also have raised \rhv\ above the values typical of a well-developed TDS. We note that a mixed debris--hydrometeor population cannot be excluded for TS2, and the signature is therefore interpreted as indicating debris lofting with hydrometeor contamination rather than a pure debris signature. Despite the differences between the \rhv\ characteristics, it is clear that both tornadic supercells possessed near-ground debris signatures and satisfy the criterion for tornado detection as stipulated in \citet{Ryzhkov2005} i.e., \ZH\ $>$ 45 dBZ, \ZDR\ $<$ 0.5 dB, \rhv\ $<$ 0.8. Furthermore, a region of reduced \rhv\ (between 0.8 and 0.9), slightly positive \ZDR, and \ZH\ $>$ 50 dBZ was also depicted and can be due to the presence of wet hail, or rain mixed with hail, within the forward flank core of the supercell thunderstorms. Such characteristics were also observed in a recent study by \citet{Clark2024} on the Jersey tornadic supercell on midnight of 1--2 November 2023 where reduced \rhv\ in the TDS was observed coinciding with the reduced \ZDR\ (near-zero or even negative) and high \ZH.

\begin{figure*}[!t]
\centering
\includegraphics[width=\textwidth]{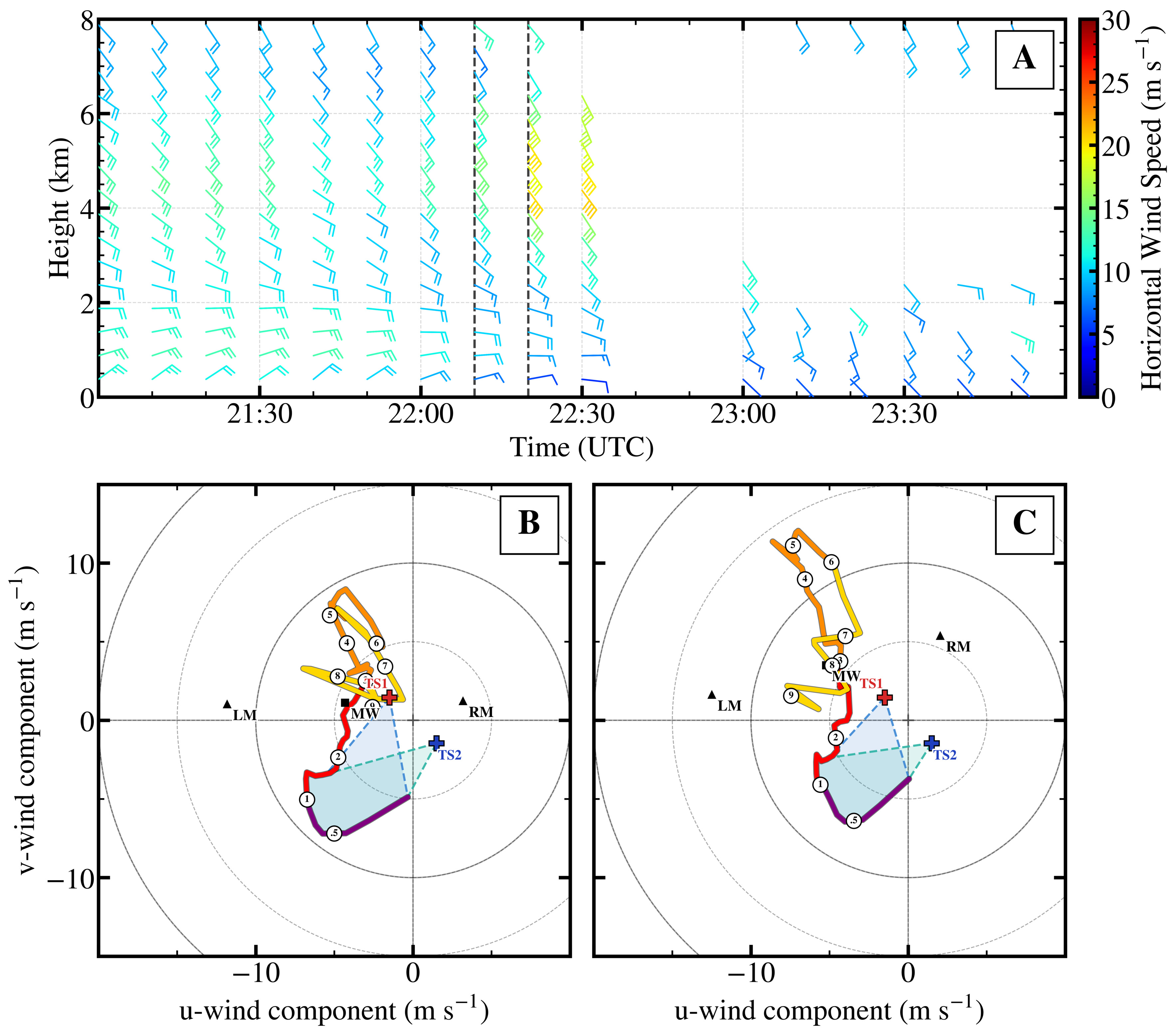}
\caption{(a) VAD Profile from 21 to 23 UTC derived from S-DAT. (b--c) Radar-derived Hodograph at 2210 and 2220 UTC (marked as black dashed line in panel a) at the time of the localized tornado outbreak event. Bunkers Storm Motion components and observed storm motion of TS1 (red plus) and TS2 (blue plus) were added.}
\label{fig_13}
\end{figure*}

The radar-derived kinematics, shown in Figure~\ref{fig_13}, broadly corroborate the ERA5 environment in the lowest 3 km. Derived from the VAD, the hodograph exhibits LLS and MLS (5.6--6.4 and 7.9--8.7 \ms) that are within $\sim$1 \ms\ of the ERA5 values (6.1--7.1 and 8.0--9.5 \ms). Also, the $\omega_{s01}$ associated with TS1 (0.0060--0.0065 \si) is comparable to that in ERA5-derived soundings (0.007--0.008 \si). The two datasets diverge, however, in the deep layer as the DLS increased from 10.0 to 14.6 \ms\ between 2210 and 2220 UTC, exceeding the ERA5 values (6.8--8.5 \ms) by up to a factor of $\sim$1.7. This enhancement coincides with the strengthening of the easterly flow as MCS approaches inland (Fig.~\ref{fig_13}a), a mesoscale feature that the $\sim$31-km hourly reanalysis is unlikely to resolve, although storm-scale departures from the linear wind field assumed by the VAD may also contribute.

The storm-relative parameters are lower in the radar-derived hodograph (Fig.~\ref{fig_13}b and \ref{fig_13}c), with 0--1 km SRH of 29--59 \mss\ and $\widetilde{\omega}_{s01}$ of 57--79\%, compared with 68--75 \mss\ and 91--93\% in ERA5. For SRH, this difference arises largely from the choice of storm motion. Specifically, when the radar profiles are evaluated with the same \BRM\ used for ERA5, 0--1 km SRH increases to 57--67 \mss, close to the ERA5 values, and 0--3-km SRH (107--109 \mss) exceeds them (93--100 \mss). However, the streamwiseness remains modest (65--71\%) for the same storm motion, indicating that the radar-observed LLS was less aligned with the \VSR\ than in ERA5. Both storms moved toward the northwest (TS1 toward 323\dg\ at 8.0 \ms; TS2 toward 333\dg\ at 3.9 \ms), to the right of and slower than the 0--6 km mean wind (toward 309\dg--315\dg\ at 9.7--12.0 \ms), but with a smaller rightward deviation ($\sim$3 and $\sim$5--6 \ms, respectively) than the 7.5 \ms\ assumed by the B2K, whose right-mover estimate was directed toward the north (353\dg--359\dg).

Relative to the \BRM, TS2 ingested less SRH in every layer, whereas TS1 experienced comparable 0--1 km SRH and greater 0--500 m SRH at 2210 UTC (59 and 34 \mss\ vs 57 and 22 \mss) before falling below the Bunkers-based values at 2220 UTC. The rightward deviation of both storms increased their 0--3 km SRH (66--87 \mss) relative to that of a storm moving with the mean wind (47--51 \mss), but not their near-ground SRH, which the mean-wind motion would have maximized in this profile. Nevertheless, the low-level vorticity available to TS1 was the most streamwise of all the motions considered as its $\widetilde{\omega}_{s500}$ and $\widetilde{\omega}_{s01}$ (69--72\% and 77--79\%, respectively) and $\omega_{s01}$ (0.0060--0.0065 \si) exceeded the values obtained with the \BRM\ (41--51\%, 65--71\%, and 0.0048--0.0052 \si). As indicated earlier, the interaction between the two tornadic storms may have further enhanced the low-level rotation through the inflow--outflow balance, which was particularly relevant given the modest low-level storm-relative inflow (radar-derived \VSR\ $\sim$6--8 \ms\ / 11--16 kt). With the two storms only $\sim$3.2 km apart, the outflow and inflow of each could modify the near-storm wind field of the other, which may have helped maintain a more favorable balance and supplied additional horizontal vorticity along the interacting boundaries. The radar-derived inflow is weaker than that in ERA5 ($\sim$8.5--10 \ms\ / 17--19 kt), largely because the ERA5 values are computed relative to the \BRM. Thus, weaker inflow experienced by TS1 and TS2 reflects their slower, more rightward motion rather than a weaker environmental low-level flow. A summary of radar-derived kinematics is shown in Table~\ref{Table_4}.

\begin{table*}[!t]
\caption{Radar-derived kinematics from the velocity azimuth display (VAD) and hodograph.}
\label{Table_4}
\centering
\footnotesize
\begin{tabular}{lcccc}
\hline\hline
 & \multicolumn{2}{c}{2210 UTC} & \multicolumn{2}{c}{2220 UTC} \\
Kinematic parameter & TS1 & TS2 & TS1 & TS2 \\
\hline
\multicolumn{5}{l}{\textit{Ground-relative (common to TS1 and TS2)}} \\
BWD$_{01}$ & \multicolumn{2}{c}{6.408 \ms\ / 12.456 kt} & \multicolumn{2}{c}{5.645 \ms\ / 10.974 kt} \\
BWD$_{03}$ & \multicolumn{2}{c}{7.868 \ms\ / 15.295 kt} & \multicolumn{2}{c}{8.673 \ms\ / 16.859 kt} \\
BWD$_{06}$ & \multicolumn{2}{c}{9.968 \ms\ / 19.376 kt} & \multicolumn{2}{c}{14.637 \ms\ / 28.451 kt} \\
\BRM & \multicolumn{2}{c}{6.168 \ms\ / 11.989 kt} & \multicolumn{2}{c}{10.397 \ms\ / 20.209 kt} \\
\multicolumn{5}{l}{\textit{Storm-relative (reference: Observed Storm Motion)}} \\
Observed Storm Motion & 7.970 \ms\ / 15.492 kt & 3.910 \ms\ / 7.600 kt & 7.970 \ms\ / 15.492 kt & 3.910 \ms\ / 7.600 kt \\
\VSR & 8.249 \ms\ / 16.035 kt & 7.405 \ms\ / 14.394 kt & 6.994 \ms\ / 13.595 kt & 5.900 \ms\ / 11.469 kt \\
SRH$_{500}$ & 34.187 \mss & 13.677 \mss & 23.749 \mss & 5.602 \mss \\
SRH$_{01}$ & 58.877 \mss & 39.809 \mss & 46.588 \mss & 29.139 \mss \\
SRH$_{03}$ & 72.761 \mss & 87.063 \mss & 65.832 \mss & 75.475 \mss \\
$\omega_{s500}$ & 0.0079 \si & 0.0040 \si & 0.0063 \si & 0.0021 \si \\
$\widetilde{\omega}_{s500}$ & 0.7166 / 71.66\% & 0.3875 / 38.75\% & 0.6941 / 69.41\% & 0.2595 / 25.95\% \\
$\omega_{s01}$ & 0.0065 \si & 0.0050 \si & 0.0060 \si & 0.0040 \si \\
$\widetilde{\omega}_{s01}$ & 0.7686 / 76.86\% & 0.6475 / 64.75\% & 0.7932 / 79.32\% & 0.5671 / 56.71\% \\
\hline
\end{tabular}
\end{table*}

\subsubsection{Vertical Cross Section of TS1 (Magang Tornado)}\label{sec342}

\begin{figure*}[!t]
\centering
\includegraphics[width=\textwidth]{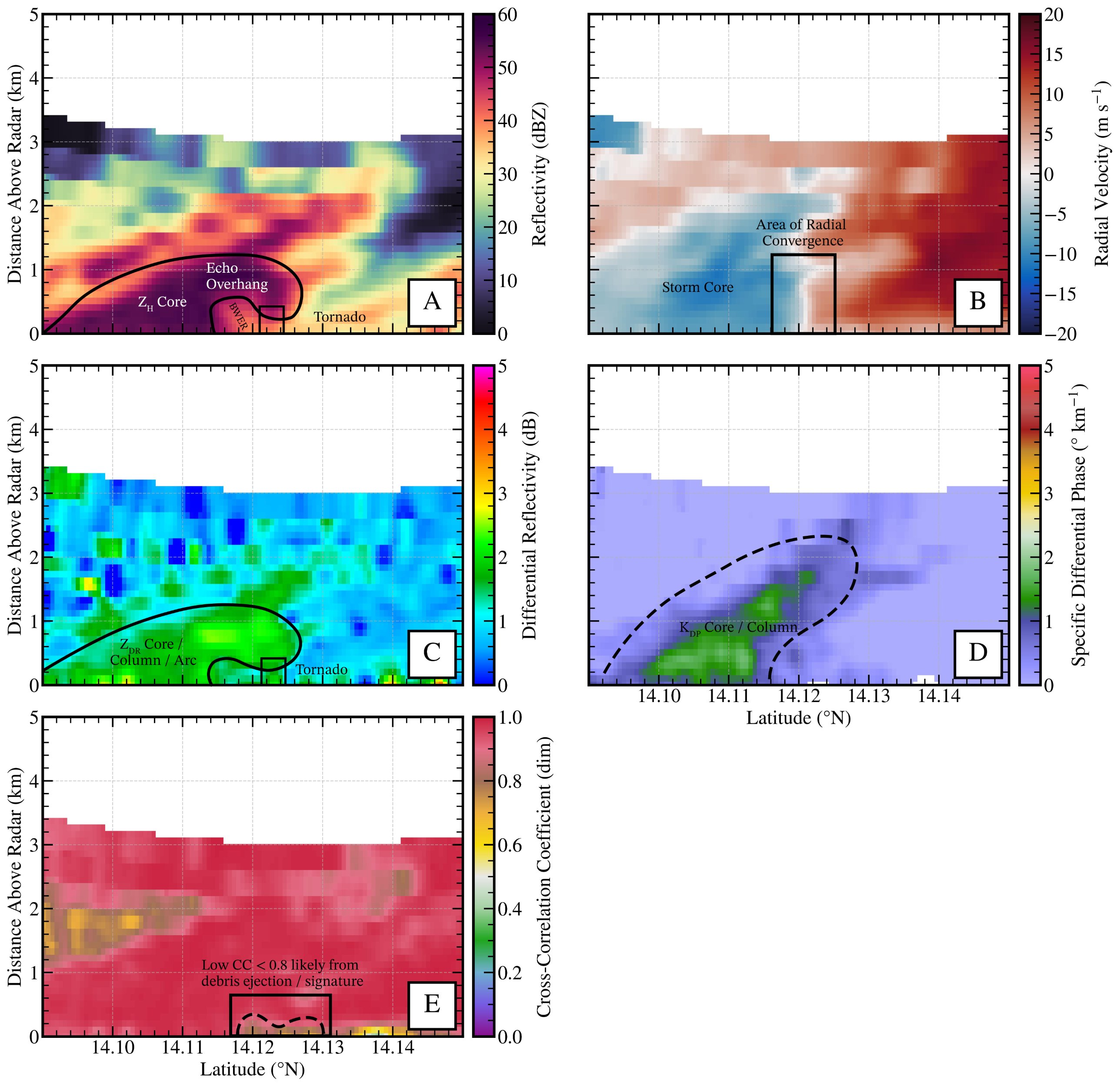}
\caption{Vertical cross section of TS1 at 0610 LST (2210 UTC) along the line shown in Fig.~\ref{fig_11}a6 from 14.13\dg N, 122.945\dg E to 14.07\dg N, 122.942\dg E. (a) Reflectivity (dBZ), (b) Radial Velocity (\ms), (c) Differential Reflectivity (dB), (d) Specific Differential Phase (\dg\,km$^{-1}$), and (e) Cross-Correlation Ratio (dimensionless). Annotations are included in each sub-panel.}
\label{fig_14}
\end{figure*}

The vertical structure of the tornadic storm that produced the Magang Tornado was further examined using radar vertical cross-section analysis from the S-DAT radar. As illustrated in Figure~\ref{fig_14}, it provides details of the severe storm's vertical structure, oriented perpendicular to the storm structure (from the FFD to the TDS i.e., north--south through the TDS). A \ZH\ maximum of 64 dBZ (solid line, Fig.~\ref{fig_14}a) extending close to 2 km delineating the small core. Shown in dashed lines, there is also a signature of a Descending Reflectivity Core (DRC) within the rear-flank composed by heightened \ZH\ values ($>$ 60 dBZ) connected within TS1's \ZH\ core. First observed by \citet{Rasmussen2006}, DRCs are often found as a precursor of tornadogenesis and can form as a manifestation brought by the intensification of the low-level rotation \citep{Byko2009}. Cross-sections of the \VR\ field (Fig.~\ref{fig_14}b) further substantiate the presence of a mesocyclonic circulation. A distinct velocity couplet structure, marked by juxtaposed inbound and outbound velocities, becomes particularly evident within the timing of the tornadogenesis. This feature is a classical indicator of low-level rotation and suggests the presence of a cyclonic vortex. This pair of opposite \VR\ was located near the surface and coincided with the DRC identified, affirming the intensification of the near-surface vortex.

Shown in Fig.~\ref{fig_14}c, dual-polarimetric analysis of the TS1 reveals a \ZDR\ core ($\sim$2 dB), with the highest signal of the \ZDR\ arc ($>$ 3 dB) also depicted within the sectional slice. This \ZDR\ signature can be identified as a \ZDR\ column where high values of \ZDR\ indicate the presence of large, oblate hydrometeors, presumably either large raindrops or water-coated hailstones. Warm-rain collision and coalescence processes are the possible origin of these hydrometeors \citep[e.g.,][]{Caylor1987,Illingworth1987,Tuttle1989,Meischner1991} or from back-sheared anvil that are caught in low-level inflow and ingested back into the updraft \citep[e.g.,][]{Conway1993,Hubbert1998,Loney2002}. In addition, \citet{Snyder2017} found that the DSD profile of supercells' \ZDR\ columns are associated with larger median rain drop sizes. The location of TS1's \ZDR\ core/column is found on the inflow side of the storm within the updraft and is consistent in observational studies conducted by \citet{Kumjian2008}. A hydrometeor fallout is also recorded viewed as slightly enhanced \ZDR\ values of $\sim$1.5 dB in the FFD $<$ 500 mARL. This is consistent with \citet{Conway1993} and \citet{Loney2002}, who found the maximum values of \ZDR\ to be coincident with weaker vertical velocities in the convective updraft.

Aside from multiple \ZDR\ structures revealed, there is also the presence of a \KDP\ column with a noticeable tilted structure likely due to the ambient wind shear seen in Fig.~\ref{fig_14}d. \citet{Schlatter2003HookEcho} asserted that convective storms that formed in a sheared environment can induce spatial offset between the \ZDR\ and \KDP\ columns, and it can be the same for this case given the modest DLS $\sim$ 9.5 \ms\ within tornadogenesis time was reported. This is further supported by modeling efforts of dual-polarimetric parameters of \citet{Snyder2017} who found high correlations between the characteristics of \KDP\ columns and the aforementioned supercell updraft properties in an environment composed by sufficient wind shear (such as in a curved hodograph). In Figure~\ref{fig_14}d, a small \KDP\ core of $\sim$1.3\,\dg\,km$^{-1}$ was recorded within the radar structure coinciding with high values of \ZH\ (as discussed before) along the supercell updraft column. This finding is in line with the results of \citet{Schlatter2003HookEcho} and \citet{Kumjian2013DP1} wherein the \KDP\ column often coincides with the \ZH\ column, thus can be indicative that the updraft is composed of high concentration of mixed-phase hydrometeors at various sizes. Furthermore, authors such as \citet{Rasmussen1987} and \citet{Hubbert1998} noted that this radar signature demarcates the presence of a region of drops shedding off hail due to the high liquid water content and positive temperature perturbation.

Finally, near-surface \rhv\ fields (Fig.~\ref{fig_14}e) reveal a continuity of the `\rhv\ depression' composed by lowered \rhv\ values. These low \rhv\ measurements ($<$ 0.80) were likely due to the non-hydrometeor species such as debris being lofted. This further validates the radar scan earlier in Fig.~\ref{fig_11} and \ref{fig_12}, with the occurrence of TDS/TVS within the area of Magang; including the DRC structure and radial convergence in Fig.~\ref{fig_14}a and \ref{fig_14}b. This is also coincident with the `unusual' \ZDR\ signature along with the tornado-relevant radar parameters discussed earlier where $\sim$1--1.5 dB were recorded which we believe is the signature of the Magang tornado (or its DRC) given its very close distance to the radar site. As indicated, this was unusual, as typical TDS are characterized by near-zero or slightly negative \ZDR\ values due to the chaotic, random orientation of tumbling lofted materials and non-Rayleigh scattering effects from large debris \citep{DenBroeke2015,Umeyama2018}. Thus, the observed positive \ZDR\ values may suggest an atypical scenario where heavy precipitation; specifically large, horizontally oblate raindrops with a high dielectric constant, was tightly entrained within the near-surface circulation, overpowering the radar return of the dry debris itself \citep{Kumjian2008,Bodine2013,Wakimoto2020}.

\subsubsection{Vertical Cross Section of TS2 (Cahabaan Tornado)}\label{sec343}

\begin{figure*}[!t]
\centering
\includegraphics[width=\textwidth]{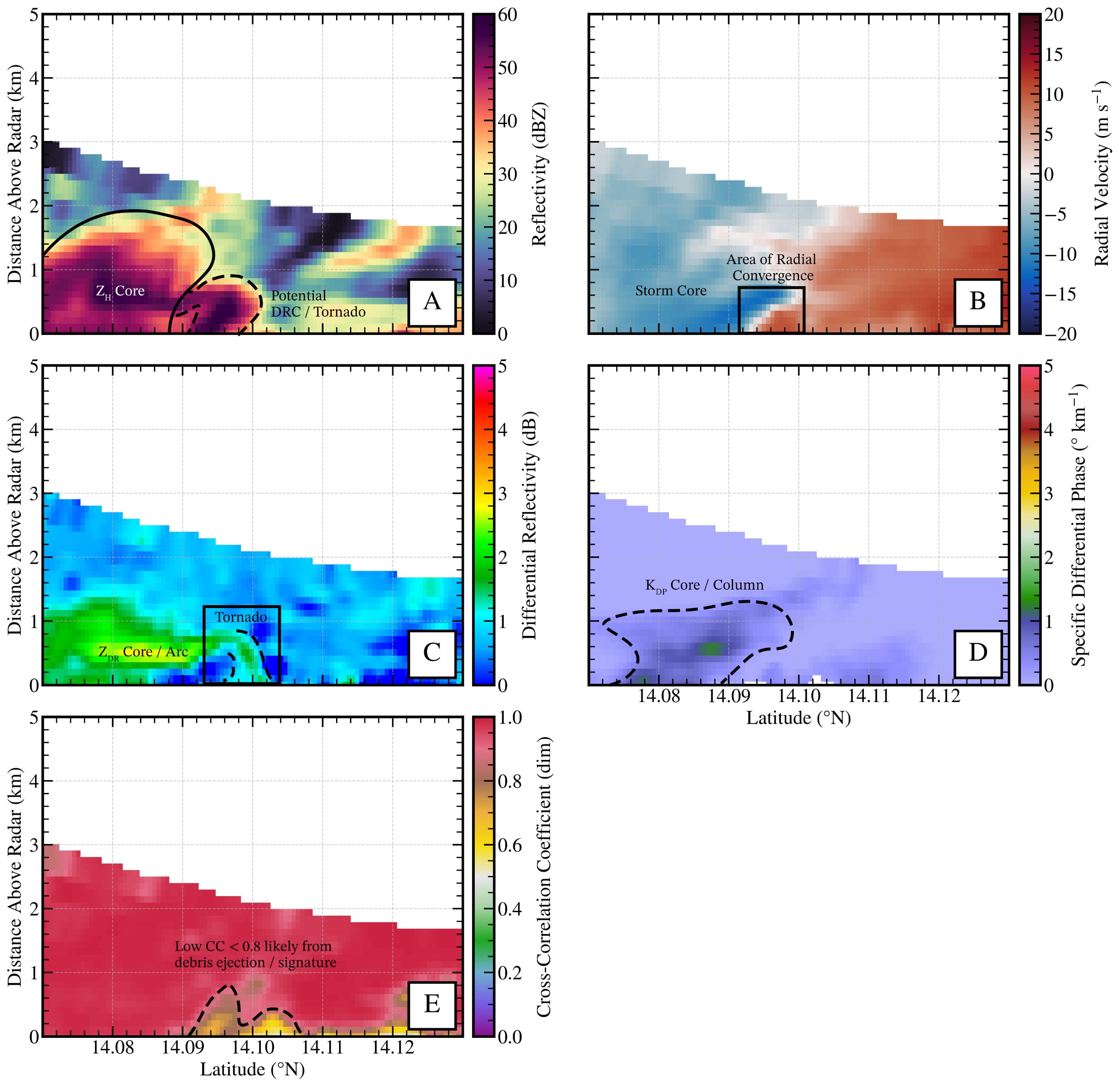}
\caption{Similar to Figure~\ref{fig_14} but for TS2 at 0620 LST (2220 UTC) along the line shown in Fig.~\ref{fig_11}a7 from 14.15\dg N to 14.09\dg N along 122.910\dg E. (a) Reflectivity (dBZ), (b) Radial Velocity (\ms), (c) Differential Reflectivity (dB), (d) Specific Differential Phase (\dg\,km$^{-1}$), and (e) Cross-Correlation Ratio (dimensionless). Annotations are included in each sub-panel.}
\label{fig_15}
\end{figure*}

Unlike TS1 and the Magang Tornado (identified as T2 in Section~\ref{sec31}), TS2 and the Cahabaan Tornado (identified as T3 in Section~\ref{sec31}) were not recognized at the time of occurrence and remained unreported until the preparation and write up of this study. Despite extensive data scraping efforts for this tornadic supercell, no contemporaneous reports or damage documentation were recovered for this event, limiting opportunities for validation. Nevertheless, based on the radar-derived signatures presented in Section~\ref{sec341}, and considering their apparent association with the localized tornado outbreak, this supercell warrants further examination and discussion.

A vertical slice was retrieved along the tornadic storm that produced the Cahabaan Tornado from the S-DAT radar. Figure~\ref{fig_15} showcases the details of the severe storm's vertical structure, similar to the slice orientation in the Magang Tornado analysis. A \ZH\ maximum of 63 dBZ (solid line, Fig.~\ref{fig_15}a) extending to $\sim$1.5 km in altitude was observed. The slice of the volume scan reveals the presence of a clear Bounded Weak Echo Region (BWER), an indicator of supercell storms and a signature for strong vertical motion. This signature is characterized by an area of low \ZH\ surrounded by higher \ZH\ aloft i.e., echo overhang and the \ZH\ column which is laterally indicative of an extremely strong updraft that lifts hydrometeors to a significant height. This feature is found in the inflow region of a thunderstorm, similar to the observations conducted by \citet{Markowski2002} and \citet{CottonAnthes1989}. In fact, BWER is a representative of a local storm that develops in a sheared environment and tends to a steady-state circulation. Similar to TS1, there is also a DRC signature, marked by \ZH\ region $\sim$ 51 dBZ, attached in the rear-flank appendage of the supercell's BWER.

Similar to TS1, the vertical cross section of \VR\ (Fig.~\ref{fig_15}b) reveals a distinct mesocyclonic circulation. This TVS becomes particularly pronounced coincident with the timing of tornadogenesis, serving as a classical indicator of intense, low-level cyclonic rotation \citep{French2013}. The localized velocity couplet was concentrated near the surface and extended upward to approximately 1 km. Notably, this circulation was slightly displaced to the west of the DRC. This spatial offset is consistent with established supercell kinematics, where the tornadic circulation typically resides along the strong \ZH\ gradient and density boundary adjacent to the RFD and its associated DRC \citep{Rasmussen2006,Byko2009}.

As illustrated in Figure~\ref{fig_15}c, TS2 exhibits a distinct \ZDR\ core and column ($\sim$1.5--2 dB), with the highest signal of the \ZDR\ arc ($\sim$2.8 dB) also depicted within the sectional slice. Although this overall \ZDR\ signature is slightly less intense than the one observed in TS1, it indicates the presence of sparsely distributed, large, highly oblate raindrops. The generation of these large hydrometeors is strongly driven by vigorous warm-rain microphysical processes; specifically rapid collision and coalescence, facilitated by favorable low-level kinematics and rich moisture advection into the convective updraft \citep[e.g.,][and references therein]{Meischner1991,Loney2002}. Furthermore, the spatial positioning of TS2's \ZDR\ core/column on the inflow flank of the storm, collocated with the primary updraft interface, is highly consistent with established conceptual models of supercell microphysics and hydrometeor size sorting \citep{Kumjian2008,Kumjian2009}.

Furthermore, similar to TS1, a \KDP\ column is also present, exhibiting a noticeable tilted structure that is likely induced by the strong ambient vertical wind shear (Fig.~\ref{fig_15}d). Sufficient vertical wind shear, such as in this case, can induce such tilt and induce a spatial offset between the \ZDR\ and \KDP\ columns \citep{Schlatter2003HookEcho,Snyder2017}. While both severe storms feature distinct \KDP\ cores and columns, the \KDP\ signature in TS2 is substantially larger in spatial coverage than in TS1. In TS2, \KDP\ values exceeding $\sim$2.0--2.5\,\dg\,km$^{-1}$ were recorded within the lowest 1 km. This \KDP\ signature, collocated with the low-level updraft and high values in the \ZH\ column, indicates ambient concentration of liquid water content and mixed-phase hydrometeors at various sizes \citep{Schlatter2003HookEcho,Kumjian2013DP1}.

Finally, similar to TS1, the near-surface \rhv\ fields (Fig.~\ref{fig_15}e) reveal a distinct `\rhv\ depression'. While these measurements (ranging from 0.70--0.75) are indicative of non-hydrometeors being lofted into the radar volume, they are notably higher than the values observed in TS1 (which dropped as low as 0.34; single-gate minimum). This difference is consistent with lofting of predominantly lighter, vegetative debris (such as grass, leaves or light agricultural material) rather than denser structural debris, although a mixed debris-hydrometeor population cannot be excluded. Polarimetric radar studies that dive into TDSs support this interpretation, demonstrating that vegetative debris typically yields higher \rhv\ values (between 0.60--0.80) due to its lower dielectric constant and less chaotic scattering properties compared to the highly irregular, highly reflective nature of structural materials like metal, roofing, and fragmented wood \citep{Bodine2013,DenBroeke2015}. The presence of this specific debris signature further corroborates the earlier radar analyses (Figs.~\ref{fig_11} and \ref{fig_12}), which identified a TVS/TDS along Cahabaan. Furthermore, it is consistent with the observed DRC, BWER, and radial convergence presented in Figs.~\ref{fig_15}a and \ref{fig_15}b.

\subsubsection{Multiparameter Debris Signature Analysis}\label{sec344}

\begin{figure*}[!t]
\includegraphics[width=\textwidth]{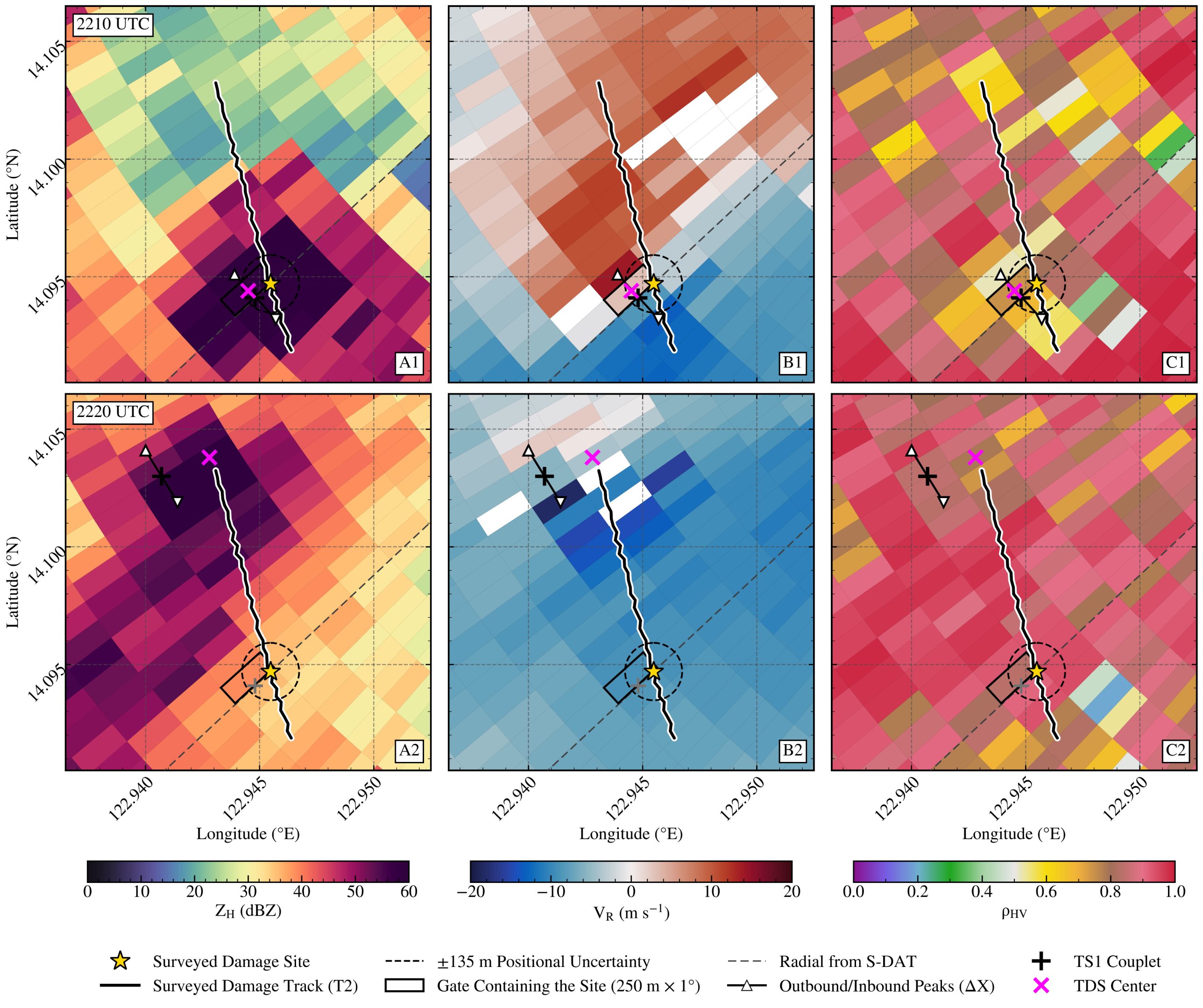}
\caption{S-DAT 1.5\dg\ elevation scans (Sweep 1) of TS1 at 2210 UTC (A1--C1) and 2220 UTC (A2--C2), showing; (a) Reflectivity (\ZH; dBZ), (b) Radial Velocity (\VR; \ms), and (c) cross-correlation ratio (\rhv). The surveyed damage track of the Magang tornado (T2) is shown as the solid white-edged line. The gold star marks the surveyed damage site in Magang, Daet and the dashed circle around it marks the $\pm$135 m combined positional uncertainty (range gate, azimuthal sampling, antenna pointing and survey georeferencing). The outlined box is the radar gate containing the damage site, and the dashed gray line is the radial from S-DAT through the site.}
\label{fig_16}
\end{figure*}

The evolution of the polarimetric TDS is examined as the tornado affects the town of Magang, Daet. For each time and elevation angle, data is saved within a 1 km radius of the center of the TVS. No \ZH\ threshold is imposed to include areas of debris within the weak-echo hole or in areas of adjacent fallout \citep[e.g.,][]{Griffin2020}. Each profile is constructed over the $\sim$1 km $\times$ 1 km box enclosing the debris signature and extracting the value at every elevation angle. After repeated sampling of the same range gate is removed, each level is computed from 56--59 distinct gates at 2210 UTC and 62--64 at 2220 UTC, approximately uniform between the lowest sampled level (45--52 m ARL) and the 0.9 km top of the plotted profiles.

Prior to examining its vertical evolution, the TDS was first located relative to the surveyed damage of the Magang tornado (Fig.~\ref{fig_16}). At 2210 UTC, the TS1 couplet at the 1.5\dg\ elevation scan ($\sim$150 mARL) was centered only 80 m from the surveyed T2 damage track and 99 m from the damage site documented in Figs.~\ref{fig_3}d--f and \ref{fig_4}a--f, while the TDS center lay 102 m from the track (Fig.~\ref{fig_16}a1--c1). These offsets fall within a single radar resolution volume at this range ($\sim$5.5 km), where the beam is $\sim$160 m wide, the azimuthal spacing is $\sim$95 m and the gate length is 250 m, giving a combined positional uncertainty of $\sim$135 m. Moreover, the offset lies almost entirely along the radial rather than across it, as expected where the range gate dominates the error. The gate directly above the damage site recorded \ZH\ $>$ 68 dBZ, \ZDR\ $\sim$ 0.26 dB and \rhv\ $<$ 0.80, with near-zero \VR\ marking the vortex center between the inbound and outbound peaks, thus satisfying the TDS criteria of \citet{Ryzhkov2005}. These criteria were also met at the 2.4\dg\ and 3.4\dg\ elevation scans, indicating that the debris column directly above the damage extended to at least $\sim$340 mARL. By 2220 UTC (Fig.~\ref{fig_16}a2--c2), the couplet had moved $\sim$1.1 km north-northwest along the damage track, and the TDS center lay only 69 m from its northern end. Meanwhile, the gate above the damage site had decreased to \ZH\ $<$ 40 dBZ with no appreciable rotation. Thus, the radar signatures lie along the surveyed path at both ends of the analysis period, independently supporting the attribution of the T2 damage to the TS1 vortex and the TDS to debris lofted from it.

\begin{figure*}[!t]
\includegraphics[width=\textwidth]{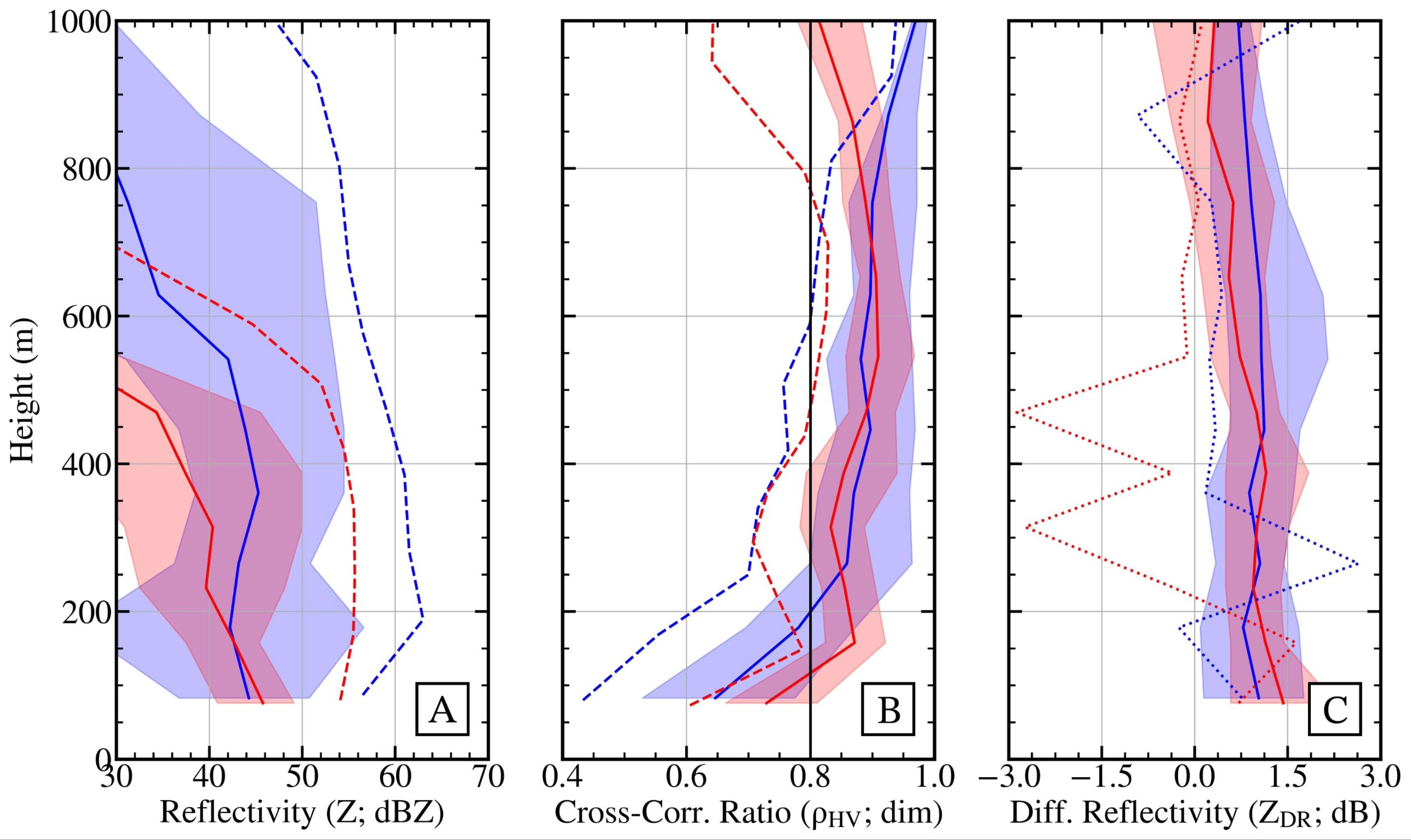}
\caption{Angle-average vertical profiles of polarimetric variables in the TDS/TVS of TS1 (Magang Event) between 0610 LST (2210 UTC; Blue profile) and 0620 LST (2220 UTC; Red profile). (a) Reflectivity (dBZ), (b) Cross-Correlation Ratio (dimensionless), (c) Differential Reflectivity (dB). For panels (a) and (b), the solid line corresponds to the mean profile with 25th and 75th percentile contours, while the dashed lines are the 90th percentile (for \ZH) and 10th percentile (for \rhv). Meanwhile, for panel (c), the dotted lines represent the skewness of the Differential Reflectivity profile.}
\label{fig_17}
\end{figure*}

In Fig.~\ref{fig_17}a and \ref{fig_17}b, the change between volumes of polarimetric signatures reveals a decrease in both the mean and 90th percentile of \ZH, accompanied by an increase in the mean and 10th percentile of \rhv, from the onset to the dissipation stage of the debris ball signal after 2220 UTC. These trends suggest that even prior to clear manifestation of TDS/TVS, there is already a discernible increase in \ZH\ and corresponding decrease in \rhv. This behavior is consistent with previously documented polarimetric characteristics of TDS in tornadic storms \citep[e.g.,][]{Wakimoto2025} and aligns with expected scattering responses to enhanced debris concentration and size variability \citep[e.g.,][]{Cross2023}. During the tornadogenesis phase (2210--2220 UTC), both \ZH\ and \rhv\ exhibit pronounced vertical gradients, with \ZH\ decreasing and \rhv\ increasing with height. This vertical structure is characteristic of debris-laden vortices, where larger and more numerous scatterers are concentrated near the surface. However, by 2220 UTC, a slight reduction in \ZH\ alongside an increase in \rhv\ becomes evident, indicating the initial weakening of the TDS and the subsequent dissipation of the Magang tornado. Quantitatively, the mean \ZH\ $>$ 40 dBZ at 2210 UTC extended up to $\sim$550 mARL, but by 2220 UTC, this height decreased to below 250 mARL. Concurrently, the mean \rhv\ increased from $\sim$0.46 at the onset to $\sim$0.61 at 2220 UTC. Similarly, the 10th percentile of \rhv\ (0.362 $<$ 0.80) reached heights of up to 600 mARL at 2210 UTC, but decreased to $\sim$450 mARL by 2220 UTC. It is possible that increased low-level debris loading associated with the vortex contributed to this enhancement in low-level debris concentrations. This is similar to \citet{Griffin2020} who found that low-level debris loading may temporarily result in a vertical gradient in polarimetric variables, and the vertical gradient decreases with time if debris lofting is sustained to create a more homogeneous vertical distribution of debris. However, unlike from the previous authors cited, no subvortices could be identified here, given the limited documentation of the Magang tornado beyond its damage extent.

The vertical structure of the \ZDR\ field exhibits behavior consistent with the observed debris orientations. Mean \ZDR\ (Fig.~\ref{fig_17}c) is positive and above 1 dB during the times where large debris are evident, and this profile never changed. However, the 25th percentile of \ZDR\ profile by 2210 UTC is closer to zero, than it is in 2220 UTC. Since the mean \ZDR\ is positive, this is indicative that the TDS is mostly composed of randomly oriented debris and/or hydrometeors that have an offsetting influence on the negative \ZDR. However, since there is a relatively small amount of large, airborne debris, negative or positive \ZDR\ associated with visible debris may not heavily influence the mean \ZDR\ in the TDS. Furthermore, TDS can be affected by precipitation entrainment. \citet{Bluestein2007a} observed that precipitation entrainment into the tornado may cause positive \ZDR\ values, despite the low \rhv, and it can be the same for the Magang Tornado and its associated TDS. Still, local regions of positive or negative \ZDR\ could skew the profile distribution.

To examine the skewness of the \ZDR\ distribution, the Pearson's moment coefficient of skewness is calculated to examine a possible positive or negative skewness of the \ZDR\ distribution during periods of debris alignment (dotted lines; Fig.~\ref{fig_17}c). The \ZDR\ exhibits some negative skewness in the lowest 200 mARL for most of the period. This is consistent with the low altitudes of the debris that are associated with negative \ZDR\ shown in Fig.~\ref{fig_12}a6. Between 200--400 mARL, \ZDR\ exhibits positive skewness, even $>$ 400 mARL by 2220 UTC just within the dissipation phase. \citet{Wakimoto2025} asserted that positive skewness would be expected if the debris field contains some large, horizontally oriented scatterers, and these are evident from the lofted roof structures or trees at this time (Fig.~\ref{fig_4}). Strong vertical motions in the vortex maintain an orientation of the debris that tends to be perpendicular to the wind, consistent with simulations by \citet{Umeyama2018} that found some degree of debris alignment perpendicular to the wind direction (e.g., more horizontal alignment in areas of strong vertical motions). In summary, combining the visible evidence and statistical analysis of the TDS, some degree of common debris alignment appears to occur locally within subregions of the TDS where large debris are present. However, the lack of mean negative \ZDR\ suggests that these scatterers are not present everywhere, at least in this case and may have been affected by precipitation entrainment as described by \citet{Bluestein2007a}, therefore yielding positive \ZDR\ that lean to small Rayleigh scatterers instead.


\section{SUMMARY AND CONCLUSION}\label{sec4}

Compared with other tornadic events and severe weather events documented in the Philippines over the past decade \citep{Capuli2024a,Capuli2026}, the 13 September 2025 Camarines Norte outbreak is distinguished less by the occurrence of multiple tornadoes than by the quality of the observational record available for it: a well-defined radar signatures of tornadic circulations, including tornadic debris signatures, combined with a ground damage survey and satellite-derived damage tracks. To our knowledge, this is the first Philippine tornado outbreak for which this combination of evidence exists. We emphasize, however, that the national record of tornado occurrence remains incomplete, and that comparably intense or more impactful events may have occurred without being documented. Thus, the present case should be understood as unusually well observed rather than as demonstrably unprecedented.

As in other tornado case studies, this study has several limitations. First, the NDVI-based damage assessment is constrained by the 55-day separation between the pre- and post-event PlanetScope scenes. Any process acting within that interval, whether convective rainfall and non-tornadic wind from subsequent storms, harvesting or other agricultural activity, and normal phenological change in rapidly growing tropical crops, can produce NDVI reductions unrelated to the tornado, whereas regrowth over the same interval can mask genuine damage. Differences in illumination and viewing geometry, atmospheric correction, and radiometric calibration among individual PlanetScope sensors introduce further variability in absolute NDVI that is not attributable to surface change, as can residual haze, cloud, and cloud shadow, which are common in tropical scenes. In addition, the 3-m pixel size is comparable to the width of the narrower damage paths, so many affected pixels are mixed, and the derived path widths are uncertain by at least one pixel on each edge. Therefore, the NDVI change is used as corroborating evidence, consistent in location and orientation with the radar-inferred tracks, rather than as independent confirmation of tornado occurrence. This interpretation follows the cautions of \citet{Kingfield2017}, who showed that the detectability of tornado damage in optical imagery depends strongly on land cover.

For the S-DAT, rotational velocities reported here should be interpreted as lower bounds on the true tornadic wind speed. Radar observations recover only the radial component of the wind, so the full horizontal vector could not be retrieved without a second radar. At the ranges of the documented tornadoes, the radar sampling volume is comparable to or larger than the circulations themselves, and azimuthal averaging within the beam systematically reduces the measured velocity difference across a couplet relative to its true value \citep{Wood1997}. Also, the lowest usable elevation angle samples well above the surface, so the reported velocities do not correspond to the near-ground winds responsible for the observed damage. In addition, the dual-PRF velocity fields required processor-error correction \citep{Alford2022} and dealiasing \citep{Louf2020} before analysis. Although these procedures remove the large majority of erroneous velocities, residual errors near strong velocity gradients cannot be fully excluded, and their effect on the retrieved shear could be of either sign. Within 25 km of the radar, however, fewer than ten gates required processor-error correction, and the TS1 and TS2 values at 3.4\dg\ quoted in Section~\ref{sec341} were unchanged when isolated residual errors were masked, indicating that the couplet analysis is not materially affected by processing artifacts. Independently of data processing, radar-derived estimates of tornado intensity are systematically biased low, because the radar resolution volume and azimuthal sampling interval smooth and under-sample the peak velocities of a vortex whose core is comparable to or smaller than the beam \citep{Wood1997}. This bias is reduced, but not removed, by the short range of the couplet ($\sim$5--6 km), where the 1\dg\ azimuthal sampling spans only $\sim$0.08--0.12 km. Radar-based intensity estimates are therefore best interpreted as lower bounds.

Another key limitation was the absence of photographic documentation for several tornadoes analyzed in this study. A ground damage survey was completed only for the Magang tornado. The Cahabaan and Napilihan tornadoes were not surveyed, and no photographic documentation of either exists; only the Lag-on and Magang events had visual evidence. Their identification and intensity assignment therefore rest on radar signatures and remote sensing rather than on inspected damage. Because the IF Scale is by construction a damage-based scale, ratings derived in this way are not strictly IF-compliant and should be regarded as lower bounds informed by storm-scale evidence rather than as DI/DoD-derived determinations. This limitation arose from accessibility constraints and insufficient personnel, which prevented full coverage of all identified tracks. Even in regions such as the United States, where storm damage surveys are routinely conducted, local National Weather Service offices often require several days to document long-track tornadoes; in the Philippine context these constraints are amplified. Strengthening citizen science initiatives, encouraging public reporting, and improving public awareness of proper severe weather documentation would substantially augment official datasets and reduce reliance on remotely sensed inference.

Finally, the absence of high-resolution numerical modeling, such as simulations using the Weather Research and Forecasting (WRF) model or Cloud Model 1 (CM1) could have provided deeper insight into the complex processes governing the tornadic environment. Instead, the analysis relied on available datasets, particularly ERA5 reanalysis, to characterize the large-scale atmospheric conditions. While ERA5 offers valuable meteorological context, its spatial and temporal resolution is insufficient to explicitly resolve storm-scale convection and finer mesoscale features that cloud-resolving models like WRF or CM1 can capture. For instance, \citet{Scheffknecht2017} employed high-resolution modeling to examine the life cycle of a supercell on the northern flank of the Alpine mountain range, demonstrating how the interaction between southwesterly synoptic flow and thermally driven plain-to-mountain circulations created a favorable environment for supercell initiation. Such detailed dynamical processes remain beyond the capability of ERA5. Nevertheless, reanalysis datasets remain robust tools for representing broader synoptic and mesoscale environments. In this study, ERA5 proved essential for characterizing the pre-, during-, and post-storm conditions of the event (around 2210 UTC / 0610 LST), consistent with other severe weather case studies that utilize similar datasets \citep[e.g.,][]{Oliveira2022,Lopes2025}. Future work would benefit from incorporating high-resolution numerical simulations (or even LES modeling) to better resolve tornadic storm dynamics and to more thoroughly assess the role of topography in modulating storm structure and intensity.

\begin{figure*}[!t]
\centering
\includegraphics[width=\textwidth]{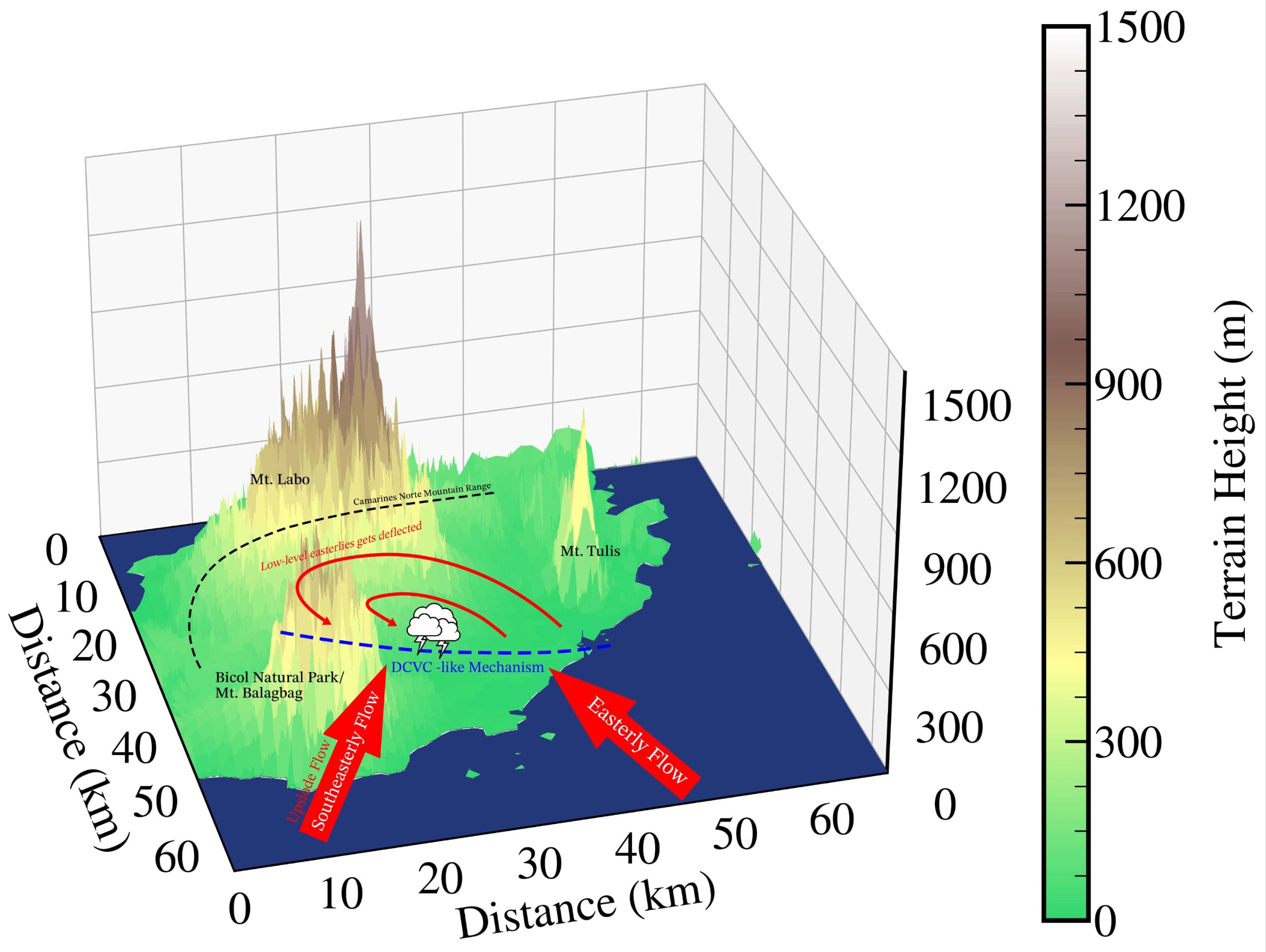}
\caption{Conceptual schematic diagram demonstrating the development of tornadic supercells along Daet, Camarines Norte. Southeasterly wind flow becomes upslope then easterly winds bends backward by the higher terrain forming a vorticity convergence zone similar to DCVZ (blue dashed line), hence with thunderstorm icon.}
\label{fig_18}
\end{figure*}

This study documents a rare early-morning tornado outbreak in the Bicol Region of the Philippines, associated with an easterly wave (inverted trough) embedded within a moist tropical environment. Synoptic-scale analysis reveals that, despite the absence of strong mid-level forcing, a persistent southeasterly flow regime and a subtle mid-level perturbation supported sustained ascent and moisture transport. The tilted structure of the tropical wave, evident from 850--700 hPa fields, enhanced low-level convergence and vorticity, establishing a favorable background for convective initiation. At the mesoscale, the environment was characterized by moderate instability (CAPE $\sim$ 2000 \Jkg), low LCLs, minimal convective inhibition, and sufficient low-level streamwise vorticity. Although low-level shear remained modest compared to classical tornadic environments, the presence of curved hodographs and enhanced storm-relative helicity supported supercell development. Importantly, terrain-induced effects; particularly flow interaction with the envelope of the elevated terrain in Camarines Norte, may have enhanced low-level convergence and vorticity through mechanisms analogous to a terrain-anchored convergence--vorticity zone and reminiscent of the DCVZ. This goes to show that the mesoscale environment can be supportive of supercell development, but tornadogenesis is unlikely to follow unless local modifications to the low-level wind field are present to increase the vorticity and convergence at spatial scales similar to that of the supercell. Topographic configurations can provide the local enhancements of wind shear and static stability that are necessary for tornadogenesis \citep{Bosart2006}. A schematic diagram of the terrain-induced enhancement through the vorticity convergence zone for this event is displayed in Figure~\ref{fig_18}.

Radar and satellite observations confirmed the presence of multiple embedded supercells within a larger MCS. Satellite analysis revealed intense convective characteristics, including OTs, AACPs, extremely cold cloud-top temperatures ($<$ $-$90\,\dg C), and intense flash rates, confirming vigorous updrafts capable of supporting severe weather. Dual-polarimetric S-DAT radar identified three supercells (TS1, TS2, TS3), each of which produced a tornado. Two additional vortices developed within non-supercellular convection, giving five vortices in total. TS1 (Magang) exhibited a well-defined TDS, velocity couplet/TVS, and polarimetric signatures including \ZDR\ arcs, \KDP\ columns, and a DRC feature, consistent with classical tornadic supercell structure. TS2 (Cahabaan) displayed similar but slightly weaker signatures, including a BWER and debris signatures possibly indicative of lighter, vegetative material. However, because these rely on observations from a single radar, sampled at a scale comparable to the tornado core and the beamwidth, the Cahabaan tornado is described as identified rather than confirmed. The evolution of polarimetric variables in the `debris ball' of Magang tornado highlighted vertical gradients in \ZH\ and \rhv, consistent with debris lofting and subsequent weakening trend. Based on combined damage assessment and radar-derived signatures, the Magang tornado was rated IF2.5 (approx. EF3-equivalent). Both the Cahabaan and Napilihan tornadoes were rated IF1 (approx. EF1-equivalent) identified primarily through radar analysis\footnote{For the Napilihan tornado (TS3), no path length was derived, as the circulation was brief and seen only in the 2220 UTC volume scan, not in successive scans. The radar presentation was factored in; thus, the IF rating is a provisional lower bound.}. In contrast, the Lag-on tornado and a potential landspout detected via S-DAT were rated IF0 (approx. EF0-equivalent) as these were likely brief and passed over vegetation and light structures without damaging them.

Therefore, this study sought to emphasize and encourage post-storm damage assessments in the Philippines, particularly for severe weather events (SWEs). Environmental reanalysis, high-resolution satellite imagery (beyond the geostationary imagery used operationally), and polarimetric radar have become essential complements to ground-level damage surveys, providing crucial information in sparsely populated areas and over forest and agricultural land. The limitations encountered in this analysis, however, point to several concrete and immediately actionable measures.

First, we recommend that DOST-PAGASA establish a rapid-response damage survey team, equipped with UAVs and staffed by personnel formally trained in IF-Scale application, activated whenever a tornadic vortex signature or debris signature is detected operationally or a credible public report is received, and deployed within 24--48 hours. The time constraint is not administrative but physical. However, in tropical agricultural landscapes, the vegetative damage that constitutes the primary evidence is erased within days by regrowth, harvesting, or subsequent convective rainfall. Training should be formalized rather than incidental, covering the DI/DoD inventory and its correct application, the distinction between tornadic, downburst, and gust-front damage patterns, UAV survey planning and photogrammetric processing, and the circumstances under which an event must be left unrated, with periodic recertification and cross-checking between surveyors so that ratings remain consistent between events and between assessors. Each deployment should then produce a standardized survey report within 72 hours containing a georeferenced damage-path polygon, UAV photogrammetry of convergent debris patterns, DI/DoD assignments with supporting imagery, and an explicit statement of confidence including, where warranted, a determination that the event is unrated for lack of damage indicators.

Furthermore, DOST-PAGASA may evaluate the deployment of X-band polarimetric radars as gap-filling and research assets in regions of demonstrated tornado occurrence. Their finer spatial resolution and rapid update cycle (order of 1--2 minutes) would resolve low-level velocity couplets and debris signatures that the operational network can only sample marginally; the associated attenuation is manageable at the short ranges over which such systems are deployed, and can be corrected using established dual-polarization methods. Because no observed proximity sounding existed within 30--50 km of this outbreak, supplementary radiosonde ascents from the nearest station during severe thunderstorm watches would also allow reanalysis-derived parameters to be used with quantified rather than assumed confidence.

Finally, local government units and disaster risk reduction and management offices are typically the first and often the only observers of these events, yet no pathway currently exists for their observations to reach a scientific archive. A low-barrier reporting mechanism; a geotagged photograph, time, and location submitted by mobile phone, combined with brief annual orientation on distinguishing tornado, downburst, and flash-flood damage, would convert this dispersed local knowledge into usable record. Reports should be ingested into a national severe weather event database under a defined confirmation workflow, with unconfirmed reports retained and revisited rather than discarded. Thus, implemented together, these measures would transform tornado documentation in the Philippines from opportunistic case study into routine practice. The number of confirmed reports would rise substantially, intensity estimates would rest on defensible evidence, and the resulting record would support a national tornado climatology of sufficient quality to inform warning policy.

\acknowledgments

The author is sincerely grateful to the residents of Camarines Norte who reported and documented the tornadoes in their communities, and whose photographs and accounts made much of this analysis possible. Thanks are also due to Dr. Funing Li (Massachusetts Institute of Technology) and Dr. John T. Allen (Central Michigan University) for valuable guidance on the use of \textit{xcape}, and to the anonymous reviewers and the editor, whose careful and constructive comments substantially improved the manuscript. This work received no external funding and was carried out on the author's own time, dedicated to severe convective storm research in the Philippines. Finally, the author thanks family and loved ones for their unwavering support throughout this research. CGHA is supported by Planet Labs PBC (Planet) through the Education and Research Program.

\contribution

\textbf{Generich H. Capuli:} Writing - original draft, Writing - review \& editing, Visualization, Methodology, Investigation, Formal Analysis, Data curation, Conceptualization, Supervision.

\datastatement

Data used in this paper were derived from the ERA5 reanalysis (openly available through the \href{https://cds.climate.copernicus.eu/}{Climate Data Store}), and the individual full-disk HIMAWARI-9 bands were accessed through the \href{https://thredds.nci.org.au/thredds/catalog/catalogs/ra22/satellite-products/arc/obs/himawari-ahi/himawari-ahi.html}{THREDDS Catalog}. PlanetScope imagery was obtained from Planet Labs PBC under the Education and Research Program. The 2023 Philippines Administrative Level 0--4 shapefiles are available in the Humanitarian Data Exchange (\href{https://data.humdata.org/dataset/cod-ab-phl}{HDX}). The author's associated 1D vertical profiles of standard ERA5 model sounding data are archived through Project SWAP and its \href{https://doi.org/10.5281/zenodo.19178931}{DR4}. The Digital Elevation Model (DEM) is the Copernicus GLO-30 DEM, distributed through \href{https://doi.org/10.5069/G9028PQB}{OpenTopography}. Lightning data from the PLDN and radar data from S-DAT are operational holdings of DOST-PAGASA and are not publicly redistributable; they may be requested from DOST-PAGASA through METTS (radar), and the corresponding author will facilitate such requests on reasonable inquiry. Proper attribution is required for these datasets.
\\
\\
This paper has made use of the following Python packages: \verb|Cartopy|, \verb|GeoPandas|, \verb|Matplotlib|, \verb|MetPy|, \verb|NumPy|, \verb|Pandas|, \verb|Rasterio|, \verb|rioxarray|, \verb|SciPy|, \verb|SounderPy| v3.1.0 \citep{Gillett2025}, \verb|Py-ART| \citep{Helmus2016}, \verb|OmniCloudMask| \citep{WRIGHT2024}, \verb|xarray|, and \verb|xcape| \citep{lepore2021}. Velocity dealiasing used \verb|UNRAVEL| \citep{Louf2020}.

\interest

The author declares no competing interests.

\appendix
\appendixtitle{Supplementary Analysis}

The Supplementary Damage Analysis related to the Magang Tornado Event (including Table S1 and Figs. S1--S16) and the \VR\ correction statistics are provided in the supplementary file of this manuscript.

\bibliographystyle{ametsocV6}
\bibliography{references}

\end{document}